\pdfoutput=1
\documentclass[article,preprint,superscriptaddress]{revtex4-1}
\usepackage{graphicx}
\usepackage{dcolumn}
\usepackage{amssymb}
\usepackage{amsmath}
\usepackage{bm}
\usepackage{pifont}
\usepackage{epsfig}
\usepackage{psfrag}
\usepackage[usenames]{xcolor}

\begin{document}

\title{Deconstructing the glass transition through critical experiments on colloids}

\author{Shreyas Gokhale$^*,$\footnote{Present address: Physics of Living Systems group, MIT, 400 Technology Square, NE46-629, Cambridge, Massachusetts 02139, USA}} 
\affiliation{Department of Physics, Indian Institute of Science, Bangalore 560012, India}
\author{A. K. Sood$^*$}
\affiliation{Department of Physics, Indian Institute of Science, Bangalore 560012, India}
\affiliation{International Centre for Materials Science, Jawaharlal Nehru Centre for Advanced Scientific Research, Jakkur, Bangalore 560064, India}
\author{Rajesh Ganapathy$^*,$}
\affiliation{International Centre for Materials Science, Jawaharlal Nehru Centre for Advanced Scientific Research, Jakkur, Bangalore 560064, India}
\affiliation{Sheikh Saqr Laboratory, Jawaharlal Nehru Centre for Advanced Scientific Research
Jakkur, Bangalore, 560064, INDIA}

\begin{abstract}
The glass transition is the most enduring grand-challenge problem in contemporary condensed matter physics. Here, we review the contribution of colloid experiments to our understanding of this problem. First, we briefly outline the success of colloidal systems in yielding microscopic insights into a wide range of condensed matter phenomena. In the context of the glass transition, we demonstrate their utility in revealing the nature of spatial and temporal dynamical heterogeneity. We then discuss the evidence from colloid experiments in favor of various theories of glass formation that has accumulated over the last two decades. In the next section, we expound on the recent paradigm shift in colloid experiments from an exploratory approach to a critical one aimed at distinguishing between predictions of competing frameworks. We demonstrate how this critical approach is aided by the discovery of novel dynamical crossovers within the range accessible to colloid experiments. We also highlight the impact of alternate routes to glass formation such as random pinning, trajectory space phase transitions and replica coupling on current and future research on the glass transition. We conclude our review by listing some key open challenges in glass physics such as the comparison of growing static lengthscales and the preparation of ultrastable glasses, that can be addressed using colloid experiments.
\end{abstract}

\maketitle

\section*{Keywords}
glass transition; colloids; microscopy; holographic optical tweezers; random first-order transition theory; dynamical facilitation; mode coupling theory; geometric frustration; dynamical heterogeneity; Stokes-Einstein relation; ellipsoids; crossovers; random pinning; replica coupling; trajectory space phase transitions; ultrastable glasses

\section*{Corresponding Authors}
\noindent Shreyas Gokhale\\
Email: gokhales@mit.edu\\
\noindent A. K. Sood\\
Email: asood@physics.iisc.ernet.in\\
\noindent Rajesh Ganapathy\\
Email: rajeshg@jncasr.ac.in\\

\tableofcontents

\section{Introduction}
It is rather ironic that despite our considerable prowess in manipulating glass and harnessing it to suit our needs, the basic physics underlying its formation still eludes our grasp \cite{angell1995old,debenedetti2001supercooled,ngai2007glass,berthier2011theoretical}. The question that places the nature of the glass transition among the greatest unsolved problems in condensed matter physics is deceptively simple: Why are glasses mechanically similar to solids in spite of being structurally similar to liquids? Unlike the thermodynamic transition from liquid to crystal, where symmetry breaking naturally leads to the onset of rigidity, the amorphous arrangement of atoms within a glass is decidedly liquid-like, which makes it difficult to apply conventional statistical mechanical tools to understand its formation. For instance, unlike in liquid-crystal transitions, the glass transition is accompanied by negligible changes in the structure factor, which makes it difficult to identify the order parameter, or even determine whether one exists. It has long been recognized that the nature of the glass phase can be better understood by studying the behavior of the liquid phase on approaching the glass transition. If a substance is cooled sufficiently rapidly from the liquid state below the freezing point so as to bypass crystallization, it enters a metastable `supercooled' regime, where it exists as an equilibrium ergodic liquid. On reducing the temperature further, this supercooled liquid becomes increasingly viscous. Moreover, this increase in viscosity, or relaxation time, becomes increasingly rapid with decreasing temperature until the liquid eventually falls out of equilibrium to form a non-ergodic glass. An enormous body of research over the past fifty years has been devoted to explaining this precipitous increase in the relaxation time, in the hope of solving the glass transition problem. 

The reason why the glass transition problem still remains unsolved is the fact that the growing relaxation time makes it impossible to equilibrate a supercooled liquid arbitrarily close to the glass transition \cite{cavagna2009supercooled}. Nonetheless, experiments on atomic and molecular liquids have generated a wealth of information on the variation of quantities such as viscosity \cite{angell1984relaxation,angell1988perspective,angell1991relaxation} and specific heat \cite{birge1985specific,birge1986specific} on approaching the glass transition over an enormous range of relaxation times. For instance, nuclear magnetic resonance (NMR) experiments have furnished relaxation time data over fourteen orders of magnitude for prototypical glass-forming liquids like ortho-terphenyl \cite{chang1994translational}. These data have in turn spurred the development of various competing theoretical frameworks aimed at explaining the apparent divergence in relaxation time. One might think that the availability of data over this immensely broad range should be sufficient to identify the correct theoretical scenario for glass formation. Unfortunately, distinct theoretical formulations that differ significantly in terms of their underlying physics fit the available data equally well. It is important to note that while experiments on atomic and molecular liquids can probe ensemble averaged quantities over an extensive dynamical range, they lack the resolution to detect subtle changes in local structure and dynamics that accompany glass formation. It is therefore evident that as far as identifying the correct theory of glass formation is concerned, the devil is most certainly in the details.   

In contrast to atomic experiments, the ability to probe the structure and dynamics of glass-formers in real space with single-particle resolution is the hallmark of numerical simulations and experiments on dense colloidal suspensions \cite{allen1989computer,hunter2012physics,lu2013colloidal}. As a consequence, simulations, in particular those based on molecular dynamics (MD), as well as colloid experiments are far better suited to testing the microscopic predictions of various theories of glass formation. Moreover, MD simulations and colloid experiments are intrinsically complimentary approaches. Colloids provide a real world test-bed for numerical predictions, whereas MD simulations can easily apply theoretical constructs that are prohibitively difficult, or even impossible to realize in colloid experiments. Unfortunately, both numerical simulations and colloid experiments can access a far more limited dynamical range, typically the first 5-6 decades in relaxation time from the high temperature liquid side, compared to molecular experiments, which can cover more than fourteen.

From the discussion in the preceding paragraphs, it would appear that we have reached an impasse. On one hand, it is clear that relaxation time data alone are insufficient to identify the correct theoretical scenario for glass formation. On the other hand, real space approaches that are capable of identifying subtle structural and dynamical changes en route to forming glass are limited to temperatures or volume fractions that seem to be too far away from the glass transition to provide a faithful picture of relaxation close to it. These limitations of experiments and simulations make it extremely difficult to invalidate particular theories. A number of distinct formulations have garnered numerical as well as experimental support over the limited dynamical range available to MD simulations and colloid experiments. Naturally, this has led to the formation of various schools of thought that advocate one or the other theoretical framework. The literature is replete with partisan review articles that present vociferous arguments in favor of one school of thought or another. This slew of contradictory and often obscurely technical arguments makes it extremely difficult for colloid experimentalists to identify key questions that can be addressed using the tools and techniques at their disposal. Indeed, the prevailing ambivalence has led many experimentalists to sidestep the canonical glass transition problem in favor of a more exploratory approach aimed at studying the influence of factors such as particle shape, interactions and confinement on glass formation. As a result, it is not clear in what way colloid experiments could potentially contribute towards a deeper understanding of the glass transition.   

The central purpose of this review is to demonstrate that colloid experiments can indeed provide useful insights into glass formation through a critical comparative assessment of competing theoretical scenarios. This approach highlights the pivotal role of crossovers in the dynamics of glass-forming liquids that have been identified both in experiments and simulations. Since these dynamical crossovers are presumably associated with changes in the dominant mechanism of structural relaxation, they may help us identify dynamical regimes over which particular theoretical frameworks are valid. The ultimate hope is that an extensive exploration of these dynamical crossovers would allow us to unambiguously invalidate certain theoretical scenarios, thereby narrowing down the search for the correct theory of glass formation. The present review has a twofold purpose. It aims to outline the key challenges in uncovering the physics of glass formation to colloid experimentalists and more importantly, how these challenges can be tackled using available experimental tools. The review is also aimed at providing theoreticians and computational physicists with a clear understanding of the potential of colloid experiments, so as to stimulate them to develop new testable predictions. 

The glass transition problem has a long and colorful history that has been chronicled in numerous comprehensive, informative and insightful review articles \cite{berthier2011theoretical,debenedetti2001supercooled,cavagna2009supercooled,langer2014theories,angell1988perspective,angell1995formation}. Choosing the subject matter for the present article therefore poses a daunting challenge. Our choice of material is guided by our focus on the potential of critical colloid experiments, with a special emphasis on dynamical crossovers as means to distinguish between competing theories. As such, this review can be viewed as a bridge between theoretical reviews, which focus on technical aspects of various frameworks, and experimental ones, which 
tend to highlight measurement techniques. On the experimental front, therefore, we have chosen to highlight the results, rather than the techniques employed. Another aspect of our review is the complementarity between simulations and colloid experiments and important numerical results therefore feature prominently in this review. However, this review does not aim to provide a comprehensive account of the contribution of simulations in understanding the glass transition problem and we have therefore restricted the discussion on simulations only to those works that have been or are likely to be of relevance to colloid experiments. Our treatment of various theoretical scenarios also follows this principle. We have limited our discussion to those theoretical frameworks whose predictions have either already been tested or can be tested readily in colloid experiments in the near future. Even within the theoretical approaches that we have discussed, our perspective is that of colloid experimentalists and consequently, we have only focussed on those theoretical aspects that are testable using colloid experiments. Wherever applicable, we have augmented the discussion on theoretical ideas with landmark simulations that enabled these ideas to be tested in experiments. Finally, we note that our aim is to enhance the rapport between experimentalists and theoreticians so that we may proceed together in a definitive and directed manner towards a complete understanding of the glass transition problem. 

Keeping these guiding principles in mind, we have organized the rest of the review as follows. We begin section 2 with a brief discussion on the salient features that make colloidal suspensions useful model systems to study atomistic phenomena in general and glass formation in particular. We then provide a concise account of the phenomenology of glass formation and introduce various quantities that are central to glass physics. In section 3, we discuss the experimental evidence supporting various competing theories of glass-formation. In particular, we focus on the mode coupling theory (MCT), the random first-order transition theory (RFOT), dynamical facilitation (DF) and geometric frustration-based approaches. It will become clear in this section that apart from MCT, which in its idealized version is known to fail beyond a certain temperature or volume fraction, existing data support many aspects of RFOT, DF as well as frustration-based models. The need for new approaches that can distinguish between these competing theories will thereby become apparent. In section 4, we present an overview of various dynamical crossovers associated with relaxation time and dynamical heterogeneity and demonstrate how these crossovers may be employed as tools to distinguish between predictions of competing theories. In particular, we review recent experiments on colloidal suspensions that have successfully exploited the presence of a dynamical crossover to ascertain the relative importance of distinct relaxation mechanisms on approaching the glass transition. In section 5, we emphasize the importance of alternate ways of approaching the glass transition in determining the correct theoretical scenario for vitrification. We outline various opportunities and challenges in realizing these approaches using colloid experiments. In section 6, we discuss promising research avenues for future colloid experiments. Finally, we present our conclusions in section 7. 

\section{Preliminaries}
\subsection{Colloidal suspensions as model atomic systems}
Colloidal suspensions, along with gels, polymers, emulsions and liquid crystals belong to a family of materials that are collectively labelled as soft matter \cite{jones2002soft, israelachvili2010intermolecular}. Compositionally, colloidal suspensions consist of particles whose size ranges from a few nanometers to a few microns, dispersed in a solvent. A hallmark feature of these systems is that they exhibit Brownian motion. Colloids are crucial ingredients in numerous technological applications \cite{cosgrove2010colloid} such as ink-jet printing, e-book readers, protective coatings, paints and photonic band gap materials \cite{singh2010inkjet, kim2011self, vlasov2001chip}. Interestingly, colloids are not only important to industry, but also of immense value to pure science, where they serve as model systems to shed light on a wide array of atomistic phenomena \cite{sood1991structural,poon2004colloids,terentjev2015oxford}. The chief reason why colloids are good mimics of atomic systems is that they are thermalized by Brownian motion, and their statistical mechanical properties are therefore analogous to those of atomistic materials. The advantage of colloidal systems is that unlike atoms, the dynamics of colloids can be probed in real time with single-particle resolution due to their large size and slow dynamics, which allows one to establish the link between macroscopic behavior and the microscopic processes that give rise to it. Crucially, due to the presence of Brownian motion, thermally activated events can be readily characterized using colloidal systems, which provides substantial insights into statistical mechanical phenomena occurring in analogous atomic systems. Yet another important feature is that colloidal systems exhibit various phases of matter such as crystals, liquids and glasses, which make them versatile model systems that can elucidate a broad class of condensed matter physics problems ranging from nucleation and growth to the glass transition \cite{lu2013colloidal}. 
\begin{figure}
\centering
  \includegraphics[width=0.75\textwidth]{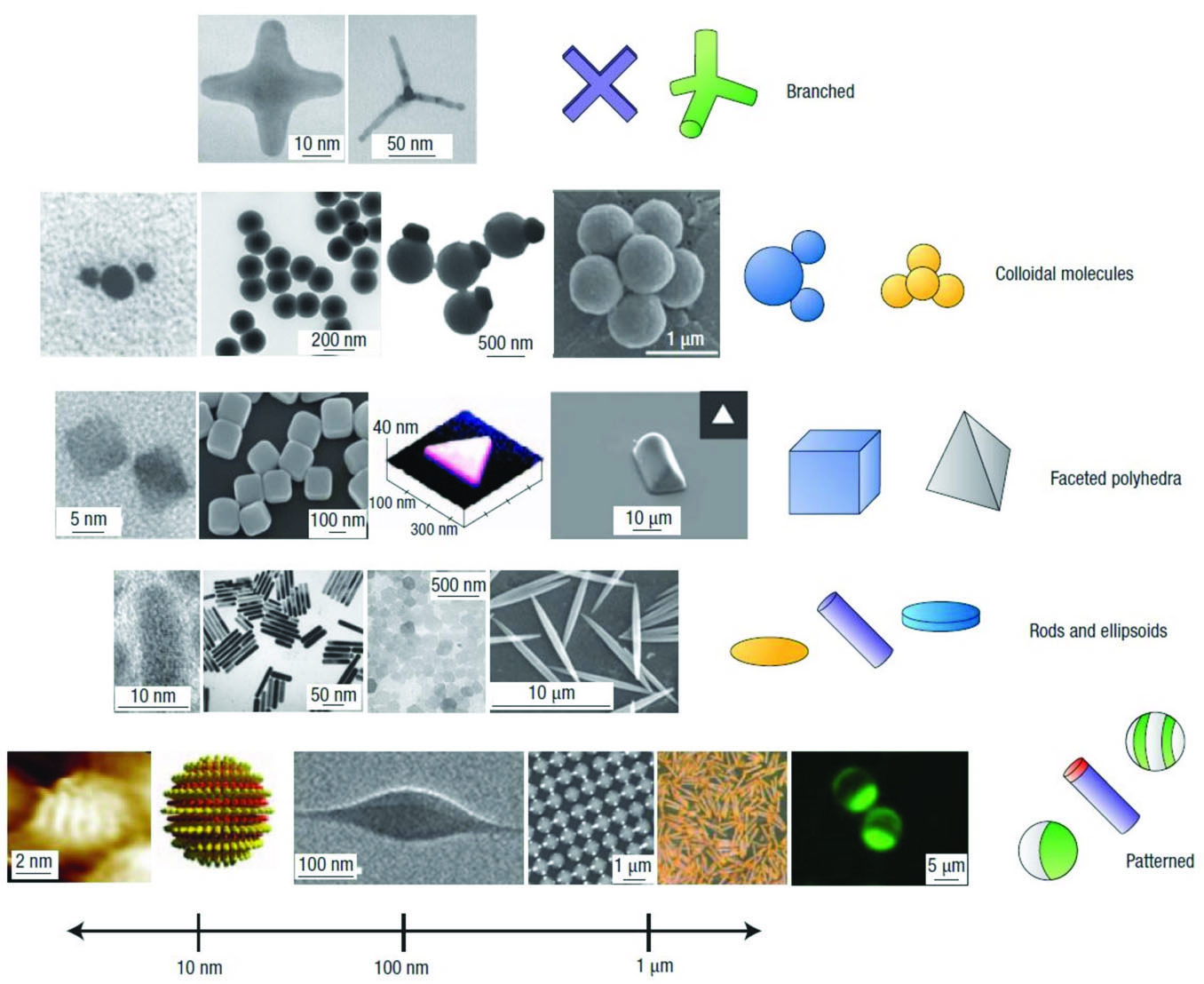}
  \caption{Representative examples of colloids with anisotropic shapes and interactions. Particles are classified in rows based on the nature of anisotropy. The top four rows show examples of particles with anisotropic shapes whereas the bottom row illustrates particles with anisotropic interactions. From left to right, the bottom row shows striped spheres, biphasic rods, patchy spheres with valence, gold-platinum composite nanorods and Janus spheres. Adapted from \cite{glotzer2007anisotropy}.}
  \label{Figure1}
\end{figure}

\subsubsection{Inter-particle interactions and phase diagrams}
Early studies that employed colloids to model atomic phenomena made use of suspensions of simple spherical particles with isotropic short-ranged repulsive interactions. However, owing to rapid advances in colloidal chemistry, scientists have established numerous protocols that enable us to tune the size, shape and interactions of particles with tremendous precision \cite{glotzer2007anisotropy, sacanna2011shape, li2011colloidal} (See Fig. \ref{Figure1} for examples). Over the last decade or so, researchers have synthesized particles with anisotropic shapes such as ellipsoids \cite{snoeks2000colloidal}, cubes \cite{rossi2011cubic}, dimples \cite{sacanna2010lock}, rods \cite{van2004synthesis}, clusters of spheres \cite{meng2010free} and many more \cite{johnson2005synthesis, kraft2008self, hong2008clusters, ahmadi1996shape, greyson2006tetrahedral, zhang2005decoration}. Enormous strides have also been made in controlling inter-particle interactions. In particular, recent studies that focus on imparting specific directional interactions have created a lot of excitement in the field. Typically, directionality is introduced either by creating `patches' on the surface of spherical particles \cite{pawar2010fabrication, hong2006clusters, 
pawar2009multifunctional, kraft2009colloidal, cayre2003fabrication} or by coating particles with DNA and exploiting its sequence specific bonding properties \cite{huo2006asymmetric, xu2006asymmetric}. In a recent dramatic illustration of this type, Pine and co-workers \cite{wang2012colloids} have used DNA patches to synthesize colloids that possess valences similar to those of hybridised atomic orbitals. In the process, they have realized colloid analogues of the molecules methane and ethylene. Studies like these show that in the coming years, colloids will not only be considered as helpful models for atomic systems, but will also serve as useful mimics of molecules as regards mechanical and thermal properties in the classical regime. In particular, they will play a major role in elucidating the physics of complex phenomena such as liquid-liquid phase transitions \cite{smallenburg2014erasing} and protein crystallization \cite{fusco2013crystallization}. 

Despite these significant advances in particle synthesis, generating large scale self-assembled structures with long range order and non-trivial symmetries using complex colloids continues to be a challenge. Nevertheless, suspensions comprising of even the simplest isotropic spherical colloids have helped provide valuable insights into several atomic phenomena such as nucleation and growth \cite{van2003colloidal}, pre-melting at crystal defects \cite{alsayed2005premelting}, super heating \cite{wang2012imaging}, crystal-crystal transitions \cite{peng2015two}, epitaxy \cite{ganapathy2010direct} and friction \cite{bohlein2012observation}. Typically, colloidal particles interact via short-ranged repulsive interactions that arise either due to charged ions in the suspension or from steric repulsions between polymeric brushes grafted on the particles' surface. In the case of electrostatic interactions, the long ranged Coulomb repulsions are usually screened by counter-ions present in the solvent. The phenomenology of these systems is very similar to the simpler, but instructive case of hard spheres. The hard sphere (HS) potential is of the following form 
\begin{eqnarray}
U_{HS} & = &\infty \quad \text{if} \quad 0 < r < \sigma\\
U_{HS} & =& 0 \quad \text{if} \quad r > \sigma 
\label{HS}
\end{eqnarray} 
Here, $r$ is the distance between the colloids and $\sigma$ is the diameter of the particles. Although this potential is somewhat idealized, it is used extensively to explore various phenomena such as glass formation and jamming \cite{parisi2010mean}. Inter-atomic interactions typically contain a long-ranged attractive part in addition to a short-ranged repulsive part and are better described by forms such as the Lennard-Jones potential, given by 
\begin{equation}
U_{LJ}(r) = 4\epsilon \Bigg [ \Bigg ( \frac{\sigma}{r} \Bigg )^{12} - \Bigg ( \frac{\sigma}{r} \Bigg )^{6} \Bigg ]
\label{LJ}
\end{equation} 
where $r$ is the inter-particle distance, $\epsilon$ is the depth of the potential well and $\sigma$ is the distance at which the potential first crosses zero. While the HS potential provides a reasonable approximation of the short-ranged repulsive part, i.e the behavior of $U_{LJ}$ for $r \leq \sigma$ is reasonably well captured by $U_{HS}$, it completely ignores long-ranged attractions. However, in a seminal work, Weeks, Chandler and Andersen have shown that the equilibrium structure of liquids is determined purely by the short-ranged repulsive part of the interaction potential and the attractive interactions only lead to a mean field contribution that becomes smaller with increasing density \cite{weeks1971role}. It is therefore not surprising that HS systems provide a satisfactory description of high density phases such as atomic liquids crystals and glasses. An important distinction is that in atomic systems, the phase diagram is governed by two thermodynamic variables, for example temperature and pressure, and phase transitions typically emerge from a competition between energy and entropy. For hard spheres, the phase diagram is controlled entirely by entropy, which in turn depends only on the volume fraction 
\begin{equation}
\phi = \frac{\pi\sigma^3N}{6V}
\end{equation}
where $\sigma$ is the particle diameter, $N$ is the number of particles and $V$ is the volume.  
\begin{figure}
\centering
  \includegraphics[width=0.8\textwidth]{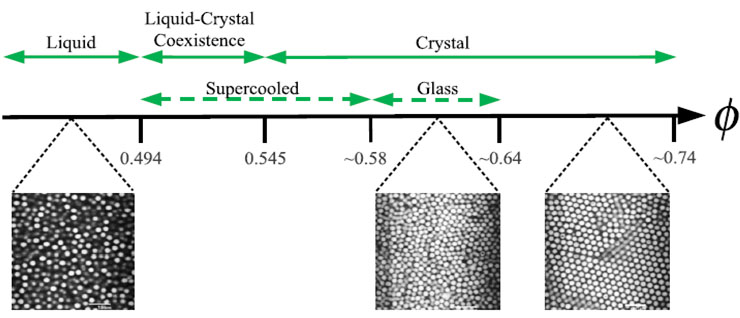}
  \caption{Phase Diagram of a system interacting via hard sphere repulsions. The snapshots are confocal images corresponding to liquid, glass and crystal phases. Adapted from \cite{hunter2012physics}. For images of Bragg diffraction from 3D colloidal crystals, see \cite{pusey1986phase,pusey2009hard}.}
  \label{Figure2}
\end{figure}
Interestingly, despite the simplistic interaction potential, hard spheres exhibit extremely rich phase behavior (See Fig. \ref{Figure2} for an illustration). For $\phi <$ 0.495, the system is in the fluid phase. For 0.494 $< \phi <$ 0.545, there is a coexistence between the fluid and a body centred cubic (bcc) crystalline phase, whereas beyond $\phi =$ 0.545, the equilibrium phase is purely crystalline. Interestingly, at $\phi \sim$ 0.58, there is a disordering transition and the system exists as an amorphous solid, or glass, till $\phi \sim$ 0.64. Finally, beyond $\phi =$ 0.64, the system reverts to a crystalline phase with cubic closed packed symmetry. This hard sphere phase diagram was first realized in experiments by Pusey and van Megen \cite{pusey1986phase}, who performed experiments on nearly HS-like PMMA colloids that interact via extremely short-ranged steric repulsions. Since then, PMMA as well as electrostatically stabilized colloids like silica, polystyrene and poly N-isopropylacrylamide (PNIPAm) have been used extensively to model various atomic phenomena \cite{weeks2000three,kegel2000direct,gasser2001real,han2008geometric,ganapathy2010direct,bohlein2012observation,gokhale2013grain,yunker2014physics}. 

Although the phase diagram of hard spheres is rich, it is unable to capture many features that are commonly observed in atomic systems. Perhaps the most stark of these, is the liquid-gas transition, which is absent in HS systems, due to the lack of attractive interactions. However, attractions can be easily introduced in colloidal systems and this is usually achieved by adding small non-adsorbing polymers to the colloidal suspension. The addition of these polymers induces an effective attraction between the colloidal particles that is purely entropic in nature, and is known as the depletion interaction. The effective interaction potential was first computed theoretically by Asakura and Oosawa \cite{asakura2004interaction}, and takes the form
\begin{eqnarray}
U_{AO}(r) &=& -\frac{\pi k_BTN_d}{12V}[2(\sigma + \sigma_d)^3 - 3(\sigma + \sigma_d)^2r + r^3], \quad \sigma \leq r \leq \sigma + \sigma_d \nonumber 
\\
&=& 0, \quad \sigma + \sigma_d < r
\end{eqnarray}
Here, $r$ is the distance between centres of two colloidal particles of diameter $\sigma$, $\sigma_d$ is the diameter of the non-adsorbing polymer, known as the depletant, and $N_d$ is the number of polymer molecules. It is evident from this form that the range of interaction is equal to $\sigma_d$ and the strength of interaction is proportional to the concentration of the depletant. Physically, when the distance between the surfaces of two colloidal particles becomes smaller than $\sigma_d$, polymer molecules can no longer enter the region between the two particles. This exclusion, or `depletion' of molecules from the space between two colloids creates an osmotic pressure difference across the surface of the colloids. The force arising from this pressure difference drives the particles closer to each other, effectively creating an attractive interaction between them. In typical experimental situations, the depletant molecules are much smaller in size compared to the colloids, i.e. $\sigma_d << \sigma$, and the AO potential is therefore much shorter in range than attractive interactions between atoms. Nonetheless, depletion interactions have a profound impact on the phase behavior of hard spheres. They give rise to equilibrium phenomena such as phase separation \cite{sanyal1992phase}, liquid-gas transitions and critical points for $\sigma_d/\sigma \leq$ 0.3 \cite{calderon1993experimental} as well as three phase coexistence and crystal-gas phase transitions \cite{ilett1995phase,poon2002physics,anderson2002insights}. Further, non-equilibrium phenomena such as glass-glass transitions \cite{voigtmann2011multiple} and gelation \cite{wilson1995gelation,dinsmore2002direct} can also occur. Interestingly, depletion interactions are also known to induce reentrant glass transitions \cite{pham2002multiple,mishra2013two}, which we shall discuss in more detail in the course of this review. 

\subsubsection{Summary of condensed matter problems addressed using colloids}
Colloidal systems have been used as models to understand a wide variety of phenomena in crystals. After the discovery of a transition from fluid to crystalline solid in simulations on hard sphere systems by Wood and Jacobson  \cite{wood1957preliminary} and Alder and Wainwright \cite{alder1957phase} in 1957, the experimental evidence for the same was provided almost three decades later by using colloidal systems \cite{pusey1986phase, pusey1989structure}. Following this seminal work, several studies realized complex self assembled structures \cite{velikov2002layer, shevchenko2006structural, van2003colloidal, dziomkina2005colloidal}. For instance, by using a mixture of positively and negatively charged particles van Blaaderen and co-workers have realized self-assembled structures analogous to NaCl and CsCl crystals \cite{leunissen2005ionic}. Even more complicated structures, such as the kagome lattice have also been realized experimentally by Chen et.al. using tri-block Janus particles, i.e. particles with three patches \cite{chen2011directed}. In addition, it has been shown that crystals with different lattice constants and crystal symmetries can be grown \cite{ramsteiner2010stiffness, savage2013entropy} using micropatterend templated surfaces in combination with sedimentation \cite{vogel2012soft, yin2001template, dziomkina2005colloidal}. Physical phenomena associated with the formation of these crystals, such as epitaxial growth \cite{ganapathy2010direct}, vibrational properties \cite{kaya2010normal} and solid-solid phase transitions \cite{peng2015two} have also been studied. In addition, non-equilibrium phenomena such as shear-induced melting and crystallization of colloidal suspensions have been explored extensively \cite{ackerson1981shear,lindsay1985shear,lahiri1994shear,imhof1994shear,wu2009melting}. 

Studies that used colloids to model atomistic phenomena gained impetus with the rapid development in real space imaging techniques such as confocal microscopy \cite{wilson1990confocal,chestnut1997confocal,dinsmore2001three,prasad2007confocal}. Concomitantly, the development of particle tracking algorithms by Crocker and Grier \cite{crocker1996methods} enabled the generation of particle trajectories, from which crucial dynamical information could be extracted. Pioneering work using colloids to investigate microscopic processes in crystals was carried out by Weitz and co-workers. In one study, the authors provided direct real space visualization of crystal nucleation and growth \cite{gasser2001real}. Real space visualization of dislocation nucleation and their subsequent dynamics has also been achieved \cite{schall2004visualization, schall2006visualizing, suresh2006crystal} using a combination of laser diffraction microscopy \cite{schall2009laser} and confocal microscopy. In one of these ingenious experiments \cite{schall2006visualizing}, the authors indented a colloidal crystal with a sewing needle, essentially performing the colloid analogue of nano-indentation experiments on atomic crystals, to study the formation of dislocations, quantify their nucleation rate and map the associated strain field. Remarkably, these works have provided evidence that the defect dynamics in these systems can be well described by the continuum approach which is used to describe defect dynamics in atomic crystals \cite{cotterell1953dislocation, frank1950report}. These studies have set the stage for using colloids to probe phenomena that are prohibitively difficult to study in atomic experiments.

While the foregoing studies have largely been restricted to single crystals, colloids have also been used extensively to explore a variety of equilibrium and non-equilibrium phenomena in polycrystalline materials \cite{gokhale2013grain}. The first systematic study of grain growth was performed by Palberg et al \cite{palberg1995grain}. Scientists have developed numerous protocols to control the average grain size in a colloidal polycrystal. These include changing the inter particle interactions \cite{palberg1995grain}, applying external electric fields or shear deformations \cite{palberg1992continuous, palberg1995grain, gokhale2012directional}, adding impurities \cite{de2005colloidal,de2009grain,yoshizawa2011exclusion,ghofraniha2012grain} and by changing the cooling rate in a thermoresponsive colloidal system \cite{gokhale2012directional}. The melting of crystals has been studied extensively using colloids. Yodh and co-workers  provided the first experimental evidence for pre-melting at dislocations and grain boundaries \cite{alsayed2005premelting}. It has also been shown that the dynamics of particles at grain boundaries share remarkable similarities with those of glass-forming liquids \cite{nagamanasa2011confined}. Colloids offer the advantage that global phenomena such as grain growth and melting can be studied in real time with single-particle resolution. For instance, grain boundary properties such as stiffness and mobility have been extracted from equilibrium \cite{skinner2010grain} as well as shear-induced non-equilibrium \cite{gokhale2012directional} grain boundary fluctuations  using the capillary fluctuation method \cite{foiles2006computation, trautt2005direct}.

Colloid experiments have also made valuable contributions to the physics of amorphous systems. Experiments on sheared colloidal glasses provided the first direct evidence for shear transformation zones \cite{schall2007structural}. Various aspects of amorphous solids such as the density of states \cite{ghosh2010density,chen2010low}, ageing \cite{yunker2009irreversible}, yielding \cite{nagamanasa2014experimental} and shear-induced melting \cite{eisenmann2010shear} have also been investigated using colloids. In addition to all these studies, an enormous body of research on the glass transition using light scattering as well as microscopy techniques has accumulated over the years, since the observation of a glass transition \cite{pusey1986phase,pusey1987observation} in hard-sphere colloids. These studies have contributed significantly to our understanding of important aspects of glass transitions phenomenology such as heterogeneous dynamics. Moreover, several unanswered questions pertaining to glass formation can be addressed using colloids. We shall discuss the contributions of colloid experiments thus far as well as promising future directions in subsequent sections of this review.

\subsection{Glass transition phenomenology}
The iconic aspect of glass formation is the colossal increase in the liquid's viscosity $\eta$ on decreasing the temperature. Indeed, the viscosity can increase by as many as fifteen orders of magnitude, when the temperature is lowered by a mere 30\% relative to the freezing point temperature $T_m$. Conventionally, the laboratory glass transition temperature $T_g$ is defined as the temperature at which the viscosity of the liquid reaches a value of $10^{13}$ poise. This value of viscosity is so large that below $T_g$, the supercooled liquid no longer flows over experimental time scales and for all practical purposes, behaves like a solid. To better understand the similarities and differences between glass formation in different liquids, it is highly instructive to plot $\text{log}(\eta)$ as a function of $T_g/T$. 
\begin{figure}
\centering
  \includegraphics[width=0.5\textwidth]{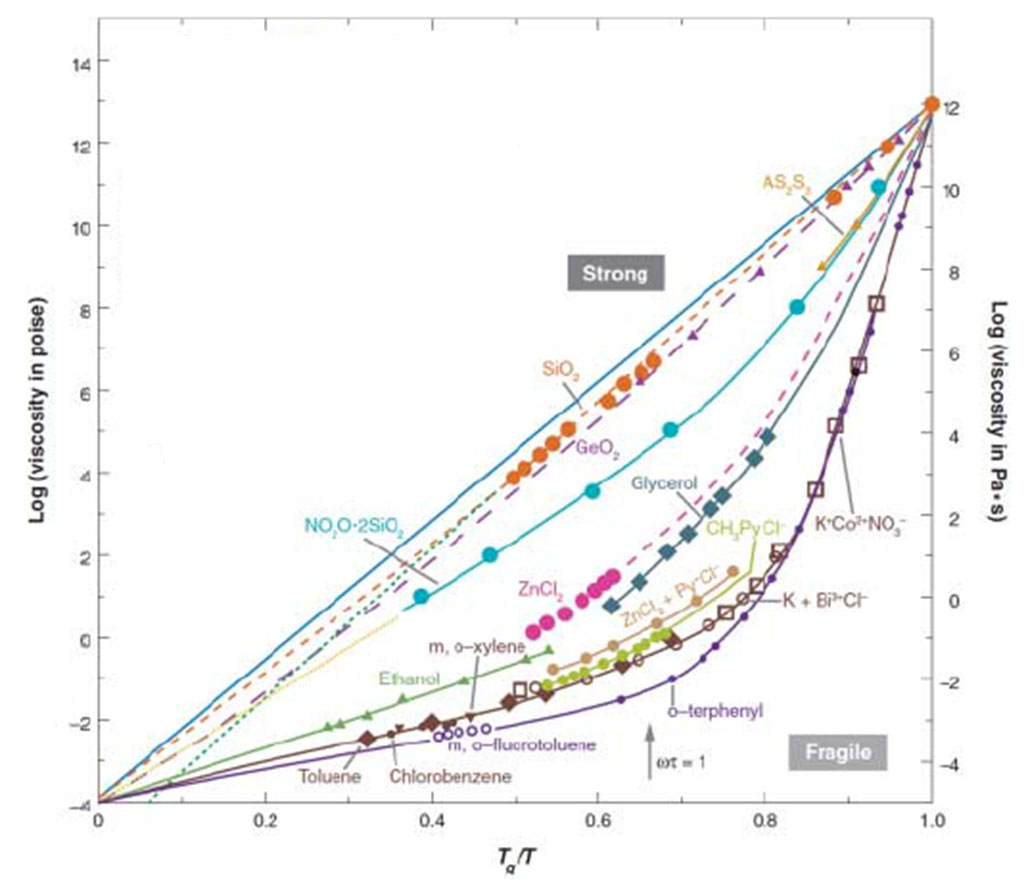}
  \caption{The Angell plot for various glass-forming liquids. Logarithm of the viscosity $\eta$ plotted as a function of $T_g/T$, where $T_g$ is the glass transition temperature. Adapted from \cite{lubchenko2015theory}.}
  \label{Figure3}
\end{figure}
This representation, first introduced by Angell \cite{angell1995formation} shows that the shape of the viscosity curves for different liquids are strikingly different. For instance, liquids like silica (SiO$_2$) exhibit a near-Arrhenius dependence of viscosity on temperature and are termed `strong'. By stark contrast, those such as ortho-terphenyl exhibit significant deviations from the Arrhenius form and are labelled `fragile'. The extent of deviation from Arrhenius dependence can be quantified by the kinetic fragility $m$, defined as the slope of the Angell plot at $T = T_g$ \cite{novikov2005correlation}, i.e.
\begin{equation}
m = \Bigg(\frac{\partial\text{log}_{10}\eta}{\partial(T_g/T)}\Bigg)_{T=T_g}
\end{equation}
The variation in viscosity with temperature shown in Fig. \ref{Figure3} is well-captured by an empirical relationship of the form $\eta = \eta_0\text{exp}(E(T)/T)$, where $E(T)$ is an effective temperature dependent activation barrier. For strong liquids, $E(T)$ is only weakly dependent on temperature, suggesting that relaxation in these liquids is primarily governed by a single activation barrier. A closer inspection reveals that the interatomic interactions in strong liquids such as Si, SiO$_2$ and GeO$_2$ are strongly directional and covalent in nature. Atomic rearrangements in these liquids therefore entail the crossing of a large microscopic activation barrier, which determines the relaxation time, and hence the viscosity. The large value of $m$ for fragile liquids on the other hand points towards an activation barrier that becomes steeper with decreasing temperature. Intuitively, therefore, one imagines that the relevant activation barrier for relaxation in fragile liquids is associated with cooperative reorganization involving a large number of constituent particles. 

A second fact evident from Fig. \ref{Figure3} is that a vast majority of glass-forming liquids can be categorized as fragile, or in other words, the effective activation barrier $E(T)$ varies significantly with temperature. Moreover, the apparent activation barrier for fragile liquids like o-terphenyl near $T_g$, as extracted from the slope of the Angell plot, can be about 500 kJ/mol \cite{ediger2000spatially}, which is significantly larger than typical bond energies in organic liquids. This suggests that collective relaxation processes are important in fragile glass-forming liquids. Not surprisingly, the bulk of the theoretical research on vitrification has aimed at uncovering the nature of the glass transition in fragile liquids. Clearly, a necessary condition for any theory of glass formation is that it should be able to describe the Angell plot quantitatively. Several such theories, motivated by distinct physical ideas have been formulated over the last few decades. In their endeavor to elucidate the physics of glass formation, and particularly to derive a unique functional form for the temperature dependence of viscosity, these theoretical approaches have identified a number of characteristic temperatures in the context of vitrification. As these temperatures are associated with striking qualitative changes in the dynamics or thermodynamics of glass-forming liquids, they provide an engaging account of the glass formation process itself. We shall therefore provide a brief discussion on the significance of characteristic temperatures in the next section.

\subsubsection{Characteristic temperatures associated with glass formation} 
Although not important from a theoretical perspective, the most obvious characteristic temperature associated with glass formation is the freezing point $T_m$ of the liquid. Normally, when cooled sufficiently slowly below $T_m$, the liquid freezes into a crystalline state, which corresponds to the thermodynamic free energy minimum. A necessary condition for glass formation is for the crystallization  process to be avoided. In atomic and molecular liquids, this is usually achieved by cooling the liquid rapidly below $T_m$. If the cooling rate is faster than the crystal nucleation rate, the liquid can avoid crystallization and enter the metastable supercooled regime. In colloidal liquids, the control variable is the volume fraction, which is analogous to inverse temperature. In colloidal suspensions therefore, a rapid reduction of temperature corresponds to a rapid increase in the volume fraction, which can be achieved through centrifugation \cite{weeks2000three}. A more frequently used approach employs a binary mixture comprising of colloidal particles of two different sizes. In this case, the disparity in particle sizes provides sufficient geometric frustration to prevent crystallization. 
\begin{figure}
\centering
  \includegraphics[width=0.45\textwidth]{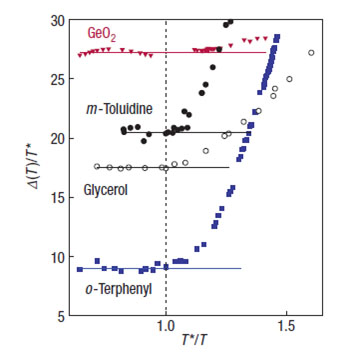}
  \caption{The effective activation barrier $E(T)$ for four molecular glass-forming liquids as a function of $T^{*}/T$, where $T^{*}$ is the high temperature onset of glassy dynamics. Ortho-Terphenyl is a fragile glass-former whereas GeO$_2$ is a relatively strong one. Adapted from \cite{kivelson2008search}.}
  \label{Figure4}
\end{figure}
A second characteristic temperature $T^{*}$ signals the high temperature onset of glassy dynamics. Experimentally, one can define this onset temperature $T^{*}$ as the temperature below which the liquid first exhibits a non-Arrhenius temperature dependence of viscosity. In other words, it is the temperature below which the effective activation barrier $E(T)$ begins to increase with temperature (Fig. \ref{Figure4}). Depending on the theoretical scenario under consideration, $T^{*}$ may or may not have thermodynamic significance. For instance, according to a thermodynamic approach based on geometric frustration \cite{tarjus2005frustration}, the onset temperature corresponds to an avoided critical point. This critical point, typically located about the freezing point, i.e. $T^{*} \geq T_m$, is associated with the onset of local order characterized by some structural motif, such as icosahedra \cite{coslovich2007understanding}. The `avoided' nature of this critical point stems from the fact that the locally preferred structural motif is incapable of tiling space. The resulting geometric frustration prevents the divergence in the correlation length typically associated with equilibrium critical phenomena, and the system organizes itself into a patchwork of domains whose size is limited by the extent of geometric frustration. 
\begin{figure}
\centering
  \includegraphics[width=0.5\textwidth]{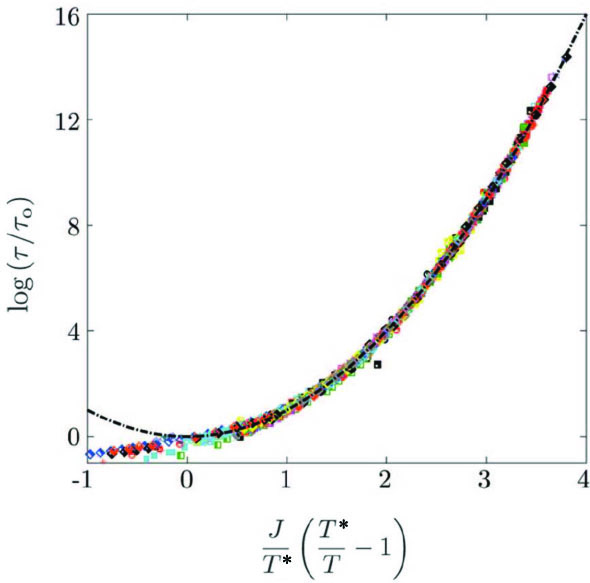}
  \caption{The temperature dependence of relaxation time for several glass-forming liquids. The fit corresponds to the parabolic form in Eqn. \ref{DFFit}. Adapted from \cite{elmatad2009corresponding}.}
  \label{Figure5}
\end{figure}
On the other hand, according to the dynamical facilitation (DF) approach, which is purely kinetic in nature, $T^{*}$ merely signals a dynamic crossover associated with the onset of caging of particles by their nearest neighbors. Within this framework, the regime $T < T^{*}$ is characterized by a separation of time scales between the short time rattling motion of particles within their cages and the rare large sporadic particle displacements associated with the escape of particles from their cages. According to the DF theory, the viscosity, or the relaxation time $\tau$, exhibits a temperature dependence of the form
\begin{equation}
\tau = \tau_0 \text{exp}\Bigg[ J^2\Bigg(\frac{1}{T}-\frac{1}{T^{*}}\Bigg)^2 \Bigg]
\label{DFFit}
\end{equation}
This form satisfactorily fits the relaxation time data for several glass-forming liquids in the regime $T < T^{*}$ (Fig. \ref{Figure5}). For $T > T^{*}$, the separation of time scales, which is central to the DF approach, is no longer present, and Eqn. \ref{DFFit} does not fit the data.  

\begin{figure}
\centering
  \includegraphics[width=0.7\textwidth]{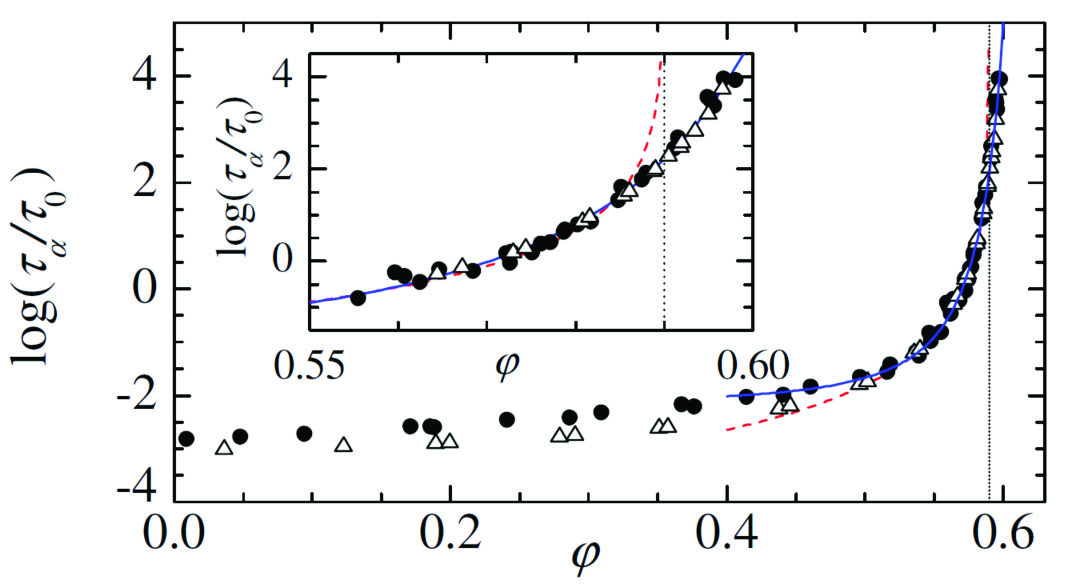}
  \caption{Variation of the relaxation time with volume fraction $\phi$. The black circles correspond to data from dynamic light scattering (DLS) experiments on poly-methylmethacrylate (PMMA) colloids of average diameter 260 nm, whereas the hollow triangles correspond to data from Monte Carlo simulations on a binary mixture of hard spheres. The dotted red curve is an MCT fit of the form $\tau = \tau_{\infty}(\phi_c - \phi)^{-\gamma}$ with $\phi_c =$0.59 and $\gamma =$2.5. The blue curve is a fit to the DLS data and has the form $\tau = \tau_{\infty}\text{exp}(A/(\phi_0 - \phi)^{\delta})$ with $\delta =$2 and $\phi_0 =$0.637. Adapted from \cite{brambilla2009probing}.}
  \label{Figure6}
\end{figure}

As the temperature is decreased further below $T^{*}$, the supercooled liquids' dynamics becomes increasingly sluggish. The motion of the liquid's constituent particles, be they atoms, molecules or colloids becomes increasingly constrained due to the caging effect of nearest neighbors. These cages become stronger with decreasing temperature, thus making it difficult for particles to escape. The mode coupling theory (MCT) of the glass transition \cite{leutheusser1984dynamical,bengtzelius1984dynamics,das1985hydrodynamic,das1990glass,gotze1999recent,das2004mode,reichman2005mode,gotze2008complex, das2011statistical} essentially employs this simple idea and predicts a dynamic glass transition at temperature $T_c$. In particular, the theory states that at $T_c$, the nearest neighbor cages constrain the particle so strongly that it becomes impossible for the particle to escape. Since this caging is experienced by all particles, it leads to a complete freezing of the liquid's dynamics, resulting in a non-ergodic glass phase below $T_c$. MCT predicts a power law divergence in the relaxation time \cite{fuchs1992primary,kob2003course,das2004mode} of the form 
\begin{equation}
\tau = \tau_0(T - T_c)^{-\gamma}
\label{MCT}
\end{equation}
Interestingly, the singularity predicted by MCT is not observed in real glass-formers \cite{berthier2011theoretical,brambilla2009probing,kob2012non}. It has been observed that the relaxation time continues to remain finite well below the MCT glass transition temperature $T_c$, or above the corresponding volume fraction $\phi_c$ (See Fig. \ref{Figure6} for an illustration of the failure of MCT in colloidal as well as simulated hard sphere systems).
This implies the presence of ergodicity restoring mechanisms that facilitate structural relaxation below $T_c$ (or above $\phi_c$). A fairly elegant thermodynamic explanation of the avoided singularity at $T_c$ as well as the restoration of ergodicity below it is given by the random first-order transition theory (RFOT). Within RFOT, the avoided MCT transition can be viewed as a dynamical crossover that signals a change in the liquid's free energy landscape \cite{berthier2011theoretical}. Above $T_c$, the free energy landscape is characterized by a single minimum corresponding to the homogeneous density profile of the high temperature liquid phase. For $T < T_c$ on the other hand, the free energy landscape is fragmented into an exponentially large number of metastable minima. In this regime, ergodicity is restored by thermally activated events that transport the liquid from one free energy minimum to another. These thermally activated events, which are absent within MCT, reduce the MCT singularity at $T_c$ to a crossover. 

The restoration of ergodicity below $T_c$ also implies that $T_g < T_c$, where $T_g$ is the laboratory glass transition temperature. While $T_g$ is extremely important from the point of view of practical applications, it is irrelevant as far as the basic physics of glass formation is concerned. Indeed, quite unlike the freezing point $T_m$, which corresponds to a thermodynamic phase transition, $T_g$ itself depends on the rate of cooling, albeit very weakly \cite{moynihan1974dependence}. Thus, $T_g$ only serves as a practical measure of how deeply one can supercool a liquid before it falls out of equilibrium. While the difference between $T_g$ and $T_c$ is quite substantial for molecular liquids, colloid experiments and numerical simulations have only recently succeeded in equilibrating the system beyond the MCT crossover.

\begin{figure}
\centering
  \includegraphics[width=0.5\textwidth]{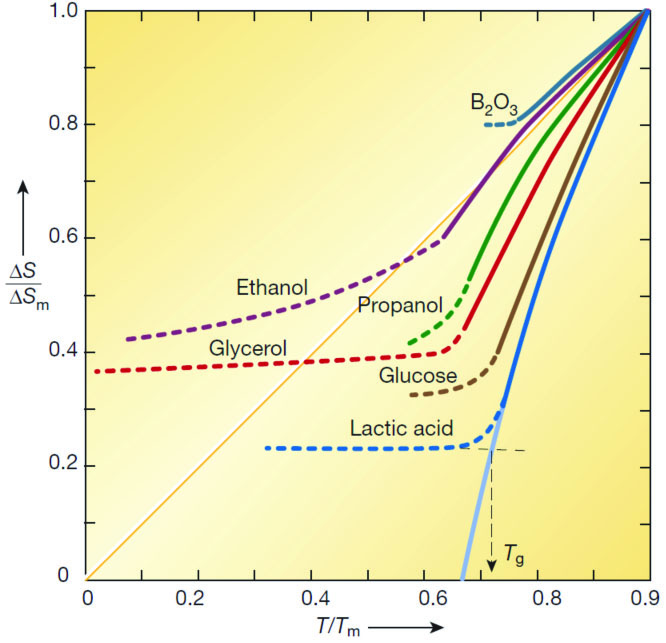}
  \caption{Difference in entropy $\Delta S$ between the supercooled liquid and the corresponding thermodynamically stable crystal scaled by the entropy of melting $\Delta S_m$ for various glass-forming liquids as a function of $T/T_m$. The solid portion of the curves represents the supercooled liquid phase and the dotted portion corresponds to the glass phase. For lactic acid, the solid light blue curve is an extrapolation of $\Delta S/\Delta S_m$ from the supercooled liquid phase and reaches zero at the Kauzmann temperature $T_K$. The intersection of the extrapolated curves for supercooled liquid and glass phases yields the laboratory glass transition temperature $T_g$. Adapted from \cite{debenedetti2001supercooled}.}
  \label{Figure7}
\end{figure}

Although $T_g$ imposes a practical limit on the degree of supercooling, two characteristic temperatures below $T_g$ are of great significance in the context of the glass transition. The first of these is motivated purely by the empirical observation that the relaxation time of glass-forming liquids can be well-described by the Vogel-Fulcher-Tammann (VFT) \cite{vogel1921law,fulcher1925analysis,tammann1926abhangigkeit} form
\begin{equation}
\eta = \eta_0\text{exp}\Bigg( \frac{DT_0}{T-T_0}\Bigg)
\label{VFT}
\end{equation}    
where $D$ is a measure of the fragility \cite{angell1976thermodynamics,angell1995formation}and $T_0$ is a characteristic temperature at which the relaxation time diverges. For all glass-forming liquids, $T_0 < T_g$, and hence, in practice, the liquid falls out of equilibrium long before $T_0$ is reached. While $T_0$ is motivated purely by experiments, the other characteristic temperature below $T_g$ has a much deeper thermodynamic significance. If one compares the temperature dependence of  the entropy of the supercooled liquid and that of the corresponding crystal, one observes that the entropy of the liquid decreases faster than that of the crystal upon cooling (Fig. \ref{Figure7}). 
In particular, if one extrapolates this difference in entropy $\Delta S$ below $T_g$, one finds that below a finite temperature $T_K$, the entropy of the liquid becomes smaller than the entropy of the crystal. Further extrapolation to $T = 0$ would then imply that the supercooled liquid has negative entropy at the absolute zero of temperature, which is in clear violation of the third law of thermodynamics. This `entropy crisis' was first articulated by Kauzmann \cite{kauzmann1948nature} and goes by the name of the Kauzmann paradox. In practice, the entropy crisis is always averted by the fact that the supercooled liquid falls out of equilibrium before the Kauzmann temperature $T_K$ is reached. In theory, various resolutions to the Kauzmann paradox have been proposed. For instance, Tanaka has argued that the supercooled liquid will crystallize before the Kauzmann temperature is attainted, i.e. the lowest temperature upto which the supercooled liquid can exist as a metastable state is higher than $T_K$ \cite{tanaka2003possible}. Stillinger \cite{stillinger1988supercooled} and Johari \cite{johari2000equilibrium} on the other hand argued that the form of $\Delta S$ below $T_g$ changes in such a way that the excess entropy of the liquid over the crystal goes smoothly to zero only at $T = 0$. 

A third resolution of the Kauzmann paradox is provided by RFOT, which claims that the supercooled liquid undergoes a thermodynamic phase transition known as an `ideal glass transition' at $T_K$. The transition is associated with the vanishing of the liquid's configurational entropy $s_c$. $s_c$ is related to the number of metastable free energy minima that the liquid can sample at a given temperature. According to RFOT, the number of these minima becomes sub-extensive below $T_K$, which leads to a vanishing of $s_c$. Alternatively, $s_c$ corresponds to the difference in the total entropy of the liquid compared to its vibrational entropy. For a deeply supercooled liquid, the vibrational entropy is comparable to that of the crystal and hence, it is reasonable to assume that $s_c \approx \Delta S$, although anharmonic effects can cause discrepancies in the two quantities \cite{johari2000contributions,johari2002entropy}. 
\begin{figure}
\centering
  \includegraphics[width=0.5\textwidth]{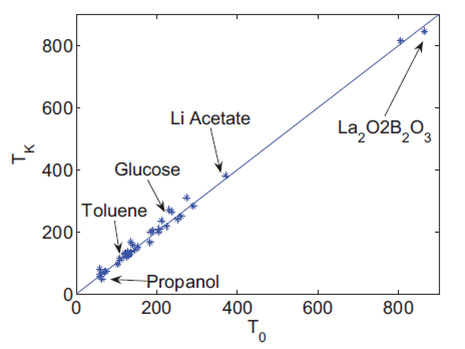}
  \caption{Correlation between the Kauzmann temperature $T_K$ obtained by extrapolating $s_c$ and the VFT temperature $T_0$ obtained from fits to the viscosity data of several glass-forming liquids. Adapted from \cite{lubchenko2015theory}.}
  \label{Figure8}
\end{figure}
More strikingly, RFOT predicts that the temperature dependence of the relaxation time obeys a VFT form, and the VFT temperature $T_0$ can therefore be identified with $T_K$. Indeed, direct measurements of these two temperatures for several glass-forming liquid certainly build a favorable case for RFOT (Fig. \ref{Figure8}). Nonetheless, as argued in \cite{hecksher2008little}, several distinct functional forms appear to fit available viscosity and relaxation time data equally well over the dynamical range accessible to present day experiments (See Fig. \ref{Figure9} for a comparison of fits to relaxation time data corresponding to functional forms given in Eqns. \ref{DFFit} (DF) and \ref{VFT} (RFOT)). One must therefore move beyond fits to macroscopic ensemble averaged quantities in order to ascertain the validity of various theoretical scenarios. 
\begin{figure}
\centering
  \includegraphics[width=0.4\textwidth]{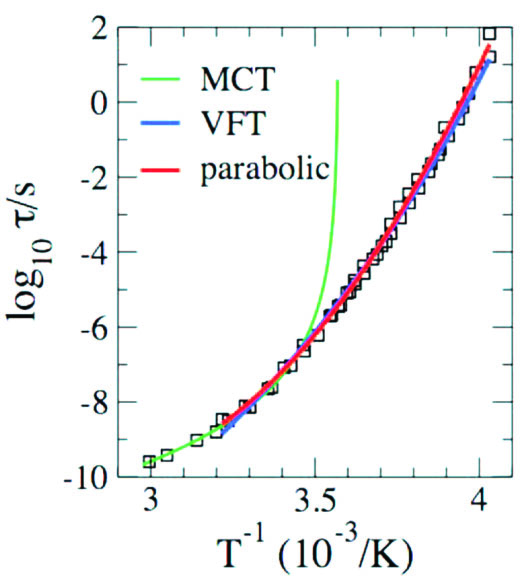}
  \caption{Comparison of parabolic (Eqn. \ref{DFFit}), VFT (Eqn. \ref{VFT}) and MCT (Eqn. \ref{MCT}) fits to relaxation time data for ortho-terphenyl. Adapted from \cite{biroli2013perspective}.}
  \label{Figure9}
\end{figure}

The foregoing discussion has outlined the significance of various characteristic temperatures associated with glass formation. These temperatures demarcate different regimes of glass formation and encourage us to examine glassy dynamics in these regimes in the context of competing theoretical scenarios. Further, by clarifying the limits of applicability of experiments and simulations, they can serve as useful pointers for future research. It is evident that experiments on atomic and molecular liquids can probe glassy dynamics over a broad range corresponding to $T^{*} \geq T \geq T_g$. Unfortunately, it is also evident that this range is not sufficiently broad to distinguish between the predictions of competing theories (Fig. \ref{Figure9}). Numerical simulations and colloid experiments on the other hand can most easily probe a dynamical range that corresponds to $T^{*} \geq T \geq T_c$, although a few forays into the $T<T_c$ regime have also been made \cite{brambilla2009probing,nagamanasa2015direct}. On the theoretical side, different approaches advocate the predominance of different relaxation mechanism over various ranges in temperature. For instance, the DF theory claims that dynamical facilitation is the dominant relaxation mechanism for $T \leq T^{*}$, and further claims that no thermodynamic phase transition at $T_K$ exists. RFOT on the other hand claims that the $T^{*} \geq T \geq T_c$ regime is satisfactorily explained by MCT, whereas cooperative activated events govern the $T_c \geq T \geq T_K$ regime. Thus, even within the regime accessible to colloid experiments, different theories differ significantly in terms of their advocated microscopic relaxation mechanisms. In principle therefore, a critical comparative analysis of these processes using the microscopic real space approach championed by colloid experiments is a promising way forward in determining the relative accuracy of various theoretical predictions. However, as we shall see in the next section, distinct theoretical formulations have garnered considerable experimental as well as numerical support, even at the level of microscopic structure and dynamics. Establishing the validity of one formulation over another is therefore fraught with challenges. Fortunately, a possible solution has appeared in the form of the recent discovery of a number of dynamical crossovers within the range $T^{*} \geq T \geq T_c$ in simulations \cite{kob2012non,flenner2014universal,hocky2014crossovers} as well as experiments \cite{nagamanasa2015direct,gokhale2014growing,mishra2015shape}. Elucidating the physical implications of these crossover and the nature of their relationship to the avoided MCT transition at $T_c$ therefore form an integral part of subsequent sections of this review article. 

\subsubsection{Dynamical heterogeneity}
\begin{figure}
\centering
  \includegraphics[width=0.5\textwidth]{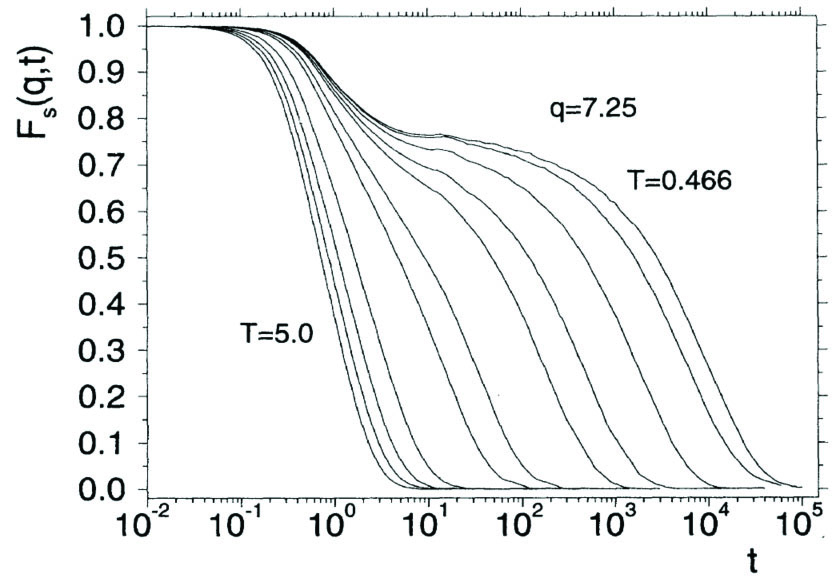}
  \caption{Self-intermediate scattering function $F_s(q,t)$ for a simulated binary glass-former composed of particles of type A and B interacting via the Lennard Jones potential. Data shown are for particles of type A only. $F_s(q,t)$ has been evaluated at $q = $7.25, which corresponds to the first peak of the static structure factor. Adapted from \cite{kob1995testing}.}
  \label{Figure10}
\end{figure}

\begin{figure}
\centering
  \includegraphics[width=0.5\textwidth]{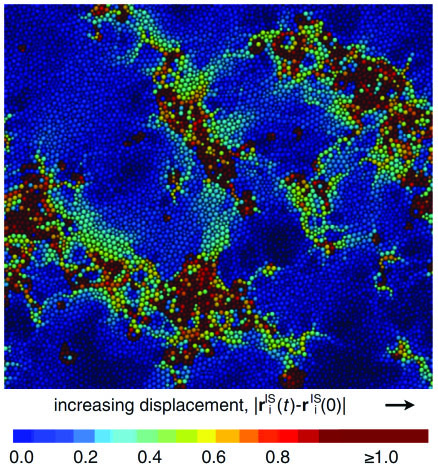}
  \caption{Dynamical heterogeneity in a simulated glass-former. The colorbar represents displacement in terms of energy minimized inherent structure coordinates. Blue indicates overlap with initial positions whereas maroon indicates a displacement of at least one particle diameter, as indicated in the key below the image. The displacements have been plotted over a time interval of $\tau_{\alpha}/$10. Adapted from \cite{keys2011excitations}.}
  \label{Figure11}
\end{figure}

\begin{figure}
\centering
  \includegraphics[width=0.5\textwidth]{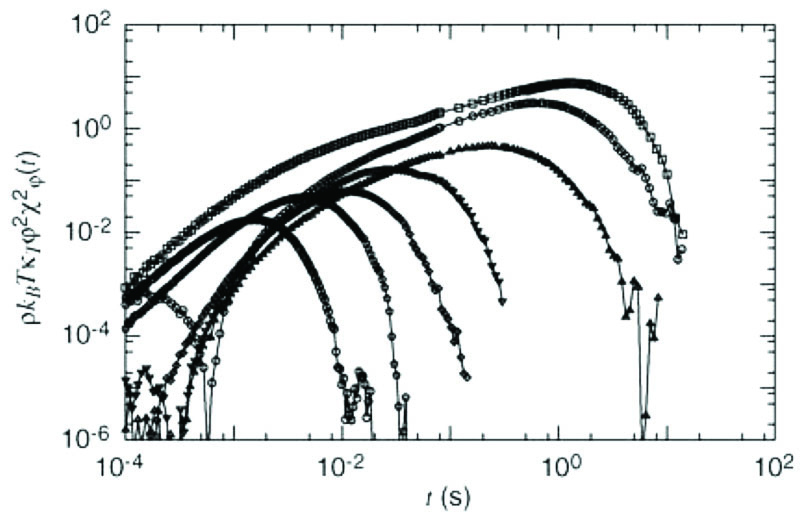}
  \caption{Dynamic four point susceptibility $\chi_{4}(t)$ from DLS experiments for a hard sphere-like colloidal glass-former for various volume fractions $\varphi$. $\varphi$ varies from 0.18 to 0.5. The data shown correspond to the approximation $\chi_4(t) = \rho k_BT \kappa_T \varphi^2 \chi_{\varphi}^2(t)$, where $\rho$ is the density and $\kappa_T$ is the isothermal compressibility. The approximation becomes increasingly accurate with increasing $\varphi$. Adapted from \cite{berthier2005direct}.}
  \label{Figure12}
\end{figure}

The non-Arrhenius temperature dependence of the viscosity of glass-forming liquids intuitively suggests the presence of multiple relaxation times. This intuition is further vindicated by the shape of relaxation functions. Structural relaxation in glass-forming liquids is often characterized by the self-intermediate scattering function $F_s(q,t)$, defined as  
\begin{equation}
F_{s}(q,t)= \Bigg \langle \frac{1}{N} \sum_{i=1}^N \exp \lbrace i\mathbf{q}. \left[ \mathbf{r}_{i}(t+t_{o}) - \mathbf{r}_{i}(t_{o}) \right] \rbrace \Bigg \rangle
\end{equation}
Here, $\mathbf{r}_{i}(t)$ is the position of the $i^{th}$ particle at time $t$ and the angular brackets indicate ensemble averaging. $F_{s}(q,t)$ probes structural relaxation over a length scale corresponding to the inverse of the magnitude of the wave vector $\mathbf{q}$. The typical choice for $q$ is $q_{\text{max}}$, which corresponds to the first peak of the static structure faction $S(q)$. Fig. \ref{Figure10} shows $F_s(q,t)$ data for a simulated binary glass-former for various temperatures. 
We see that at high temperatures ($T>1$), $F_s(q,t)$ decays to zero exponentially. The temperature values quoted are in units of the Lennard-Jones energy parameter \cite{kob1995testing}. This signifies that in the high temperature liquid phase, the decay of $F_s(q,t)$ is characterized by a single microscopic timescale associated with the rearrangement of individual particles. At lower temperatures ($T<1$) on the other hand, the shape of $F_s(q,t)$ changes in two important ways. First, a plateau begins to develop at finite values of $F_s(q,t)$, which implies that the supercooled liquid's dynamics is transiently arrested, and it retains memory of its structure over intermediate timescales. The sudden appearance of this plateau at a finite value of $F_s(q,t)$ also implies a discontinuous jump in the non-ergodicity parameter and suggests that the structural glass transition has first-order like character \cite{kirkpatrick1987p,lubchenko2015theory}. This is in stark contrast with spin glasses, where the non-ergodicity parameter, also known as the Edwards-Andersen order parameter, increases continuously from zero in the spin glass phase, signalling a bonafide critical phenomenon \cite{edwards1975theory,sherrington1975solvable,parisi1980sequence}. However, the system still remains ergodic, as evidenced by the fact that $F_s(q,t)$ decays to zero and long times. Nonetheless, as the temperature is decreased, the plateau in $F_s(q,t)$ extends to longer and longer times, showing that it becomes increasingly harder for particles in the liquid to reorganize into different configurations. More interestingly, in the low temperature regime, the final decay of $F_s(q,t)$ is no longer exponential. In fact it is often described by the so-called Kohlrauch-Williams-Watts (KWW) or stretched exponential form $\text{exp}(-(t/\tau)^{\beta})$. This immediately suggests the presence of a broad distribution of relaxation times in the glass-forming liquid. As noted by Ediger \cite{ediger2000spatially}, the broad spectrum of relaxation times can emerge due to two physically distinct situations. The first scenario suggests that the supercooled liquid comprises of a heterogeneous set of regions such that relaxation within each region is exponential, but the timescale varies across different regions. The non-exponential decay of $F_s(q,t)$ can then be attributed to ensemble averaging over different regions. The other scenario suggests that relaxation is intrinsically non-exponential even at the molecular level. While experimental evidence for both scenarios exists \cite{ediger2000spatially}, it has been established beyond doubt that the dynamics of glass-forming liquids over time scales of the order of the structural relaxation time $\tau_{\alpha}$ are heterogeneous in space \cite{bohmer1998nanoscale,sillescu1999heterogeneity,ediger2000spatially,glotzer2000spatially,berthier2011dynamical,karmakar2014growing}. In other words, glass-forming liquids comprise of regions that are significantly more or less mobile compared to the average dynamics (Fig. \ref{Figure11}). 
The presence of dynamical heterogeneity raises several important questions. For instance, since supercooled liquids are ergodic, all immobile regions become mobile and vice versa and hence, over timescales much longer than $\tau_{\alpha}$, the dynamics are homogeneous. It is therefore evident that heterogeneities are characterized by a finite lifetime. The spatial extent of mobile and immobile regions is another important characteristic of heterogeneous dynamics. From a theoretical perspective, it is important to determine how the size as well as lifetime of dynamical heterogeneities evolve on approaching the glass transition. Naturally, several attempts have been made to quantify these properties. One of the first efforts in this direction was made by Dasgupta, Ramaswamy and coworkers \cite{dasgupta1991there} who defined a four-point correlation function to examine the possible growth of a dynamic length scale in glass-forming liquids. Their work was motivated by earlier research on spin glasses. In equilibrium statistical mechanics, a static correlation length is usually defined through the decay of the spatial correlation of the order parameter. For spin glasses, the order parameter $\hat{o}$, first proposed by Edwards and Anderson \cite{edwards1975theory} is itself defined as the long time limit of a two-point correlation function. 
\begin{equation}
\hat{o} = \lim_{t\to\infty} \frac{1}{N}\sum_{i}\langle\sigma_i(t_0)\sigma_i(t_0+t)\rangle = \frac{1}{N}\sum_i\sigma_{i}^{2}
\end{equation}    
The spatiotemporal correlation of the order parameter is therefore a four-point function. Extending this notion to structural glass-formers, a four-point dynamic correlation function $G_4(r,t)$ \cite{dasgupta1991there} was defined as
\begin{equation}
G_4(\mathbf{r},t) = [\langle \rho(\mathbf{r_0},t_0)\rho(\mathbf{r_0}+\mathbf{r},t_0)\rho(\mathbf{r_0},t_0+t)\rho(\mathbf{r_0}+\mathbf{r},t_0+t) \rangle]
\end{equation} 
where $\rho(\mathbf{r},t)$ is the local density, $\langle\rangle$ denotes averaging over the reference time $t_0$ and $[..]$ denotes averaging over the reference position $\mathbf{r_0}$ in space. A dynamic correlation length $\xi_4$ can be extracted from $G_4(\mathbf{r},t)$, by defining a four-point susceptibility $\chi_4(t)$  
\begin{equation}
\chi_4(t) = \int d\mathbf{r} G_4(\mathbf{r},t)
\end{equation} 
The static correlation length $\xi_4$ is related to $\chi_4(t)$ through the relation $\chi_4(t) \propto \xi_{4}^{\zeta}$, where the exponent $\zeta$ depends on the shape of the dynamically correlated regions. For compact correlated regions in three dimensions, $\zeta = 3$. While no evidence for a growing dynamical correlation length was found in \cite{dasgupta1991there}, subsequent research has uncovered indisputable evidence for $\xi_4$ that increases on approaching the glass transition. Colloid experiments have played a vital role in garnering this evidence, through DLS measurements \cite{berthier2005direct} as well as video microscopy \cite{narumi2011spatial}. In light scattering experiments, $\chi_4(t)$ can be quantified approximately from fluctuations in the autocorrelation of the scattered intensity \cite{pine1988diffusing,mackintosh1989diffusing,weitz1993diffusing,maret1997diffusing,cipelletti2003time,duri2005time,berthier2005direct,ballesta2008unexpected}. Representative results for $\chi_4(t)$ for dense suspensions of hard sphere-like colloids \cite{berthier2005direct} are shown in Fig. \ref{Figure12}.
$\chi_4(t)$ first increases with time, reaches a peak value $\chi_{4}^{*}$ at a time $t \sim \tau_{\alpha}$ and then decreases again for larger times. This implies that the dynamics are maximally correlated over a timescale of the order of the structural relaxation time. Accordingly, the dynamic correlation length is defined over the timescale at which $\chi_4(t)$ exhibits a maximum. The fact that $\chi_4(t)$ decays to zero at long times implies that unlike in spin glasses, one cannot use its long time limit to extract a diverging static length scale \cite{binder1986spin}. Nonetheless, the observed evolution of $\chi_4(t)$ with volume fraction $\phi$ (Fig. \ref{Figure12}), or temperature, suggests that a growing dynamic length scale might be able to account for the precipitous growth in the relaxation time. 

\begin{figure}
\centering
  \includegraphics[width=\textwidth]{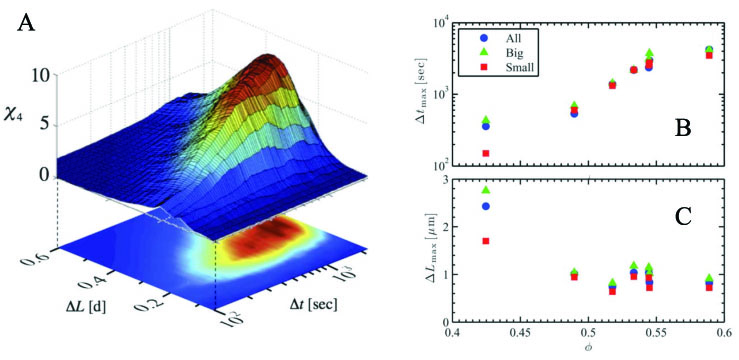}
  \caption{Dynamic four point susceptibility $\chi_4(t)$ from confocal microscopy experiments on dense binary suspensions of PMMA colloids. The large particles have an average radius of 1.55$\mu$m and small particles have an average radius of 1.18$\mu$m. (A) $\chi_4(\Delta L,\Delta t)$ for large PMMA colloids for $\phi =$ 0.52. (B) The time $\Delta t_{\text{max}}$ at which $\chi_4$ shows a maximum as a function of volume fraction $\phi$. (C) The probe length scale $\Delta L_{\text{max}}$ that maximizes $\chi_4$ as a function of $\phi$. Adapted from \cite{narumi2011spatial}.}
  \label{Figure13}
\end{figure}

$\chi_4(t)$ can also be evaluated by directly analysing particle displacements. This method is feasible not only for simulations but also for colloid experiments that employ video microscopy to simultaneously follow trajectories of thousands of particles. To define $\chi_4(t)$, one first computes the self-overlap function $Q_{p,t}(a,\tau)$ \cite{candelier2009building}, defined as 
\begin{equation}
Q_{p,t}(\Delta L,\Delta t) = \text{exp}\Bigg( \frac{-|\Delta \vec{r_p}(t,t+\Delta t)|^2}{2\Delta L^2} \Bigg)
\end{equation} 
where $\Delta \vec{r_p}(t,t+\Delta t)$ is the displacement of particle p between times $t$ and $t+\Delta t$ and $\Delta L$ is the probing length scale. Clearly, small displacements give rise to a large self-overlap and vice versa. $\chi_4(\Delta L,\Delta t)$ is defined through the fluctuations of the self-overlap.
\begin{equation}
\chi_4(\Delta L,\Delta t) = N(\langle Q_t(\Delta L,\Delta t)^2 \rangle - \langle Q_t(\Delta L,\Delta t) \rangle^2)
\label{Chi4Micro}
\end{equation}
where $N$ is the number of particles and $Q_t(a,\Delta t) = (1/N)\sum_p Q_{p,t}(a,\Delta t)$. Notice that in this case, the susceptibility depends on the probe length scale $\Delta L$. In practice, one computes $\chi_4$ for several values of $\Delta L$ and $\Delta t$ and determines the values $\Delta L_{\text{max}}$ and $\Delta t_{\text{max}}$ that maximize it. Fig. \ref{Figure13} shows data for $\chi_4$ from confocal microscopy experiments on dense binary suspensions of PMMA colloids \cite{narumi2011spatial}. 
It is quite instructive to analyse the dependence of $\Delta t_{\text{max}}$ and $\Delta L_{\text{max}}$ on the volume fraction $\phi$. On increasing $\phi$, i.e. on approaching the glass transition, $\Delta L_{\text{max}}$ decreases from $\sim$ 1$\sigma$ to $\sim$ 0.3$\sigma$, $\sigma$ being the average particle diameter. Moreover, while $\Delta t_{\text{max}}$ certainly increases with $\phi$, it does not increase as rapidly as the structural relaxation time $\tau_{\alpha}$. Rather, $\Delta t_{\text{max}}$ compares quite well with $t^{*}$, the typical time taken by a particle to break out of the cage formed by its nearest neighbors \cite{narumi2011spatial}. This is in contradiction with many numerical and experimental studies of the glass transition which found that $\chi_4$ peaks at the structural relaxation time $\tau_{\alpha}$. This difference stems from that fact that in these studies the probe length scale is fixed \textit{a priori}, and usually corresponds to the first peak of the static structure factor, whereas in \cite{narumi2011spatial}, the relevant length and time scales are those that correspond to the global maximum of $\chi_4$ in the $\Delta L - \Delta t$ plane. Further, $\Delta L_{\text{max}}$ is much smaller than the lengthscale corresponding to the first peak of the static structure factor, which is of the order of the particle diameter. Hence, the corresponding susceptibility peaks at shorter times ($\sim t^{*}$) rather than the structural relaxation time $\tau_{\alpha}$. 

These differences demonstrate that the two definitions of $\chi_4$ probe distinct aspects of heterogeneous dynamics. $\Delta L_{\text{max}}$ can be viewed as the length scale that distinguishes particle motion associated with rattling within cages from that associated with cage rearrangements. Indeed, with increasing $\phi$, particles are more tightly constrained by their neighbors and hence, the cage size decreases on approaching the glass transition \cite{weeks2002properties}. The decrease in $\Delta L_{\text{max}}$ with $\phi$ is completely consistent with this picture. The fact that $\Delta L_{\text{max}}$ and $\Delta t_{\text{max}}$ maximize $\chi_4$ implies that dynamical correlations are most significant over the length and time scales corresponding to cage rearrangements. This is plausible given that in a dense system, particles cannot break cages without perturbing other particles in their vicinity. As a result, cage rearrangements can be thought of as cooperative events in which groups of particles exhibit collective bursts of mobility over a relatively short period of time. Thus, $\chi_4(\Delta L, \Delta t)$ as defined in Eqn. \ref{Chi4Micro} essentially probes spatiotemporal correlations between mobile particles. On the other hand, in most experiments and simulations, the probe lengthscale is of the order of a particle diameter, which makes these measurements sensitive to correlations over long time scales. Since displacements associated with cage rearrangements are much smaller that a particle diameter, dynamical correlations of mobility decay over time scales much smaller than the relaxation time $\tau_{\alpha}$. The long time correlations probed in many experiments and simulations therefore correspond to immobile particles. This explains why the peak in $\chi_4(t)$ generally correlates well with $\tau_{\alpha}$ rather than $t^{*}$ \cite{glotzer2000time,lavcevic2003spatially}.

\begin{figure}
\centering
  \includegraphics[width=\textwidth]{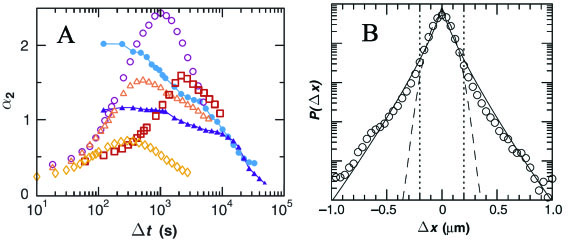}
  \caption{Non-Gaussian displacements in dense suspensions of PMMA colloids. (A) Non-Gaussian parameter $\alpha_2(t)$ for various $\phi$. (B) Probability distribution of displacements along the X-axis, $P(\Delta x)$ for $\phi =$ 0.56, computed over $t^{*} = $ 1000 s, the time at which $\alpha_2(t)$ exhibits a maximum. The dotted lines separate the top 5\% most mobile particles from the remaining slow particles. The dashed curve is a Gaussian fit to the region near the peak of the distribution and solid curves are stretched exponential fits to the tails of the distribution. Adapted from \cite{weeks2000three}.}
  \label{Figure14}
\end{figure}

Although four-point susceptibilities are useful to quantify spatiotemporal correlations between mobile as well as immobile particles, more transparent measures of these correlations can be defined by analysing particle trajectories in real space. This real space approach, first employed in molecular dynamics simulations, has become immensely popular in colloid experiments, especially since the advent of confocal microscopy. The first direct experimental demonstration of the existence of correlations between mobile particles was provided independently by Kegel and van Blaaderen \cite{kegel2000direct} and Weeks, Weitz and coworkers \cite{weeks2000three}. The quantity evaluated in these studies is the kurtosis of the distribution of particle displacements, also known as the non-Gaussian parameter $\alpha_2(t)$. In one dimension, the non-Gaussian parameter is given by 
\begin{equation}
\alpha_2(t) = \frac{\langle \Delta x^4 \rangle}{3\langle \Delta x^2 \rangle^2} - 1
\end{equation}
The evolution of $\alpha_2(t)$ for dense suspensions of PMMA colloids is shown in Fig. \ref{Figure14}A. 
At short times, the dynamics of the liquid are homogeneous and the particle displacements show a nearly Gaussian distribution. As a result, $\alpha_2(t)$ is small. At longer times, particles are transiently trapped in cages formed by nearest neighbors, which leads to an increase in $\alpha_2(t)$. Further, at a characteristic timescale $t^{*}$ corresponding to the typical residence time of a particle in a given cage, the dynamics are maximally non-Gaussian owing to the presence of anomalously large particle displacements associated with cage rearrangements, as evidenced by the displacement distribution $P(\Delta x)$ (Fig. \ref{Figure14}B). At times much longer than $t^{*}$, most particles undergo multiple cage rearrangements, which results in particle displacements becoming increasingly Gaussian. This is manifested as a decrease in $\alpha_2(t)$. As the glass transition is approached, caging becomes stronger, and cage rearrangements become more and more infrequent. As a consequence, $t^{*}$ as well as $\alpha_2(t^{*})$ increase with $\phi$. Beyond the glass transition at $\phi_g \approx $ 0.58, $\alpha_2(t)$ decreases sharply, since the dynamics are frozen and the only motion possible is rattling within cages. 
\begin{figure}
\centering
  \includegraphics[width=0.6\textwidth]{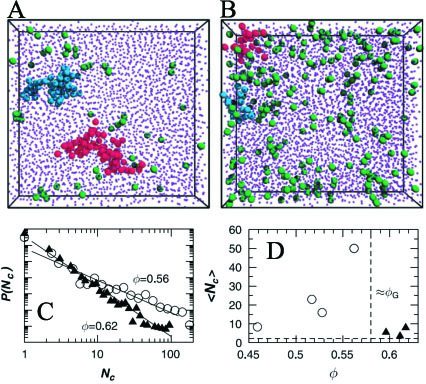}
  \caption{Clusters of mobile particles. (A-B) Representative clusters formed by top 5\% most mobile particles (green, red and blue) defined over $t^{*}$ in (A) the supercooled liquid regime ($\phi = $ 0.56) and (B) the amorphous regime ($\phi = $ 0.61). In (A), the red cluster contains 69 particles and the blue cluster contains 50 particles. In (B), the largest (red) cluster contains 21 particles. (C) Distribution of cluster sizes $P(N_c)$ for a supercooled liquid ($\phi = $ 0.56, open circles) and a glass ($\phi = $ 0.62, filled triangles). The solid lines are power law fits to the data. (D) Average cluster size $\langle N_c \rangle$ as a function of $\phi$. Data points in the supercooled liquid regime are shown as open circles and those in the glassy regime are shown as filled triangles. Adapted from \cite{weeks2000three}.}
  \label{Figure15}
\end{figure}
The spatial organization of particles that contribute to non-Gaussian displacements sheds further light on the nature of dynamical heterogeneities. In \cite{weeks2000three}, the authors considered the top 5\% most mobile particles (see Fig. \ref{Figure14}B) over the time interval $t^{*}$ and clustered them based on nearest neighbor connectivity using Delaunay triangulation. In the supercooled liquid regime, these mobile particles often form large clusters, whereas in the amorphous regime beyond $\phi \sim$ 0.58, the clusters are much smaller (Fig. \ref{Figure15}A-B). This fact is also illustrated by the distribution of cluster sizes $P(N_c)$ (Fig. \ref{Figure15}C). The decrease in cluster size beyond the glass transition indicates that the small displacements associated with cage rattling in glasses are far less cooperative than the cage rearrangement events in supercooled liquids. The average cluster size, defined as $\langle N_c \rangle = \frac{\sum N_{c}^{2}P(N_c)}{\sum N_cP(N_c)}$ shows a significant increase on approaching the glass transition but drops sharply beyond $\phi_g$ (Fig. \ref{Figure15}C). Indeed, the drop in $\langle N_c \rangle$ is a clear signature of the fact that the liquid has fallen out of equilibrium and entered the amorphous regime. Within the supercooled liquid, however, the growth in $\langle N_c \rangle$ provides further evidence of growing dynamical correlations on approaching the glass transition. 

Another method of quantifying dynamical correlations between mobile particles is stringlike cooperative motion \cite{donati1998stringlike}. The clustering method used by Weeks et al. does not take into account the direction of motion of mobile particles. While constructing strings of mobile particles on the other hand, two particles are said to belong to the same string only if the initial position of one particle and the final position of the other are separated by a distance smaller than some threshold. The central idea of this analysis is that if one particle makes a large displacement, a neighboring particle will move in to fill the void left by the first particle. The average string length, defined analogously to the average cluster size, also grows on approaching the glass transition.
\begin{figure}
\centering
  \includegraphics[width=\textwidth]{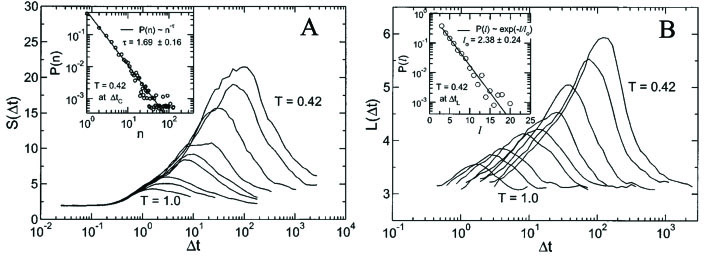}
  \caption{Dependence of average cluster size and average string length on time interval in a simulated glass-former for various temperatures. (A)Average cluster size. Inset shows the distribution of clusters sizes. The distribution shows a power law form. (B) Average string length. Inset shows the distribution of string lengths. The distribution is exponential.  Adapted from \cite{gebremichael2004particle}.}
  \label{Figure16}
\end{figure} 
In \cite{weeks2000three}, the average cluster size was computed over a fixed interval corresponding to the peak of $\alpha_2(t)$. Subsequent simulations have examined the dependence of the average cluster size and string length on the time interval $\Delta t$ itself \cite{gebremichael2004particle} (Fig. \ref{Figure16}). It was observed that the mean cluster size as well as the string length    goes through a maximum as a function of $\Delta t$, implying that there exists a characteristic time scale over which the dynamics are maximally cooperative. Further, this timescale increases on approaching the glass transition, along with the peak cluster size and string length. It has been verified in simulations \cite{gebremichael2004particle} as well as experiments \cite{gokhale2014growing} that this timescale is very close to $t^{*}$, at which the dynamics are most non-Gaussian. These studies conclusively show that the anomalously large displacements in glass-forming liquids are indeed associated with cooperative cage breaking events. 

Another important insight from studies on dynamical heterogeneity is a microscopic explanation for the observed breakdown of the Stokes-Einstein relation, i.e. the decoupling of the diffusion coefficient of the liquid from its viscosity. From the studies described above, it is evident that the motion of mobile particles is maximally correlated over $t^{*}$, whereas that of immobile particles is correlated over $\tau_{\alpha}$. The typical trajectory of a single particle involves several cage rearrangements interspersed with quiescent periods of rattling within cages. Over long times, the trajectory resembles a random walk. Since the frequency of cage jumps is dictated by $t^{*}$, this time scale is relevant for diffusion. On the other hand, $\tau_{\alpha}$ corresponds to the time scale over which the slowest particles first escape from their cages, and is therefore related to the viscosity. Since simulations as well as experiments have consistently found that $t^{*}$ grows much slower than $\tau_{\alpha}$ on approaching the glass transition, the decoupling between these two timescales offers a natural explanation for the breakdown of the Stokes-Einstein relation \cite{starr2013relationship}.

\begin{figure}
\centering
  \includegraphics[width=\textwidth]{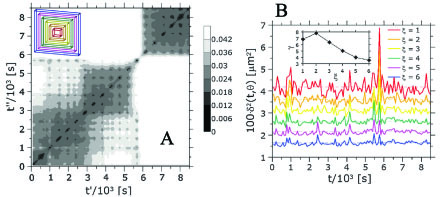}
  \caption{Metabasin transitions in a colloidal glass-former (A) The distance matrix $\Delta^2(t',t'')$ for a small portion of the imaged sample volume containing 77 particles. Dark grey areas indicate a greater similarity between configurations at $t'$ and $t''$. Inset to (A): Boxes correspond to the sizes of different portions considered for computing the distance matrix. Data shown in the main plot of (A) correspond to the smallest box.(B) Average squared displacement $\delta^2(t,\theta)$ as a function of time for different boxes $\xi$ shown in the inset to (A). The colors in (B) are identical to those in the inset to (A). Inset to (B): $\gamma = |\delta_{\xi}^2(t = 5706 s,\theta)-\langle \delta^2(t,\theta) \rangle_{\xi}|/\sigma_{\xi}$ vs $\xi$ A, where $\sigma_{\xi}$ is the standard deviation in $\delta^2$ and $t =$ 5706 s corresponds to the time at which the largest MB-MB transition occurred. Adapted from \cite{fris2011experimental}.}
  \label{Figure17}
\end{figure}

The heuristic definitions of cooperativity employed in the preceding paragraphs establish the prominence of spatiotemporal heterogeneity in the dynamics of glass-forming liquids and provide ample evidence for the presence of a growing dynamic correlation length accompanying the colossal increase in $\tau_{\alpha}$. However, since both clusters and strings of mobile particles show correlations over timescales much smaller than $\tau_{\alpha}$, their contribution to the slowdown of dynamics is not clear. In particular, they do not provide a causal connection between growing lengthscales and growing timescales. Kob and coworkers \cite{appignanesi2006democratic} adopted a different approach based on the potential energy landscape (PEL) picture \cite{debenedetti2001supercooled} first developed by Goldstein \cite{goldstein1969viscous} to determine the nature of particle dynamics that is most relevant for structural relaxation. The procedure developed in \citep{appignanesi2006democratic} was subsequently tested in colloid experiments \cite{fris2011experimental}. It has been shown numerically that the PEL of glass-forming liquids comprises of `meta-basins' (MBs), which are collections of energy minima \cite{sastry1998signatures,sastry2001relationship,debenedetti2001supercooled}. The activation barriers between various minima within an MB are fairly low, and transitions between these minima therefore correspond to small particle motions associated with the short time $\beta$-relaxation processes. Different MBs on the other hand are separated by large activation barriers and transitions between distinct MBs therefore contribute significantly to the long time $\alpha$-relaxation. Intuitively, one should expect such MB-MB transitions to be associated with large collective particle displacements that occur over a relatively short period of time. Naturally, these transitions correspond to significant changes in the supercooled liquid's configuration. To capture such configurational changes, the authors of \cite{appignanesi2006democratic} defined the distance matrix
\begin{equation}
\Delta^2(t',t'') = \frac{1}{N}\sum_{i=1}^{N}|\mathbf{r}_i(t')-\mathbf{r}_i(t'')|^2
\end{equation}
where $\mathbf{r}_i(t)$ denotes the position of the $i^{th}$ particle at time $t$. The distance matrix provides a measure of how much the configurations of the supercooled liquid at times $t'$ and $t''$ differ from each other. Fig. \ref{Figure17}A shows the plot of $\Delta^2(t',t'')$ with $t'$ and $t''$ for the experimental data analyzed in \cite{fris2011experimental}.  
The plot shows that the system resides in a given MB for times much longer than the cage breaking time $t^{*} \approx$ 1000s at the volume fraction ($\phi =$ 0.56) considered, as seen from the size of the square-like grey regions. In the simulations of \cite{appignanesi2006democratic} on the other hand, the authors found this timescale is comparable to $t^{*}$. Since the sojourn time of the system is smaller than $\tau_{\alpha}$, the authors conclude that $\alpha$-relaxation corresponds to a succession of 5-10 MB-MB transitions \cite{appignanesi2006democratic}.

To demonstrate that MB-MB transitions are associated with rapid particle motion, the authors computed the average squared displacement (ASD) $\delta^2(t,\theta)$ \cite{appignanesi2006democratic}, defined as
\begin{equation}
\delta^2(t,\theta) = \Delta^2(t-\theta/2,t+\theta/2) = \frac{1}{N}\sum_{i=1}^{N}|\mathbf{r}_i(t-\theta/2)-\mathbf{r}_i(t+\theta/2)|^2
\end{equation}
This quantity, averaged over initial time $t$, yields the usual mean squared displacement over lag time $\theta$. The curves in Fig. \ref{Figure17}B show the variation in $\delta^2(t,\theta)$ with time $t$ for various portions of the imaged volume in the colloid experiments, for a fixed $\theta$ of 72s. This choice of $\theta$ is motivated by the fact that the time taken by the system to transition from one MB to another is much smaller than the typical sojourn time within an MB, and is of the order of 70s \cite{fris2011experimental}. Hence, large displacements over timescales of this order should be indicative of MB-MB transitions. Indeed, as expected, $\delta^2(t,\theta)$ shows sharp peaks precisely at those times at which the system leaves an MB. To estimate the size of MB-MB transitions, the authors computed the quantity $\gamma = |\delta_{\xi}^2(t,\theta)-\langle \delta^2(t,\theta) \rangle_{\xi}|/\sigma_{\xi}$, where $\sigma_{\xi}$ is the standard deviation in $\delta^2$. For small system sizes $\xi$, $\gamma$ is suppressed due to large fluctuations in $\delta^2$, i.e. large $\sigma_{\xi}$, whereas for large $\xi$, $\gamma$ is lowered by particles that do not contribute to MB-MB transitions. As a result, the portion of the imaged volume that maximizes $\gamma$ is indicative of the size of the MB-MB transition. This is illustrated in the inset to Fig. \ref{Figure17}B, for the largest MB-MB transition, which occurs at $t =$ 5706s. Here, $\gamma$ shows a maximum at $\xi =$ 2, which corresponds to a portion containing 310 particles. 

The authors also computed the fraction $m(t,\theta)$ of mobile particles, i.e. those that moved by a distance greater than $r_{th} =$ 0.23$\mu$m, a value very close to the size of the cage formed by nearest neighbors \cite{fris2011experimental,weeks2002properties}, over the time interval $[t,t+\theta]$. It was observed that in accordance with simulations \cite{appignanesi2006democratic}, the fraction of mobile particles increases significantly whenever the ASD increases. Indeed, for the MB-MB transitions occurring at 5706s, more than 25\% of the particles were observed to be mobile \cite{fris2011experimental}. This shows that in stark contrast to clusters or strings of mobile particles, which contain very few particles, MB-MB transitions are associated with cooperative displacements that involve a much larger number of particles, which led such displacements to be termed `democratic' \cite{appignanesi2006democratic}. 

Most of the experimental results described in the preceding paragraphs have been obtained using relatively simple colloidal glass-formers interacting via nearly hard sphere-like interactions. However, it is quite reasonable to assume that the nature of heterogeneous dynamics is sensitive to various intrinsic factors such as particle shape and interaction potential as well as extrinsic ones such as confinement. These considerations may not be central to unravelling the physics underlying glass formation, but they do provide a wealth of information on the nature of glassy dynamics which can augment the search for the correct theory of the glass transition. We shall briefly mention the key findings of these studies here. 

As mentioned in the previous section, attractive depletion interactions can be introduced by adding small non-adsorbing polymers. Cates, Poon and coworkers demonstrated \cite{pham2002multiple} that the introduction of attractive interactions leads to multiple glassy states as well as re-entrance in glassy dynamics. The results can be intuitively understood as follows. For low strength of attractions, the glass transition is driven by particle caging. Increasing the strength of attractive interactions increases the available free volume per particle and restores ergodicity, thus pushing the glass transition towards higher volume fractions. At very large strengths of attraction, the dynamic arrest is driven by the formation of strong inter-particle bonds that prevent structural relaxation. Due to the increased resistance to bond breaking, the glass transition is once again pulled down to lower volume fractions. This explains the re-entrance in glassy dynamics as the system transitions from a caging dominated glass to a bonding dominated one with increasing strength of attractive interactions. Yodh and coworkers employed a different method that allows attractive interactions to be reversibly switched on or off \textit{in situ} \cite{zhang2011cooperative}. In their protocol, the solvent, which is a mixture of water and lutidine, undergoes a demixing transition above a critical temperature. In the vicinity of this temperature, critical fluctuations of the water-lutidine mixture lead to effective attractive interactions between suspended colloidal particles \cite{pontoni2003microstructure,lu2008liquid,bonn2009direct}. By quantifying $\chi_4(\Delta L,\Delta t)$ in this system, the authors of \cite{zhang2011cooperative} showed that in the presence of attractive interactions, dynamical heterogeneities are observed over broader length as well as time scales and their typical size is larger compared to that of repulsive glass-formers. Moreover, the authors observed that clusters of mobile particles are string-like in repulsive glass-formers but compact in attractive ones. Even for purely repulsive interactions, the softness of the potential can have a significant influence on dynamical heterogeneities. For instance, Schall and coworkers have shown \cite{rahmani2012dynamic} that softer interaction potentials have significantly larger four-point susceptibilities and and their dynamics are therefore more cooperative in nature. Further, in a seminal work, Weitz and coworkers demonstrating using DLS measurements that increasing the softness of the potential leads to a decrease in the fragility of the glass-forming liquid \cite{mattsson2009soft}.  

The simplest departure from spherical shape is an ellipsoid of revolution. While the protocol for synthesizing prolate colloidal ellipsoids of desired shape and aspect ratio $\alpha$ by uniaxially stretching polystyrene spheres was developed in the early 1990s \cite{ho1993preparation}, studies on glass formation in suspensions of such colloidal ellipsoids are fairly recent \cite{zheng2011glass,mishra2013two}. For ellipsoids of aspect ratio $\alpha =$ 6, Han and coworkers showed that in accordance with mode coupling theory \cite{de2007dynamics,pfleiderer2008glassy} and experiments on liquid crystals \cite{cang2003dynamical}, suspensions of repulsive ellipsoids have two distinct glass transitions: orientational dynamics are arrested first, followed by a freezing of the translational degrees of freedom \cite{zheng2011glass}. The authors further showed that analogous to dynamical heterogeneities in spherical colloids, mobile clusters for translational as well as rotational motion exhibit a power law distribution with a mean cluster size that grows on approaching the glass transition. Mishra et al. \cite{mishra2013two}, working with ellipsoids of a much smaller aspect ratio $\alpha =$ 2.1 showed that while the repulsive system at low aspect ratio exhibits a single glass transition, the addition of attactive depletion interactions leads to a decoupling of the two glass transitions, which is also accompanied by a spatial decoupling between translational and rotational dynamical heterogeneities. They further showed that as in the case of spherical systems, attractive interactions give rise to re-entrant glassy behavior.

Yet another factor which has profound implications for glass formation is confinement. Confocal microscopy experiments have shown that confinement leads to a slowdown of dynamics and hence, a confined supercooled liquid is closer to the glass transition than it bulk counterpart \cite{nugent2007colloidal}. Further, since the dynamic correlation length increases on approaching the glass transition, confinement effects set in at larger length scales for samples with higher volume fractions. Dynamical heterogeneities themselves are also affected by confinement \cite{edmond2012influence}. In \cite{edmond2012influence} it was observed that the presence of smooth walls leads to layering in the liquid's density profile along the direction of confinement. This in turn leads cooperatively rearranging regions, defined as clusters of mobile particles to become increasingly planar in shape. The influence of confinement on glass formation acquires significance in the light of observed differences in the dynamics of 2D and 3D glass-forming liquids. Numerous studies have revealed these differences in the context of structural relaxation \cite{flenner2015fundamental}, the Adam-Gibbs relation \cite{sengupta2012adam}, the breakdown of the Stokes-Einstein relation \cite{sengupta2013breakdown} as well as random pinning glass transitions \cite{cammarota2012ideal,cammarota2013random,cammarota2013general}. By varying the degree of confinement, it is possible to transition from a 2D system to a 3D system. It would be fascinating to investigate how various structural and dynamical quantities associated with glass formation evolve during this transition.

Thus far, we have reviewed various studies on dynamical heterogeneities that have contributed enormously to our understanding of glassy dynamics. From the point of view of identifying the correct theoretical scenario for glass formation, however, these works offer little conclusive evidence. The dependence of various measures of heterogeneous dynamics such as the dynamic susceptibility, mobile clusters, strings and democratic motion on time as well as temperature or volume fraction can be satisfactorily taken into account within various distinct theoretical formulations. Moreover, dynamical heterogeneity studies do not address the possibility of structural changes accompanying the glass transition, which is a crucial ingredient of many theories of vitrification. In the next section, therefore, we shall focus our attention on those experiments that have attempted to gather evidence in favor of various prominent theories of glass formation.   

\section{Experimental evidence for competing theories of glass formation}
Early studies on colloidal glasses focused on binary systems composed of highly charged particles \cite{lindsay1982elastic}. Further studies on the formation of these so-called Wigner glasses examined various aspects of structure \cite{sanyal1995brownianI,sanyal1995brownianII,tanaka2004nonergodic}, dynamics \cite{tanaka2005kinetics} as well as cooperative motion \cite{sanyal1996cooperative,sanyal1997cooperative}. These glasses have also been investigated within the context of the mode coupling theory (MCT) \cite{hartl1995glass} and more recently, the random first-order transition theory (RFOT) \cite{kang2013manifestation}. A majority of colloid experiments aimed at verifying specific theoretical predictions, on the other hand, have been performed on hard sphere-like colloids, using dynamic light scattering \cite{berne2000dynamic} as well as microscopy techniques and we review these in detail below. 
\subsection{Mode coupling theory}
Mode Coupling Theory (MCT) holds the distinction of being the only first principles theory of glass formation. It is a dynamic approach that predicts the average dynamics of molecular or colloidal liquids using the static structure factor as the input \cite{leutheusser1984dynamical,bengtzelius1984dynamics,gotze1991liquids,gotze1992relaxation,gotze2008complex}. The key physical aspect of MCT that is relevant to glass formation is that the non-linear feedback of density fluctuations on the microscopic dynamics results in complete structural arrest of the system below a characteristic temperature $T_c$ or above a volume fraction $\phi_c$. The central ingredient of MCT is the equation of motion for the self-intermediate scattering function $F_s(q,t)$, which is the Fourier Transform of the two-point density correlator. The equation is typically solved with the help of reasonable approximations to yield predictions for the relaxation of glass-forming liquids. The starting equation of motion for $F_{s}(q,t)$ for a single component atomic liquid is 
\begin{equation}
\frac{d^2F_s(q,t)}{dt^2} + \frac{q^2k_BT}{mS(q)}F_s(q,t) + \int_{0}^{t}d\tau M(q,\tau)\frac{d}{dt}F_s(q,t-\tau) = 0
\label{MCTEqn}
\end{equation}
Here, $S(q)$ is the static structure factor, m is the mass of the particle and $k_B$ is the Boltzmann's constant. The memory kernel $M(q,\tau)$ takes into account the influence of all degrees of freedom other than the density field on the density field itself. The next crucial step in MCT involves developing a series of approximations for the memory kernel and solving the resultant self-consistent equations for $F_s(q,t)$ \cite{zwanzig2001nonequilibrium}. For simple liquids, the memory kernel is delta correlated in time. Further, in the overdamped limit relevant to colloids, the second derivative of $F_s(q,t)$ with respect to time, which corresponds to the inertial term, can be neglected with respect to the first derivative. In this limit, Eqn. \ref{MCTEqn} yields the usual exponential decay of a liquid's dynamic structure factor \cite{terentjev2015oxford,hansen1990theory}. Using the formalism described above, MCT predicts the existence of three relaxation processes. The short time regime corresponds to relaxation on the microscopic time scale $\tau_0$ and therefore depends on the microscopic details of the system. For instance, this motion is ballistic for atomic liquids and diffusive for colloidal ones. In this regime, $F_s(q,t)$ asymptotically approaches a plateau, such that
\begin{equation}
F_s(q,t) = f_q + \frac{h(q)}{t^{a}} \qquad \text{for} \qquad t>>\tau_0
\label{MCTa}
\end{equation}
where $f_q$ is known as the nonergodic parameter. The exponent $a$ can be extracted from the equation
\begin{equation}
\frac{\Gamma^2(1-a)}{\Gamma(1-2a)} = \lambda
\end{equation}
where $\lambda$ is a number that depends on the structure factor and $\Gamma(x)$ is the gamma function. The above equation constrains the value of $a$ to lie in the interval $0 \leq a < 1/2$. Over intermediate time scales, known as the $\beta$ relaxation regime, particles are caged by their nearest neighbours, which leads to the appearance of a plateau in  $F_s(q,t)$. In this regime, MCT predicts that $F_s(q,t)$ has the form
\begin{equation}
F_s(q,t) \approx f_q + c_{\sigma}h(q)g(t/\tau_{\beta})
\label{MCTb}
\end{equation}
where $g(t) \propto t^{-a}$ for $t << \tau_{\beta}$ and $g(t) \propto t^{b}$ for $t >> \tau_{\beta}$. Further, $\sigma$ denotes the distance $(T-T_c)/T_c$ or $(\phi_c - \phi)/\phi_c$ from the MCT glass transition at $T_c$ or $\phi_c$ and $c_\sigma \propto \sqrt{\sigma}$ The exponent $b$ satisfies the equation 
\begin{equation}
\frac{\Gamma^2(1+b)}{\Gamma(1+2b)} = \lambda
\end{equation}
which restricts $b$ to the interval $0 \leq b \leq 1$. Importantly, MCT predicts that the $\beta$ relaxation time diverges
as $\tau_{\beta} \sim \sigma^{-1/2a}$ on approaching the MCT glass transition. The final decay of $F_s(q,t)$ at very long times can be described by a stretched exponential form
\begin{equation}
F_s(q,t) \propto \exp \left(-\left( t/\tau_{\alpha}\right) ^{\beta} \right)
\end{equation}
with $\beta < 1$. This regime is known as the $\alpha$-relaxation regime and the corresponding time scale $\tau_{\alpha}$ is known as the alpha relaxation time. It is this time scale that is often referred to as the structural relaxation time and scales with the viscosity of the liquid. MCT predicts that the alpha relaxation time diverges at $T_c$ or $\phi_c$ as 
$\tau_{\alpha} \sim \sigma^{-\gamma}$, where $\gamma = \frac{1}{2a} + \frac{1}{2b}$. 

\begin{figure}
\centering
  \includegraphics[width=\textwidth]{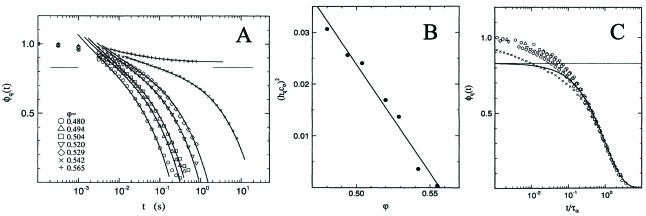}
  \caption{Verification of MCT predictions for hard sphere-like colloidal glass-formers (A) Normalized intermediate function $\phi_q(t) = F(q,t)/S(q)$ for different volume fractions $\varphi$. The solid curves are MCT fits using a combination of Eqns. \ref{MCTa} and \ref{MCTb}. (B) Verification of the square root singularity predicted by MCT. The singularity occurs at $\varphi_c =$ 0.557. (C) Scaling in the $\alpha$ relaxation regime. The solid curve is a Kohlrausch-Williams-Watts fit of the form $f_q\text{exp}(-(t/\tau_{\alpha})^{\beta})$, with $\beta =$ 0.88. Adapted from \cite{gotze1991beta}.}
  \label{Figure18}
\end{figure}

Some of the earliest experimental evidence supporting MCT was provided by van Megen, Pusey and coworkers in a series of experiments on hard sphere-like colloidal suspensions \cite{van1991dynamic,van1993dynamic,van1993glass,van1994glass,van1995crystallisation,van1998measurement}. One of the first among these experiments obtained structural relaxation data for metastable colloidal liquids composed of PMMA spheres of mean hydrodynamic radius $\approx$ 170 nm and polydispersity of $\approx$ 4.5\% using static and dynamic light scattering techniques \cite{van1991dynamic}. In particular, these experiments extracted the intermediate scattering function from the autocorrelation of the scattered intensity using the following equations
\begin{equation}
g_E^{(2)}(q,t) = \langle I(q,0)I(q,t)\rangle_E/\langle I(q,0)^2 \rangle_E = 1 + [cF(q,t)/S(q)]^2
\end{equation} 
where $\langle\rangle_E$ denotes ensemble averaging, $I(q,t)$ is the scattered intensity at time $t$ for probing wave vector $q$, $S(q) = F(q,0)$ is the static structure factor and $c \approx$ 0.8 is an experimentally determined constant \cite{oliver1974photon}. The intermediate scattering function is defined as 
\begin{equation}
F(q,t) = \Bigg\langle\frac{1}{N} \sum_{j,k=1}^{N} \text{exp}\Big[i\mathbf{q}.(\mathbf{r_j}(0)-\mathbf{r_k}(t))\Big]\Bigg\rangle
\end{equation} 
As is typically the case, most measurements were made at $q = q_m =$ 2.17$\times$10$^5$ cm$^{-1}$, which corresponds to the first peak of $S(q)$. The data obtained from these experiments were analyzed by G\"{o}tze and Sj\"{o}gren \cite{gotze1991beta} in order to test the MCT predictions for structural relaxation. Fig. \ref{Figure18}A shows the MCT fits to the normalized intermediate function $\phi_q(t) = F(q,t)/S(q)$ for various volume fractions $\varphi$. 
The functional form uses a combination of those in Eqns. \ref{MCTa} and \ref{MCTb}. In this fitting procedure, the authors used the values $a =$ 0.301 and $b =$ 0.545 which had been calculated earlier for hard spheres \cite{barrat1989liquid}. For the non-ergodicity parameter $f_q$, the authors chose the value 0.83, which is slightly smaller than the calculated value of 0.87 \cite{bengtzelius1984dynamics}, taking into account the fact that MCT tends to overestimate the propensity for glass formation \cite{gotze1991beta}. The function $h(q)$ was also previously determined and used as such for the fitting \cite{barrat1989liquid}. The data show good agreement with MCT, except at low values of $\phi_q(t)$, where the form of the long time $\alpha$ relaxation is different from that of the von Schweidler law (Eqn. \ref{MCTb}). Further, the square root dependence of $c_{\sigma}$ on $(\varphi_c - \varphi)/\varphi_c$ was also confirmed, which lead to the determination of the MCT glass transition volume fraction to be $\varphi =$ 0.557 (Fig. \ref{Figure18}B). MCT also predicts a scaling law for the long time $\alpha$ relaxation, which has the form $\phi_q(t) = f_q\Phi_q(t/\tau_{\alpha})$. Here, $\Phi_q(t)$ is a scaling function independent of volume fraction $\varphi$. Fig. \ref{Figure18}C shows that this scaling form is indeed obeyed for times much longer than the $\beta$ relaxation time $\tau_{\beta}$. Moreover, in the regime where the scaling holds, it can be well described by a Kohlrausch-Williams-Watts form $\phi_q(t) = f_q\text{exp}(-(t/\tau_{\alpha})^{\beta})$, with $\beta =$ 0.88, the value obtained by Bentzelius for a Lennard-Jones system \cite{bengtzelius1986dynamics}. In a subsequent experiment, van Megen and Underwood quantified the non-ergodicity parameter and confirmed that the values obtained are consistent with MCT predictions \cite{van1993dynamic}.

\begin{figure}
\centering
  \includegraphics[width=0.5\textwidth]{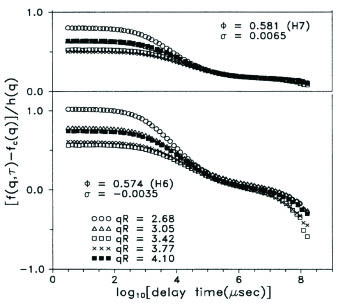}
  \caption{Evidence of the factorization property of $\beta$ relaxation in hard sphere-like colloidal suspensions. The shifted and scaled intermediate scattering function $(f(q,\tau)-f_c(q)/h(q)$ as a function of lag time $\tau$ for various values of $q$ in the supercooled liquid ($\varphi =$ 0.574) and amorphous ($\varphi =$ 0.581) regime. $\sigma$ denotes the distance from the glass transition. In the text, $f(q,\tau)$ is denoted by $\phi_q(t)$ and $f_c(q)$ by $f_q$. Adapted from \cite{van1993glass}.}
  \label{Figure19}
\end{figure}

An important prediction of MCT is the factorization property of $\beta$ relaxation. For the intermediate regime of relaxation described by Eqn. \ref{MCTb}, $\phi_q(t)$ depends on $q$ and $t$ through the product $h(q)g(t/\tau_{\beta})$. This factorization implies a collapse of $\phi_q(t)$ for various $q$ at a given volume fraction. Such a collapse was indeed observed by van Megen and Underwood \cite{van1993glass} for volume fractions in the supercooled as well as amorphous regimes (Fig. \ref{Figure19}). It is evident from the figure that the data collapse occurs only over an intermediate range of times. The lack of collapse at large times is explained by the fact that this regime is governed by the $\alpha$ relaxation process, which is quite distinct from the $\beta$ process and therefore constitutes a different dynamical regime. In the glass phase, $\alpha$ relaxation is absent and consequently, the data collapse is preserved even at long times. At very short times on the other hand, the microscopic dynamics of the system are dominant, and particles do not feel cages formed by their neighbors. 

The greatest drawback of idealized MCT is that the glass transition that it predicts does not occur in real glass-forming liquids. In real liquids, thermally activated particle rearrangements restore ergodicity beyond the dynamical glass transition predicted by MCT. As a consequence, the MCT transition becomes a crossover, and one has to be careful while fitting power laws to relaxation time data close to $T_c$ or $\phi_c$. However, in spite of the fact that it cannot describe glassy dynamics close to $T_g$, MCT is often used as the first line of attack, as it can serve as a good qualitative indicator of the nature of glassy dynamics in the moderately supercooled regime \cite{berthier2011theoretical}. More importantly, it can predict novel phenomena, such as the existence of orientational glass transitions in glass-formers composed of anisotropic particles \cite{letz2000ideal} and reentrant glass transitions in colloidal systems with short-ranged attractive interactions \cite{dawson2000higher}. Indeed, some of these predictions have been verified by colloid experiments and we shall review them briefly here. 

\begin{figure}
\centering
  \includegraphics[width=\textwidth]{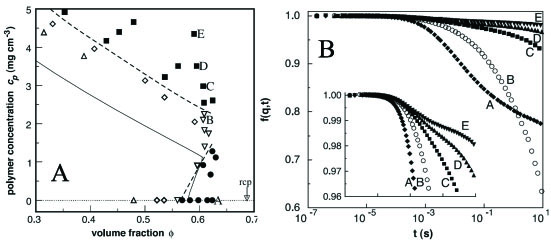}
  \caption{Re-entrant glass transition in suspensions of PMMA colloids (A) The glass phase diagram in the polymer concentration - volume fraction plane. Points labelled A through E depict various strengths of attraction for $\phi \approx$ 0.6. (B) The normalized dynamic structure factor $f(q,t)$ measured at $qR =$ 2.93 for the points labelled in (A). Inset: The same data with an enlarged Y-axis. Note the logarithmic decay of the data for point C. Adapted from \cite{pham2002multiple}.}
  \label{Figure20}
\end{figure}

In the presence of attractive interactions, MCT predicts the existence of two distinct types of glass, one dominated by the repulsive part of the potential and the other by the attractive part. MCT further predicts that over a certain region in parameter space, the repulsive and attractive glasses are separated by an ergodic liquid phase for intermediate strengths of attraction, a phenomenon known as re-entrance. The first experimental verification of these predictions was provided by Cates, Poon and coworkers using microscopy and DLS experiments \cite{pham2002multiple}. The authors observed samples of PMMA colloids of average radius $R =$ 202 nm and polydispersity of 7\% interacting via short-ranged attractive depletion interactions induced by adding non-adsorbing polymers of radius of gyration $r_g =$ 17 nm. In the absence of attractive interactions, the system exhibits a glass transition at $\phi_g \approx$ 0.58. Since the particle size is comparable to the wavelength of visible light, colloidal crystals Bragg-scatter visible light and appear as iridescent specks. The authors used this fact to infer whether the samples exhibited fully crystalline order, crystal-liquid coexistence or a complete lack of crystalline order. To investigate the possibility of re-entrance, the authors observed the samples at a fixed volume fraction $\phi \approx$ 0.6, where the equilibrium phase is crystalline, as a function of polymer concentration, i.e. the strength of attraction (Points A through E in Fig. \ref{Figure20}A). 
For the purely repulsive case A, the sample is a glass, as expected from the hard sphere phase diagram. However, for intermediate values of polymer concentration, the authors found that the samples crystallized completely. This implies that for these polymer concentrations, sample equilibration is possible and hence the glass transition has shifted to higher $\phi$. At large polymer concentrations, however, crystallization was not observed for weeks or even months, suggesting that the system had entered a second glassy regime. This result provided the first experimental indication of a re-entrant glass transition. A similar conclusion was reached independently by Eckert and Bartsch by analyzing the decay of $F_s(q,t)$ using DLS experiments \cite{eckert2002re}. The authors of \cite{pham2002multiple} proceeded to map out the entire phase diagram in the polymer concentration - volume fraction plane to locate the glass transition line. As seen in Fig. \ref{Figure20}A, the experimental results (dashed lines) are in good qualitative agreement with MCT predictions (solid lines). It is important to note that the MCT prediction shown contains no free parameters. Observed deviations from the MCT predicted transition line presumably arise due to many body effects that are not captured by the Asakura-Oosawa potential used by the authors in their MCT calculations. MCT also makes predictions regarding the asymptotic value of the dynamical structure factor, or the self-intermediate scattering function $F_s(q,t)$. In particular, according to MCT, the value of $F_s(q,t)$ as $t \rightarrow \infty$ should be much smaller for repulsive glasses than for attractive ones. To verify this prediction, the authors obtained the normalized dynamic structure factor $f(q,t) = F_s(q,t)/S(q)$ where $S(q)$ is the static structure factor, for $qR =$ 2.93 using light scattering experiments (Fig. \ref{Figure20}B). This wave vector corresponds to a lengthscale that is slightly larger than the mean inter-particle separation. Although $f(q,t)$ does not saturate over the duration of the experiment, the results are consistent with the predictions of MCT. For the purely repulsive glass, $f(q,t)$ shows a dip followed by an inflection point, indicating a possible saturation at 0.7. The attractive glasses on the other hand show a negligible decay over the same duration, suggesting that $f(q,t)$ for these data sets saturate at higher values. Furthermore, $f(q,t)$ for point C shows a logarithmic decay over one decade in time (Fig. \ref{Figure20}B inset). Such a logarithmic decay is expected in the vicinity of end points of glass-glass transitions, where the difference between the two types of glasses vanishes \cite{dawson2000higher}.  

\begin{figure}
\centering
  \includegraphics[width=\textwidth]{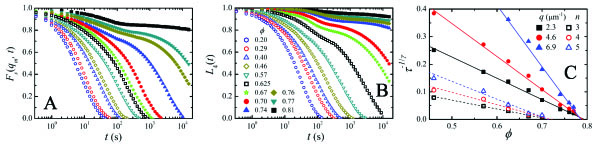}
  \caption{Orientational and translational glass transitions in suspensions of repulsive colloidal ellipsoids of aspect ratio $\alpha =$ 6. (A) Self-intermediate scattering function $F_s(q_m,t$, where $q_m =$ 2.3$\mu$m$^{-1}$ corresponds to the first peak in the structure factor and (B) Orientational correlation function $L_4(t)$ for various area fractions $\phi$. (C) $\tau^{-1/\gamma}$ vs $\phi$. Solid symbols denote translational relaxation times for various wave vectors $q$. Hollow symbols denote orientational relaxation times obtained from $L_n(t)$ for $n =$ 3,4 and 5. Adapted from \cite{zheng2011glass}.}
  \label{Figure21}
\end{figure}

In the case of anisotropic particles such as prolate ellipsoids, rotational motion must be considered in addition to translations. Intuitively, one would expect that as one perturbs the particle shape from spherical to ellipsoidal, rotations and translations would be strongly coupled for small to intermediate $\alpha$ whereas at large $\alpha$, they would be decoupled. Consistent with this intuition, MCT predicts that below an aspect ratio of $\alpha =$ 2.5, a system of ellipsoids should exhibit a single glass transition whereas for larger aspect ratios, the orientational and translational glass transition are decoupled, with orientations freezing before translations \cite{letz2000ideal,de2007dynamics,pfleiderer2008glassy}. To test these predictions, Han and coworkers performed video microscopy experiments on quasi-2D suspensions of colloidal polystyrene ellipsoids with $\alpha =$ 6. Translational relaxation can be investigated by computing $F_s(q,t)$ from the displacements of centres-of-mass of the ellipsoids. For rotational relaxation in 2D, one can compute the correlation functions $L_n(t)$, defined as
\begin{equation}
L_n(t) = \Bigg\langle \frac{1}{N} \sum_{j}^{N} \text{cos}[n(\theta_j(t)-\theta_j(0))] \Bigg\rangle
\end{equation} 
where $n$ is an integer and $\theta_j(t)$ is the orientation of ellipsoid $j$ at time $t$. Different values of $n$ yield the same glass transition point and the authors chose $n = 4$ in order to best represent the decay of orientational correlations. Fig. \ref{Figure21}A \& B show the decay of $F_s(q,t)$ and $L_4(t)$ respectively for various area fractions $\phi$. 
According to MCT, the relaxation time $\tau$ diverges at a critical volume fraction $\phi_c$ as $(\phi_c - \phi)^{-\gamma}$. As a consequence, $\tau^{-1/\gamma}$ must vanish at $\phi_c$. This ansatz can be used to extract the glass transition point for rotations as well as translations, provided the value of the exponent $\gamma$ is known. In practice, $\gamma$ can be extracted from the shape of relaxation functions over smaller timescales. Specifically, $\gamma = 1/(2a) + 1/(2b)$, where $a$ and $b$ are exponents that can be extracted by fitting appropriate sections of $F_s(q,t)$ and $L_n(t)$ with the functional forms given in Eqns. \ref{MCTa} and \ref{MCTb} respectively. Using this procedure, the authors found the exponents for translational and orientational relaxation to be $\gamma_T = 2.45 \pm 0.05$ and $\gamma_{\theta} = 2.33 \pm 0.05$ respectively. Since MCT predicts that the dynamics over all lengthscales is frozen at the critical area fraction, various values of $q$ are expected to yield the same glass transition. Moreover, $L_n(t)$ is also expected to yield the same orientational glass transition irrespective of $n$. The authors therefore extracted translational relaxation times for various values of $q$ and orientational relaxation times for $n =$ 3,4 and 5. As expected from MCT, the authors found that $\tau^{-1/\gamma}$ vanishes at two distinct values of $\phi$ for orientational and translational degrees of freedom, thus demonstrating the existence of two glass transitions in the system. Moreover, the orientational glass transition at $\phi_c^{\theta} \approx$ 0.72 precedes the translational glass transition at $\phi_c^{T} \approx$ 0.79, in concord with MCT predictions. 

\begin{figure}
\centering
  \includegraphics[width=\textwidth]{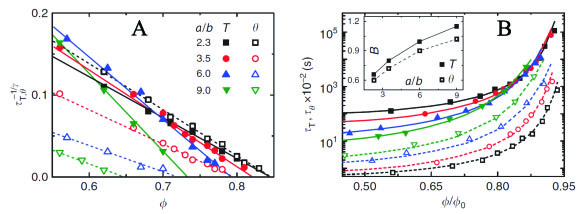}
  \caption{MCT scaling of translational (T) and rotational ($\theta$) relaxation times for PMMA ellipsoids of various aspect ratios. (A) Translational and rotational relaxation times raised to the power $-1/\gamma$ as a function of $\phi$ for various $\alpha$. (B) The absolute translational and rotational relaxation times vs $\phi$. The dashed and solid curves correspond to VFT fits of the form $\tau(\phi) = \text{exp}[B\phi/(\phi_0-\phi)]$, where $B$ is the fragility index. The inset shows the variation of $B$ with $\alpha$ for translational as well as rotational relaxation. Adapted from \cite{zheng2014structural}.}
  \label{Figure22}
\end{figure}

In a subsequent study, Han and coworkers explicitly verified the prediction from MCT that below an aspect ratio of $\alpha =$ 2.5, a system of ellipsoids exhibits only one glass transition. They performed a series of video microscopy experiments on 2D suspensions of PMMA ellipsoids of aspect ratios ranging from 2.3 to 9. From the scaling of translational and rotational relaxation times, they showed that for $\alpha =$ 2.3, the two glass transitions occur at the same area fraction, whereas they decouple for larger aspect ratios. They further showed that the extent of this decoupling increases with $\alpha$ (Fig. \ref{Figure22}A). Interestingly, by examining the dependence of relaxation times on $\phi$, the authors also found that an increase in aspect ratio leads to a decrease in the fragility of the liquid (Fig. \ref{Figure22}B). 

\begin{figure}
\centering
  \includegraphics[width=0.75\textwidth]{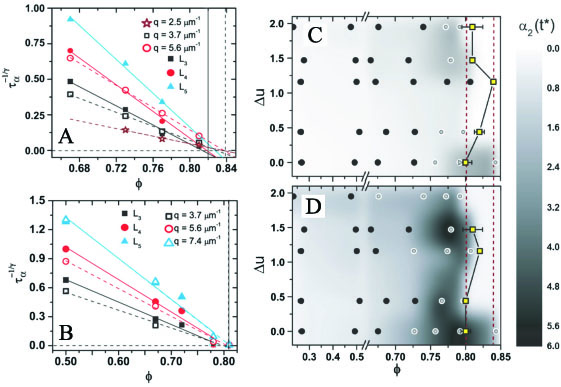}
  \caption{Orientational and translational glass transitions in suspensions of colloidal ellipsoids of aspect ratio $\alpha =$ 2.1 for various strengths $\Delta u$ of attractive interactions. (A-B) $\tau^{-1/\gamma}$ vs $\phi$ for (A) $\Delta u =$ 1.16 and (B) $\Delta u =$ 1.47. The translational and orientational relaxation times have been extracted from $F_s(q,t)$ and $L_n(t)$ for various values of $q$ and $n$ respectively. (C-D) Glass transition phase diagram in the $\Delta u$-$\phi$ plane for translational (C) and rotational (D) degrees of freedom. In (C) and (D), Black circles represent points at which relaxation functions decayed fully and white circles represent points where relaxation functions decayed partially. The yellow symbols denote glass transition points extracted from MCT scaling. The underlying colormap corresponds to the maximum value of the non-Gaussian parameter, $\alpha_2(t^{*}$. Adapted from \cite{mishra2013two}.}
  \label{Figure23}
\end{figure}

The simultaneous presence of attractive interactions and particle shape anisotropy can give rise to an even richer glass transition scenario, as shown by Mishra et al. \cite{mishra2013two}. In a series of video microscopy experiments, Mishra et al. investigated glass formation in quasi-2D suspensions of colloidal polystyrene ellipsoids of low aspect ratio ($\alpha =$ 2.1) in the presence of attractive depletion interactions. While MCT predicts a single glass transition for this system in the purely repulsive case, there are no predictions for the phase diagram of suspensions of ellipsoids in the presence of attractive interactions. Nonetheless, the authors applied the procedure adopted in \cite{zheng2011glass} to their system and extracted $\phi_c^{T}$ as well as $\phi_c^{\theta}$ for various strengths of the attractive interaction $\Delta u = \Delta U/k_BT$, where $\Delta U$ is the depth of the potential well as extracted from the potential of mean force \cite{kepler1994attractive}. In accordance with MCT, the authors indeed found a single glass transition for the purely repulsive case $\Delta u =$ 0. For $\Delta u =$ 1.16, however, the system showed two distinct glass transitions with $\phi_c^{\theta} \approx$ 0.82 and $\phi_c^{T} \approx$ 0.84 (Fig. \ref{Figure23}A). 
Note that $\phi_c^{\theta} < \phi_c^{T}$, similar to the results of \cite{zheng2011glass} for the repulsive system with high aspect ratio. On further ramping up $\Delta u$, the system once again exhibits a single glass transition. For instance, at $\Delta u =$ 1.47, $\phi_c^{\theta} \approx \phi_c^{T} \approx$ 0.81 (Fig. \ref{Figure23}B). The authors then charted the complete phase diagram in the $\Delta u$-$\phi$ plane for translational (Fig. \ref{Figure23}C) as well as orientational (Fig. \ref{Figure23}D) glass transitions. Two features stand out in these phase diagrams. First, as in the case of hard-spheres, there is a re-entrant glass transition in the translational degrees of freedom with increasing $\Delta u$. There is evidence for re-entrance even in the orientational degrees of freedom as well, although it is much weaker, given the considerable uncertainty in determining area fractions of colloidal suspensions \cite{poon2012measuring}. Secondly, for intermediate values of $\Delta u$, the orientational glass transition appears to precede the translational one. The authors explained this observation by invoking the anisotropy in depletion interactions. Since depletion interactions are entropic in nature, they maximize the free volume available to the polymer molecules. For prolate ellipsoids, this free volume is maximized if the ellipsoids are aligned with their long axes parallel to each other. This effective anisotropic interaction increases the nematic order in the suspension. For a purely repulsive system, nematic order is unimportant for small aspect ratios, but begins to set in at larger aspect ratios. Mishra et al. \cite{mishra2013two} therefore argued that ellipsoids with attractive interactions effectively behave like repulsive ellipsoids of a larger aspect ratio, for which MCT indeed predicts two glass transitions. Collectively, these results highlight the utility of MCT in describing novel glass formation in systems with complex particle shapes and interactions.

Despite its numerous successes, the original formulation of MCT has serious shortcomings \cite{reichman2005mode}. It assumes nearest neighbor cages to be local objects and therefore cannot explain the presence of spatially heterogeneous dynamics or growing dynamic length scales \cite{berthier2011theoretical}. Further, it does not offer any explanation for the breakdown of the Stokes-Einstein relation (SER) routinely observed in real and simulated glass-formers \cite{reichman2005mode}. Over the last fifteen years or so, MCT has been refined and expanded significantly, in order to incorporate a description of spatially heterogeneous dynamics. While the predictions emerging from these theoretical developments are yet to be observed in experiments, they are crucial from the point of view of our current understanding of MCT and we briefly review them here. The first advances in incorporating heterogeneous dynamics within MCT were due to Franz and Parisi \cite{franz2000non} and later Biroli and coworkers \cite{biroli2006inhomogeneous}, who argued that even within MCT, caging must necessarily involve some degree of cooperativity. This follows from the intuition that in order for neighbors to form a cage that constrains a given particle, the neighbors themselves must be constrained \cite{berthier2011theoretical}. In one of the first studies on dynamical correlations within MCT, Franz and Parisi showed that the putative MCT transition is accompanied by a diverging four-point susceptibility \cite{franz2000non}. To compute this susceptibility, they considered two `replicas' of a system coupled via a field $\epsilon$ that is conjugate to the configurational overlap between the two replicas \cite{franz1997phase}. Further, one of the replicas is quenched in the initial configuration $X_0$ whereas the other evolves according to the Hamiltonian  
\begin{equation}
H_{tot}(X) = H(X) - \epsilon q(X,X_0)
\end{equation} 
where $H(X)$ is the Hamiltonian of the individual replicas and $q(X,X_0)$ is the configurational overlap between the two replicas. Within this formulation, the four-point susceptibility, which measures fluctuations of the time-dependent configurational overlap, emerges naturally as a linear response function
\begin{equation}
\chi_4(t) = \frac{\partial \langle q(X_t,X_0) \rangle}{\partial \epsilon} 
\end{equation} 
The time evolution of the above Hamiltonian can be studies within the MCT formalism, which allows $\chi_4(t)$ to be quantified \cite{franz2000non}. Bouchaud and Biroli \cite{biroli2004diverging} then demonstrated the existence of a diverging dynamic correlation length using a field theoretic formulation of MCT. Soon after, in a landmark paper, Biroli et al. extended MCT to supercooled liquids in inhomogeneous external fields \cite{biroli2006inhomogeneous} and showed that a diverging length scale can be extracted even within the standard Mori-Zwanzig projection operator formalism of MCT \cite{gotze1992relaxation}. The central idea is to impose a weak inhomogeneous field $U(\mathbf{x}x)$ on the liquid and quantify its response, which is given by   
\begin{equation} 
\frac{\delta F(\mathbf{x},\mathbf{y},t)}{\delta U(\mathbf{z})} \Bigg |_{U=0} = \int d\mathbf{k}d\mathbf{q} e^{-i\mathbf{k}.(\mathbf{x})-\mathbf{y}) + i\mathbf{q}.(\mathbf{y}-\mathbf{z})} \chi_{\mathbf{q}}(\mathbf{k},t)
\end{equation}
where $\chi_{\mathbf{q}}(\mathbf{k},t) \propto [\delta F(\mathbf{k},\mathbf{q}+\mathbf{k},t)/\delta U(\mathbf{q})]|_{U=0}$. Unlike the conventional four-point susceptibility, $\chi_{\mathbf{q}}(\mathbf{k},t)$ is related to a three-point density correlation function in the absence of the external field. Nonetheless, $\chi_{\mathbf{q}}(\mathbf{k},t)$ can capture the growing dynamical correlation length in supercooled liquids. The reason is that fluctuations in the density correlator induced by an external field perturb the liquid's dynamics over a distance corresponding to the correlation length of spontaneous fluctuations in the absence of the field \cite{berthier2005direct,biroli2006inhomogeneous,berthier2007spontaneous,berthier2007spontaneousII}. Biroli et al. then proceeded to derive scaling forms for the susceptibility as well as the correlation length in the $\alpha$ as well as $\beta$ relaxation regimes. 

Although the procedure developed by Biroli et al. to evaluate susceptibilities appears cumbersome at first sight, it has certain advantages over the conventional estimation of four-point susceptibilities. First, four-point susceptibilities are more sensitive to the statistical ensemble in which they are computed, compared to those obtained from responses to inhomogeneous fields. Moreover, given that response functions are typically easier to quantify than correlation functions, $\chi_{\mathbf{q}}(\mathbf{k},t)$ offers a potentially facile way of experimentally measuring dynamical correlations in glass-forming liquids. This possibility is reinforced by results on inhomogeneous molecular dynamics simulations \cite{kim2013dynamic}. In this study, the authors considered a system of $N$ particles and subjected it to an external field of the form 
\begin{equation} 
U = h\rho_q(t) = h\sum_{i=1}^{N} \text{exp}[-i\mathbf{q}.\mathbf{r_i}(t)]
\end{equation}
The field induces a change in the two-point density correlator $F(\mathbf{k},\mathbf{q},t) = (1/N)\langle \rho_k(t) \rho_{-k-q}(0)\rangle_U$. The corresponding susceptibility is defined as $\chi_U(\mathbf{k},\mathbf{q},t) = -dF(\mathbf{k},\mathbf{q},t)/dh$. In practice, the strength of the field $h$ can be chosen to be very small, such that it lies in the linear response regime. The susceptibility can then be evaluated in a straightforward manner using the relation
\begin{equation} 
\chi_U(\mathbf{k},\mathbf{q},t) = -[F(\mathbf{k},\mathbf{q},t)|_h - F(\mathbf{k},\mathbf{q},t)|_{h=0}]/h
\end{equation}
where the first correlator is calculated in the presence of the small external field whereas the second one is evaluated in the absence of the field. Using this procedure, the authors verified that four-point susceptibilities as well as those obtained from inhomogeneous MCT yield the same growing dynamical correlation length. Further, the simulation results recapitulate the qualitative phenomenology of inhomogeneous MCT, but the observed critical exponents differ substantially from the predicted ones. Given these numerical observations, it is desirable to systematically test the predictions of inhomogeneous MCT in colloid experiments. Fortunately, the dynamical regime accessible to colloid experiments is precisely the one where MCT predictions are expected to hold. It should therefore be feasible to test various MCT predictions, such as those for the scaling of the correlation length \cite{kim2013dynamic} and coarsening of the three-point susceptibility after a quench to the glass phase \cite{nandi2012glassy}, in colloid experiments. A key advantage of colloidal glass-formers is that they can be subjected to inhomogeneous external fields in a controlled manner. Thus, it is in principle possible to replicate and extend the protocol used in inhomogeneous molecular dynamics simulations by generating a variety of potentials that are either periodic or localized in space, using a suitable technique such as holographic optical tweezers. 

\subsection{Dynamical facilitation}
In the last two years, there have been important theoretical developments that exploit the similarities between the MCT transition in glass-forming liquids and physics of the random field Ising model (RFIM) \cite{nandi2014critical,nandi2015spinodals}. The connection is based on the fact that the MCT transition belongs to the same universality class as the spinodal transition in the RFIM \cite{franz2011field}. In the RFIM, the spinodal transition is dominated by avalanches of spin flips. In the context of glassy dynamics, a spin flip corresponds to a local relaxation event and hence, glassy dynamics is associated with correlated cascades, or avalanches, of relaxation events. From the perspective of analysing dynamical heterogeneities in real space, therefore, one could in principle ignore the microscopic details of MCT and instead focus on the statistics of these localized relaxation events. Indeed, such an approach, has been developed independently well before the advent of inhomogeneous MCT and is known as the Dynamical Facilitation (DF) theory of glass formation \cite{garrahan2002geometrical,chandler2010dynamics}. The basic idea of facilitation is that a glass-forming liquid is largely composed of immobile regions interspersed with fewer mobile ones. In time, these mobile regions transfer their mobility to neighboring regions and themselves become immobile. In this manner, the presence of a mobile region `facilitates' structural relaxation in its neighborhood. The physics encoded in the facilitation picture can be well-captured by a class of spin systems known as Kinetically Constrained Models (KCMs) \cite{ritort2003glassy}. A crucial aspect of this family of models is that they often possess trivial thermodynamic properties and hence, no finite temperature thermodynamic phase transitions. However, by virtue of externally imposed kinetic constraints on spin flips, they can reproduce a wide array of dynamical features exhibited by glass-forming liquids including slow relaxation \cite{ritort2003glassy}, the breakdown of the Stokes-Einstein relation \cite{jung2004excitation}, spatially heterogeneous dynamics \cite{garrahan2002geometrical} and growing dynamic correlations \cite{chandler2006lengthscale}. The first models of this type were proposed by Fredrickson and Andersen \cite{fredrickson1984kinetic,fredrickson1985facilitated}. These models are typically defined on a d-dimensional hypercubic lattice and are characterized by a trivial Hamiltonian of the form
\begin{equation}
H = \sum_i n_i
\end{equation}
where $n_i$ and $n_j$ can assume one of the two values 0, i.e. `down spin state' and 1, i.e. `up spin state'. Physically, lattice sites with $n_i =$ 0 correspond to localized regions of low particle mobility and those with $n_i =$ 1 correspond to regions of high mobility. Thus, KCMs are coarse-grained models and their Hamiltonian does not capture microscopic inter-particle interactions. Obviously, the Fredrickson-Andersen model has no thermodynamic phase transitions and the equilibrium concentration of mobile regions is given by 
\begin{equation}
c_{\text{eq}} = \langle n_i \rangle = 1/(1+\text{e}^\beta)
\end{equation}
where $\beta = 1/k_BT$, $k_B$ being the Boltzmann constant. In these models, the glass transition corresponds to the complete loss of mobility, i.e. $c_{\text{eq}} =$ 0. This happens only at $T =$ 0 and hence, the model does not have a finite temperature glass transition. The main idea of Fredrickson and Andersen was that a given region in the system cannot reorganize unless it has a sufficient number of abutting mobile regions to facilitate this reorganization. In other words, the spin at the ith site cannot flip until and unless it has $f \geq$ 1 nearest neighbors in the up spin state. On decreasing the temperature, the number of up spin states also decreases, and it becomes increasingly difficult for any region to have $f$ mobile neighbors. This directly results in a slowdown of dynamics. It is fairly intuitive, that since the non-trivial dynamics of KCMs is dictated by the kinetic constraint, a change in the nature of the constraint leads to profound changes in the model's glassy dynamics. Indeed, models with vastly different constraints have been thoroughly explored \cite{ritort2003glassy}. 

The notion of mobile regions inducing motion in their neighborhood, termed dynamical facilitation, being the origin of slow dynamics forms the cornerstone of the eponymous DF theory. According to the DF theory, all features of glassy dynamics can be explained through a combination of facilitated dynamics and the reduction in $c_{\text{eq}}$ with temperature. The DF theory also makes the rather strong assumption that no motion is possible without facilitation. In other words, it claims that mobility defects are essentially conserved during relaxation. This assumption distinguishes the DF theory from other approaches such as the Random First-Order Transition theory, in which facilitation as a relaxation process is present, but plays a subordinate role \cite{bhattacharyya2008facilitation}. In order to test these assumptions in simulations and experiments, it was necessary to develop a systematic method for identifying mobile regions, henceforth termed excitations, as well as to quantify the degree of facilitation in glass-forming liquids. Owing to these daunting challenges, the DF theory remained confined to spin models for several years. One of the earliest attempts to quantify facilitation in atomistic glass-forming liquids was made by Glotzer and coworkers, who defined the mobility transfer function $M(\Delta t)$ \cite{vogel2004spatially}. Given two successive time intervals of fixed duration $\Delta t$, the mobility transfer function $M(\Delta t)$ quantifies the excess probability that a mobile particle in the second interval to be located close to a mobile particle in the first one, relative to the probability that it is located near an immobile particle. Mathematically,
\begin{equation}
M(\Delta t) = \frac{\int_{0}^{r_{min}}P_{M}(r,\Delta t)dr}{\int_{0}^{r_{min}}P_{M}^{\ast}(r,\Delta t)dr}
\label{MTF}
\end{equation} 
where $r_{min}$ corresponds to the first minimum of the radial pair correlation function $g(r)$. $P_{M}(r,\Delta t)$ is the probability that $r$ is the minimum distance between a mobile particle in the second interval and the set of mobile particles in the first interval. Similarly, $P_{M}^{\ast}(r,\Delta t)$ is the probability that $r$ is the minimum distance between a mobile particle in the second interval and a set of randomly chosen immobile particles in the first interval. If the DF approach is correct, mobile regions must facilitate relaxation in their neighborhood and hence, two sets of mobile particles in successive intervals must lie close to each other. This in turn implies that $M(\Delta t)$ must exhibit a large value. Moreover, since the dynamics of mobile particles are maximally correlated over a timescale $\Delta t_{max} = t^{*}$, one should expect $M(\Delta t)$ to exhibit a maximum at $t^{*}$. Glotzer and coworkers observed that this was indeed the case in their simulations. They further argued that the growth of $M(t^{*})$ with decreasing temperature signalled the growing importance of facilitation \cite{vogel2004spatially}. For colloid experiments, the mobility transfer function was computed much later \cite{gokhale2014growing} and as we shall see in forthcoming sections, its evolution with area fraction helps determine the dynamical range over which the facilitation approach is valid \cite{nagamanasa2015direct}. 

While the mobility transfer function is an extremely valuable tool, it does not offer any insight into the nature of localized mobile regions, or excitations, which form the building blocks of structural relaxation within the DF approach. In particular, It should be possible to visualize facilitation as the concerted motion of these elementary dynamical objects. Motivated by these considerations, Candelier, Dauchot and Biroli provided an operational definition of excitations by analysing data from experiments on driven granular media \cite{candelier2009building}. While driven granular media are intrinsically non-equilibrium systems, they do exhibit striking similarity with glass-forming liquids \cite{keys2007measurement} and can therefore provide useful insights into glassy dynamics. Recognizing the fact that particle motion in glass-forming liquids is divided into long periods of quiescent cage-rattling interspersed with rare sporadic cage-breaking events, the authors of \cite{candelier2009building} devised a procedure to detect time instants when particles escaped from their cages, or underwent `cage jumps'. These cage jumps are clearly localized in space and occur over a very short duration. Further, they are clustered in space as well as time and the authors showed that rapid chains of successive cage jumps in the form of avalanches also occur \cite{candelier2009building}. This behavior was also confirmed in a simulated supercooled liquid \cite{candelier2010spatiotemporal}. These results indicate that cage jumps are fairly promising candidates for excitations in the DF scenario. To quantify facilitation, cage jumps that occurred within a small threshold both in space and time \cite{candelier2010dynamical} were linked to form a network whose nodes are cage jumps and whose links signify correlations between neighboring cage jumps. Further, the facilitation time $\tau_{fac}$ which quantifies how long the growing network of correlated cage jumps remains connected was also investigated. Since $\tau_{fac}$ actually \textit{decreases} with increasing area fraction, the authors concluded that facilitation diminishes in importance on approaching the glass transition \cite{candelier2010dynamical}.  

\begin{figure}
\centering
  \includegraphics[width=\textwidth]{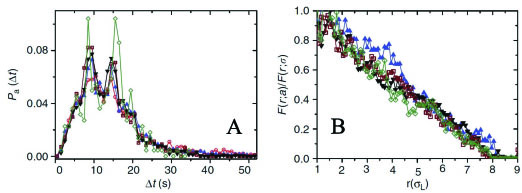}
  \caption{Excitations are localized in space and time. (A) Distribution of instanton times $P_a(\Delta t)$ and (B) $F(r;a)/F(r;\sigma)$ for $a =$ 0.5$\sigma_S$ for various area fractions $\phi$. Adapted from \cite{gokhale2014growing}.}
  \label{Figure24}
\end{figure}

While the analysis of Candelier et al. is certainly insightful, it ignores a crucial ingredient of cage-breaking events, namely the magnitude of the displacement associated with cage jumps. Correlations between cage jumps will in general depend on this magnitude and hence, one must take this information into account while quantifying facilitation. This difficulty was resolved in a landmark paper by Chandler and coworkers in 2011 \cite{keys2011excitations}. In this work, using atomistic simulations, the authors provided a rigorous procedure to identify excitations and provided evidence for deep connections between the dynamics of atomistic glass-formers and that of the East Model, a KCM with a directional kinetic constraint \cite{evans2002anomalous}. Many of these predictions have been tested in recent colloid experiments \cite{gokhale2014growing}. Much like the procedure adopted in \cite{candelier2009building}, the protocol developed by Chandler and coworkers \cite{keys2011excitations} takes advantage of the fact that particle trajectories in glass-forming liquids can be decomposed into periods corresponding to cage-rattling and cage-breaking. In this procedure, a particle is said to be associated with an excitation of size $a$ and instanton time duration $\Delta t$, if it makes a jump of magnitude $a$ over $\Delta t$ and remains in its initial as well as final position for at least $\Delta t$. This procedure can be formally applied to coarse-grained particle trajectories $\bar{\textbf{r}}_{i}(t)$ in order to compute the functional
\begin{equation}
h_{i}(t,t_{a};a) = \prod\limits_{t' = t_{a}/2 - \Delta t}^{t_{a}/2} \theta(|\bar{\textbf{r}}_{i}(t+t')-\bar{\textbf{r}}_{i}(t-t')| - a)
\end{equation}
where $\theta(x)$ is the Heaviside step function and the commitment time $t_{a}$ is typically chosen to be $\sim$ 3-4 times the mean value of $\Delta t$ for an excitation of size $a$ \cite{keys2011excitations,gokhale2014growing}. Coarse-graining the trajectories before computing the functional ensures that vibrational motion within cages is not spuriously counted as an excitation. $h_{i}(t,t_{a};a)$ is 1 whenever a particle is associated with an excitation and zero otherwise. It is important to establish that excitations so defined are localized both in space and time. The characteristic timescale associated with excitations is the instanton time $\Delta t$. Temporal localization of excitations demands that the distribution of these instanton times should not depend on temperature or area fraction. Moreover, the mean instanton time for large area fractions must be significantly smaller than the structural relaxation time $\tau_{\alpha}$. Both these conditions are satisfied in simulations \cite{keys2011excitations} as well as experiments \cite{gokhale2014growing} (Fig. \ref{Figure24}A). 
The spatial extent of excitations can be gauged by the distance over which the displacement field is affected by the presence of an excitation. This can be mathematically expressed using the function
\begin{multline}
\mu(r,t,t';a) = \frac{1}{\langle h_{1}(0,t_{a};a) \rangle} \Bigg\langle h_{1}(0,t_{a};a) \sum\limits_{i \neq 1}^{N}|\bar{\textbf{r}}_{i}(t')-\bar{\textbf{r}}_{i}(t)|\delta(\bar{\textbf{r}}_{i}(t) - \bar{\textbf{r}}_{1}(t) - \bar{\textbf{r}}) \Bigg\rangle
\label{MuFunction}
\end{multline}   
For $t = -t_{a}/2$ and $t' = t_{a}/2$, $\mu(r,t,t';a)$ yields the displacement density at a distance $r$ from an excitation of size $a$ located at the origin at time $t = 0$, over a time interval $t_{a}$ centred on $0$. From $\mu(r,t,t';a)$, the spatial extent of excitations was then extracted by defining the function
\begin{equation}
F(r;a) = \frac{\mu(r,-t_{a}/2,t_{a}/2;a)}{g(r)\mu_{\infty}(t_{a})} - 1
\end{equation} 
where $g(r)$ is the radial pair-correlation function and $\mu_{\infty}(t_{a}) = \langle|\bar{\textbf{r}}_{i}(t+t_{a}) - \bar{\textbf{r}}_{i}(t)|\rangle$. $F(r;a)$ decays within 8 particle diameters irrespective of area fraction $\phi$ (Fig. \ref{Figure24}B). This confirms that excitations are localized in space.

\begin{figure}
\centering
  \includegraphics[width=\textwidth]{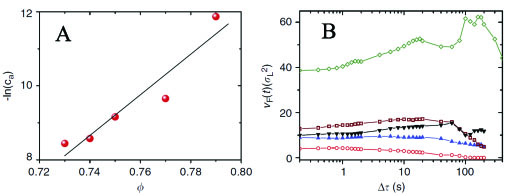}
  \caption{Concentration of excitations and facilitation volumes. (A) Concentration of excitations $c_a$ and (B) facilitation volume $v_F(t)$ for $a =$ 0.5$\sigma_S$ for various area fractions $\phi$. Adapted from \cite{gokhale2014growing}.}
  \label{Figure25}
\end{figure}

The first of two core ingredients of the DF theory is the concentration of excitations $c_a$, defined as
\begin{equation}
c_{a} = \Bigg\langle \frac{1}{Vt_{a}} \sum\limits_{i=1}^{N} h_{i}(0,t_{a};a)\Bigg\rangle
\end{equation}  
where $V$ is the volume and $N$ is the total number of particles. $c_{a}$ is expected to decrease on approaching the glass transition. Chandler and coworkers have shown in several simulated glass-formers that $c_a \propto \text{exp}[-J_a(1/T - 1/T_0)]$, where $J_a$ is the formation energy of excitations of size $a$ and $T_0$ is the high temperature onset of glassy dynamics. Although no such predictions for $c_a$ as a function of $\phi$ exist, if the DF theory is correct, one should expect a significant reduction in $c_a$ with increasing $\phi$. Fig. \ref{Figure25}A shows that this is indeed the case.

To quantify the degree of facilitation, Chandler and coworkers introduced the concept of facilitation volume $v_F(t)$, defined as 
\begin{equation}
v_{F}(t) = \int\Bigg[\frac{\mu(r,t_{a}/2,t;a)}{g(r)\mu_{\infty}(t-t_{a}/2)} - 1\Bigg]d\textbf{r}
\end{equation} 
For all temperatures and area fractions considered in the numerical and experimental work respectively, $v_{F}(t)$ initially increases with time, reaches a maximum value $v_{F}^{max}$ at time $t_{max}$ and then decreases at longer times (Fig. \ref{Figure25}B). Further, $t_{max}$ as well as $v_{F}^{max}$ increase with $\phi$ (Fig. \ref{Figure25}B). These observations are consistent with the growing importance of facilitation on approaching the glass transition, although they do not explicitly demonstrate the conservation of mobility defects \cite{elmatad2012manifestations}.

Since dynamical heterogeneity is central to DF theory, excitation dynamics should be able to capture the patterns of heterogeneous dynamics that have been reproducibly observed in experiments and simulations. Chandler and coworkers have demonstrated numerically that string-like cooperative motion can indeed emerge from heterogeneous dynamics. In an earlier study, Glotzer and coworkers had shown \cite{gebremichael2004particle} that string-like cooperative motion can be decomposed into small units called microstrings, which comprise of a few particles and whose size does not change on approaching the glass transition. Chandler and coworkers postulated that excitation dynamics was akin to microstrings, a claim verified experimentally in \cite{gokhale2014growing}. Further, it was shown that the average string length is proportional to the mean separation between excitations \cite{keys2011excitations}. The vectorial nature of particle displacement lends a directional character to excitation dynamics, which leads the authors to postulate that structural relaxation in glass-forming liquids is consistent with the dynamics of the East model \cite{evans2002anomalous}. The apparent success of the parabolic law in fitting relaxation time data for numerous simulated glass-formers (Eqn. \ref{DFFit}) supports this conclusion \cite{keys2011excitations}. 

One of the most important numerical findings of \cite{keys2011excitations} is a logarithmic hierarchy of excitation energy scales $J_a$, which can be expressed as
\begin{equation}
J_a - J_{a'} = \gamma J_{\sigma}\text{ln}(a/a')
\end{equation}
where $\gamma$ and $J_{\sigma}$ are material dependent constants. This hierarchy ultimately leads to the parabolic temperature dependence of the relaxation time. As a result, verifying the existence of such a hierarchy in experiments would certainly build a stronger case for the DF scenario as the correct theory of glass formation. However, it is prohibitively difficult to obtain adequate statistics over a sufficiently broad dynamical range to verify this prediction directly. Moreover, as mentioned in the beginning of this article, the success of the parabolic law over the limit dynamical range accessible to experiments and simulations does not provide conclusive evidence in favor of the DF approach. 

Given the difficulty in verifying quantitative predictions even in the simplest of glass-formers, a promising alternative is to examine whether the DF theory can make qualitatively correct predictions in a more complex system. This approach was adopted by Mishra et al. \cite{mishra2014dynamical} to investigate glass formation in suspensions of colloidal ellipsoids interacting via purely repulsive as well as attractive interactions from the perspective of the DF theory. Towards this end, the authors analyzed data from experiments that showed the existence of two-step and re-entrant glass transitions in suspensions of  ellipsoids interacting via attractive depletion interactions \cite{mishra2013two}. Since ellipsoids have rotational as well as translational degrees of freedom, one must define and characterize translational as well as rotational excitations. Accordingly, by analyzing data for ellipsoids with purely repulsive interactions, Mishra et al. first defined translational and rotational excitations using a procedure identical to that employed for spheres \cite{keys2011excitations,gokhale2014growing} and demonstrated that they are localized in space and time. Further, the authors showed that the concentration of excitations decreases whereas the facilitation volume increases for rotational as well as translational degrees of freedom on approaching the glass transition \cite{mishra2014dynamical}. A vital aspect of glass formation in ellipsoids is that translational and rotational relaxation are coupled to each other. Thus, a translational excitation will influence the rotational field in its vicinity and vice versa. To understand the nature of this coupling within the framework of facilitation, the authors defined the following functions
\begin{eqnarray}
\label{MixedMu}
\nonumber\mu_{r\theta}(r,t,t';a_r) = \frac{1}{\rho\mu_{\infty}^{r}(t'-t)\langle h_{1}^{r}(0,t_{r};a_r) \rangle} \Bigg\langle h_{1}^{r}(0,t_{r};a_r) \times \sum\limits_{i \neq 1}^{N}|\bar{\theta}_{i}(t')-\bar{\theta}_{i}(t)|\delta(\bar{\textbf{r}}_{i}(t) - \bar{\textbf{r}}_{1}(t) - r) \Bigg\rangle
\\
\mu_{\theta r}(r,t,t';a_{\theta}) = \frac{1}{\rho\mu_{\infty}^{\theta}(t'-t)\langle h_{1}^{\theta}(0,t_{\theta};a_{\theta}) \rangle} \Bigg\langle h_{1}^{\theta}(0,t_{\theta};a_{\theta}) \times \sum\limits_{i \neq 1}^{N}|\bar{\textbf{r}}_{i}(t')-\bar{\textbf{r}}_{i}(t)|\delta(\bar{\textbf{r}}_{i}(t) - \bar{\textbf{r}}_{1}(t) - r) \Bigg\rangle
\end{eqnarray}
$\mu_{r\theta}(r,t,t';a_r)$ quantifies the effect of translational excitations on the rotational displacements in the neighborhood whereas $\mu_{\theta r}(r,t,t';a_{\theta})$ quantifies the impact of rotational excitations on translational displacements. An interesting finding to emerge from this analysis was that facilitation in the two degrees of freedom is not symmetric. In particular, translational excitations have a much stronger impact on rotational dynamics than vice versa \cite{mishra2014dynamical}.

\begin{figure}
\centering
  \includegraphics[width=\textwidth]{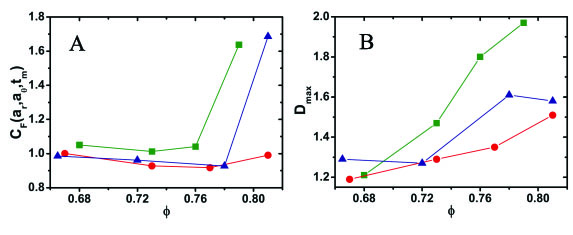}
  \caption{Decoupling of translational and rotational facilitation and dynamical heterogeneities. The coupling coefficient for facilitation, $C_{F}(a_r,a_{\theta},t_m)$, with $a_r = 0.5l$ and $a_{\theta} = 20^{\circ}$ (A) and the coupling coefficient for dynamical heterogeneities, $D_{\text{max}}$ (B), as a function of $\phi$ for $\Delta U/k_BT =$ 0 ({\color{green!50!black} $\boldsymbol \blacksquare$}), $\Delta U/k_BT =$ 1.16 ({\color{red} $\boldsymbol \bullet$}) and $\Delta U/k_BT =$ 1.47 ({\color{blue!60!black} $\boldsymbol \blacktriangle$}). Adapted from \cite{mishra2014dynamical}.}
  \label{Figure26}
\end{figure}

As described in the section on MCT, attractive interactions have a profound influence on glass formation in colloidal ellipsoids. Based on the MCT scaling of relaxation times, Mishra et al. had shown that with increasing strength of attractive interactions $\Delta u$, the single glass transition in the repulsive case first splits into two glass transitions, which then recombine with further increase in $\Delta u$. In \cite{mishra2014dynamical}, the authors found that this behavior was associated with a decoupling and subsequent re-coupling between translational and rotational facilitation, which is in turn manifested as a decoupling and re-coupling of dynamical heterogeneities in the two degrees of freedom. To demonstrate this, the authors computed two coupling coefficients, one for facilitation and one for heterogeneities. The functions $\mu_{ij}(r,t,t';a_i)$ with $i,j \in \lbrace r,\theta \rbrace)$, defined in Eqn. \ref{MixedMu} suggest a natural way to quantify the extent to which facilitation in rotational and translational degrees of freedom is coupled. The height of the first peak of these functions, denoted by $\mu_{ij}^{max}(t) = \mu_{ij}(\sigma,-t/2,t/2;a_i)$, where $\sigma$ is the first peak of $g(r)$, signifies the impact of an excitation on the translational or rotational displacement field in its immediate vicinity. If excitations in one degree of freedom affect relaxation in the other strongly, the off-diagonal functions ($i \neq j$) will have strong peaks. On the other hand, if facilitation in the two degrees of freedom is decoupled, only the diagonal functions ($i = j$), defined analogously to the off-diagonal ones in Eqn. \ref{MixedMu}, will have strong peaks. With this physical insight in mind, the authors defined the coupling coefficient for facilitation as 
\begin{equation}
C_{F}(a_r,a_{\theta},t_m) = \frac{\mu_{r\theta}^{max}(t_m)\mu_{\theta r}^{max}(t_m)}{\mu_{rr}^{max}(t_m)\mu_{\theta\theta}^{max}(t_m)}
\end{equation}
The relevant timescale for this analysis is the commitment time and hence $t_m = \text{max}(t_r,t_{\theta})$. With increasing $\Delta u$, $C_{F}(a_r,a_{\theta},t_m)$ first decreases for all $\phi$ and later begins to increase again, showing that rotational and translational facilitation get decoupled and are subsequently re-coupled (Fig. \ref{Figure26}A).
The DF theory claims that cooperative rearrangements emerge from the facilitated dynamics of excitations. If this claim is true, one should expect that the decoupling of rotational and translational facilitation should result in the decoupling of dynamical heterogeneities in the two degrees of freedom. To examine whether this is the case, the authors defined a coupling coefficient for dynamical heterogeneities \cite{mishra2014dynamical}. The authors first defined the function
\begin{equation}
D(\Delta t) = \frac{\int_{0}^{r_{min}}P(r,\Delta t)dr}{\int_{0}^{r_{min}}P^{\ast}(r,\Delta t)dr}
\end{equation}
This function is very similar to the mobility transfer function (Eqn. \ref{MTF}), except that instead of probing the temporal correlation between mobile particles in two successive time intervals, it probes the \textit{spatial} correlation of translational and rotational mobile particles in the \textit{same} interval. Specifically, for an interval of given duration $\Delta t$, $P(r,\Delta t)$ measures the probability that $r$ is the minimum distance of a translationally mobile particle from a set of rotationally mobile particles and $P^{\ast}(r,\Delta t)$ is the reference distribution of minimum distances of translationally mobile particles from a set of randomly chosen immobile particles. Just as in the case of the mobility transfer function, $D(\Delta t)$ exhibits a maximum near the cage-breaking time $t^{*}$ and hence, the maximum value $D_{max}$ is an appropriate measure of the spatial correlation between rotational and translational heterogeneities. Strikingly, $D_{max}$ first decreases with $\Delta u$ and then increases again (Fig. \ref{Figure26}B), strongly suggesting that the decoupling in facilitation leads to a decoupling of dynamical heterogeneities. 

\begin{figure}
\centering
  \includegraphics[width=\textwidth]{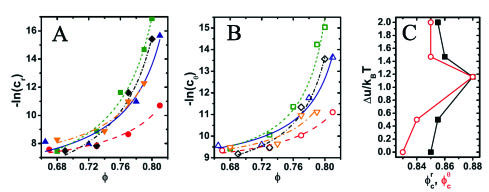}
  \caption{Prediction of re-entrant glass transitions from concentration of excitations. (A) The dependence of the concentration of translational excitations $c_r$ on $\phi$ for $a_r = 0.5l$, for $\Delta U/k_BT =$ 0 ({\color{green!50!black} $\boldsymbol \blacksquare$}), $\Delta U/k_BT =$ 0.47 ({\color{black} $\boldsymbol \blacklozenge$}), $\Delta U/k_BT =$ 1.16 ({\color{red} $\boldsymbol \bullet$}), $\Delta U/k_BT =$ 1.47 ({\color{blue!60!black} $\boldsymbol \blacktriangle$}) and $\Delta U/k_BT =$ 1.95 ({\color{orange} $\boldsymbol \blacktriangledown$}). (B) The dependence of the concentration of rotational excitations $c_{\theta}$ on $\phi$ for $a_{\theta} = 20^{\circ}$ for $\Delta U/k_BT =$ 0 ({\color{green!50!black} $\boldsymbol \square$}), $\Delta U/k_BT =$ 0.47 ({\color{black} $\boldsymbol \lozenge$}), $\Delta U/k_BT =$ 1.16 ({\color{red} $\boldsymbol \circ$}), $\Delta U/k_BT =$ 1.47 ({\color{blue!60!black} $\boldsymbol \triangle$}) and $\Delta U/k_BT =$ 1.95 ({\color{orange} $\boldsymbol \triangledown$}). In (A) and (B), the concentrations $c_r$ and $c_{\theta}$ respectively are reported in units of $l^{-2}s^{-1}$. The curves are empirical fits of the form $\phi_0 + A(\phi_c - \phi)^{-1}$. (C) The translational glass transition $\phi_{c}^{r}$ ({\color{black} $\boldsymbol \blacksquare$}) and rotational glass transition $\phi_{c}^{\theta}$ ({\color{red} $\boldsymbol \circ$}) obtained from fits to the curves in (A) and (B), for various values of $\Delta U/k_BT$. Adapted from \cite{mishra2014dynamical}.}
  \label{Figure27}
\end{figure}

Mishra et al. have shown using MCT scaling of relaxation times that suspensions of ellipsoids exhibit re-entrant glass transitions with increasing $\Delta u$ \cite{mishra2013two}. If the DF approach is valid, relaxation time is determined by the concentration of excitations and hence, in principle, one should be able to predict re-entrant glass transitions simply by observing the evolution of these concentrations with $\phi$. Fig. \ref{Figure27}A-B shows the variation of the concentration of translational excitations ($c_r$) as well as rotational ones ($c_{\theta}$) with $\phi$. The concentrations appear to vanish at finite values of $\phi$, which corresponds to a divergence in the plot of $-\text{ln}(c_{r,\theta})$. Empirical fits of the form $\phi_0 + A(\phi_c - \phi)^{-1}$ to these data allowed the authors to predict the rotational and translational glass transitions $\phi_c^{\theta}$ and $\phi_c^{r}$, respectively (Fig. \ref{Figure27}C). The shape of the glass transition lines in the $\Delta u$-$\phi$ plane clearly demonstrates the presence of re-entrant transitions. It is crucial to note that the prediction from DF theory is based solely on the concentration of spatiotemporally localized objects, whose relevant timescale is much smaller than the structural relaxation time. This result demonstrates that the DF theory can account for complex glassy phenomenology in ellipsoids with attractive interactions, thereby bolstering the case of facilitation as the correct theoretical scenario for glass formation. 

\subsection{Random first-order transition theory}
MCT and DF are purely dynamical theories that do not anticipate any structural changes to accompany the glass transition. Naturally, these theories maintain the view that glass formation is not associated with any underlying thermodynamic phase transition. There are nonetheless several good reasons to challenge this viewpoint. Perhaps the most important among these is the observed correspondence between the Kauzmann temperature $T_K$ and the VFT temperature $T_0$ (Fig. \ref{Figure8}). $T_0$ is extracted from fits to the temperature dependence of viscosity, and therefore captures the evolution of dynamics on approaching the glass transition. The Kauzmann temperature on the other hand is obtained from measurements of the excess entropy of the supercooled liquid over the thermodynamically stable crystalline state, and therefore has a well-defined thermodynamic significance. As a result, it is not unreasonable to expect the correspondence between $T_0$ and $T_K$ to be indicative of a correspondence between dynamics and thermodynamics. The first formal connection between the relaxation time and excess entropy of the liquid, also known as the configurational entropy $s_c$, was made by Adam and Gibbs \cite{adam1965temperature}. The Adam Gibbs relation reads
\begin{equation}
\tau_{\alpha} = \tau_0\text{exp}\Bigg( \frac{A}{Ts_c(T)}\Bigg)
\end{equation}
This form predicts a VFT type divergence of $\tau_{\alpha}$ at the Kauzmann temperature, if $Ts_c(T) \propto (T/T_K - 1)$.
To connect dynamics to thermodynamics, Adam and Gibbs invoked the concept of cooperatively rearranging regions (CRRs). In particular, they postulated that the supercooled liquid is composed of CRRs, which are groups of particles that move collectively. Different CRRs are assumed to reorganize independently, which leads to an inverse relation between the size of CRRs and the configurational entropy $s_c$. The theory further postulates that the glass transition is associated with a vanishing of $s_c$ at the Kauzmann Temperature \cite{kauzmann1948nature} and a concomitant divergence in the CRR size. However, the Adam Gibbs theory neither provides a precise definition of CRRs not any procedure to measure their size. Nonetheless, the ideas put forth by Adam and Gibbs were instrumental in the development of the Random First-Order Transition theory (RFOT), which placed the notions of growing cooperativity and vanishing $s_c$ on a stronger thermodynamic footing, through inputs from spin glass physics as well as MCT. While RFOT ranks among the most prominent theories of glass formation, it has far-reaching applications in a number of diverse fields such as protein folding and cancer biology \cite{kirkpatrick2015colloquium}. RFOT is a mean field theory that has its origins in the thermodynamics of a certain class of spin systems known as p-spin models \cite{kirkpatrick1987p}. In a mean field treatment, these models have two phase transitions: a dynamic transition associated with ergodicity breaking at temperature $T_c$ that has been formally identified with the MCT transition, and a thermodynamic phase transition associated with vanishing configurational entropy, which occurs at a lower temperature $T_K$. The temperature $T_K$ can be identified with the Kauzmann temperature, as in the Adam-Gibbs scenario. The two transitions can be understood in terms of the nature of the liquid's free energy landscape. In the $T>T_c$ regime, the free energy landscape is characterized by a single minimum corresponding to the homogeneous liquid phase, and the configurational entropy in therefore identically zero. For $T_c > T > T_K$, the free energy landscape fragments into a large number of metastable minima. The number of these minima is exponential in the system size, which gives rise to a finite configurational entropy. Finally, for $T<T_K$, the number of metastable minima becomes sub-extensive, leading to a vanishing of the configurational entropy in the thermodynamic limit. 

While the aforementioned mean field treatment offers a comprehensive thermodynamic picture, it does not provide a satisfactory description of slow dynamics, particularly in the regime $T_c > T > T_K$, where the energy landscape is dominated by metastable minima. Early phenomenological arguments state that in this regime, the system exists as a patchwork, or `mosaic' of metastable amorphous configurations. Structural relaxation corresponds to entropy driven nucleation-like events that transport the system from one metastable configuration to another \cite{kirkpatrick1989scaling}. The activation barrier for this process emerges from the competition between the configurational entropy gain $Ts_c(T)\xi^{d}$ due to melting and the energy gain $\Upsilon\xi^{\theta}$ due to surface tension $\Upsilon$ between the nucleating and surrounding phases. In general, $\theta \leq d-1$, although it was argued by Kirkpatrick, Thirumalai and Wolynes that $\theta =$ 3/2 \cite{kirkpatrick1989scaling}. The competition between entropy and surface tension yields the typical size of the mosaic $\xi^{*(d-\theta)} = \Upsilon/Ts_c(T)$. As the configurational entropy decreases on approaching $T_K$, the mosaic becomes increasingly coarser , and the mosaic length scale $\xi^{*}$ increases. Since the activation barrier for relaxation increases with $\xi^{*}$, the vanishing of $s_c$ at $T_K$ leads to a divergence in the mosaic length, which ultimately results in the divergence of the relaxation time. While these arguments are plausible, they neither offer a precise definition of the mosaic length scale nor prescribe a method to measure it in experiments or simulations. 

This difficulty was overcome by Bouchaud and Biroli \cite{bouchaud2004adam}, who provided both a practical way of computing the mosaic length $\xi^{*}$ as well as a clear physical interpretation, by identifying $\xi^{*}$ with the point-to-set length $\xi_{PTS}$. To evaluate $\xi_{PTS}$, the positions of all particles outside a spherical cavity of radius $R$ are frozen in an equilibrium configuration of the liquid. The dynamical evolution of the remaining free particles in the system is then investigated in the presence of the pinning field generated by the frozen particles. $\xi_{PTS}$ is defined as the minimum radius of the cavity beyond which relaxation at the centre of the cavity is unaffected by the pinning field. Bouchaud and Biroli argued that this point-to-set length in fact embodies the same physics as the mosaic length scale. In particular, for a cavity of radius $R$, one can explore a multiplicity of configurations for the free particles within the cavity, which would lead to an entropic lowering of the free energy by $-Ts_c(T)R^d$. However, all these configurations, except the one corresponding to the frozen boundary of the cavity, will have to be deformed at the boundary to satisfy the constraint imposed by the pinning field. This will lead to a gain in free energy due to the surface tension term $\Upsilon R^{\theta}$. These considerations yield a crossover length that has an identical dependence on $\Upsilon$ and $s_c(T)$ as the mosaic length. However, the Bouchaud-Biroli construction has the advantage that the pinning procedure can be realized in numerical simulations and even in colloid experiments using optical tweezers. 

While the cavity pinning geometry has a direct connection to the mosaic length, other pinning geometries have also been employed in the literature to probe the nature of static as well as dynamic correlations in glass-forming liquids \cite{berthier2012static}. One of these configurations, namely the amorphous wall geometry, is of particular significance since it can be used to probe not just the size, but also the shapes of cooperatively rearranging regions. Neither the originial treatment of Kirkpatrick, Thirumalai and Wolynes \cite{kirkpatrick1989scaling} nor the Bouchaud-Biroli construction provide a dynamical description of the `melting' of amorphous configurations and their transition to new metastable states. On the other hand, the dynamical heterogeneity studies described in the previous sections shed light on the nature of these relaxation events. An interesting and important feature that emerges from numerical as well as experimental studies on dynamical heterogeneities is that the morphology of CRRs is stringy, or fractal-like \cite{donati1998stringlike,weeks2000three}. This is in contradiction with RFOT, which assumes these CRRs to be compact in shape. To account for this discrepancy, Stevenson, Schmalian and Wolynes developed the `Fuzzy Sphere Model' that describes the evolution of the shapes of CRRs on approaching the glass transition \cite{stevenson2006shapes}. Qualitatively, the Fuzzy Sphere Model postulates that CRRs are composite objects that contain a compact core that is dressed by a more ramified string-like shell. The free energy change during a reconfiguration of this fuzzy sphere has an entropic term associated with the multiplicity of configurations accessible to the ramified string-like shell as well as an energetic contribution associated with the breaking of favourable bonds at the surface of the CRRs. At low temperatures, i.e. close to $T_g$, the energetic term dominates, resulting in compact nearly spherical CRRs that minimize the number of broken surface bonds. Conversely, at high temperatures close to the onset of glassy dynamics $T_A$, the entropic term dominates and gives rise to predominantly string-like CRRs. This change in morphology of CRRs is not predicted by other theories of glass formation and can therefore serve as a test for the validity of RFOT.

\begin{figure}
\centering
  \includegraphics[width=\textwidth]{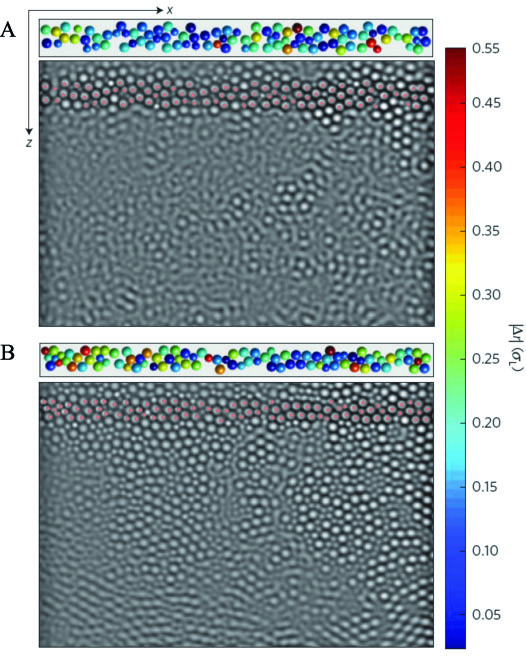}
  \caption{Experimental realization of the amorphous wall pinning geometry. The underlying grey scale images have been generated by time-averaging snapshots over 30$\tau_{\alpha}$ for $\phi = 0.68$ (A) and $\phi = 0.76$ (B). For (A) and (B), the relaxation times $\tau_{\alpha} =$ 12s and 274s, respectively. The red circles correspond to the coordinates of particles forming the amorphous wall. The spheres at the top of the images in (A-B) constitute the pattern whose fast Fourier Transform was fed into the spatial light modulator (SLM). Spheres are colour coded according to the displacement between the input coordinates for creating traps and time-averaged particle positions in units of $\sigma_L$. Adapted from \cite{nagamanasa2015direct}.}
  \label{Figure28}
\end{figure}

\begin{figure}
\centering
  \includegraphics[width=0.7\textwidth]{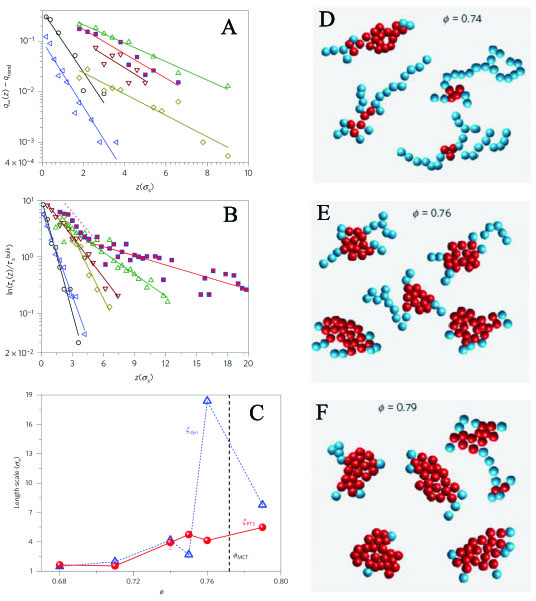}
  \caption{Static and dynamic length scales and shapes of cooperatively rearranging regions in a binary colloidal glass-former. (A) $q_{\infty}(z) - q_{rand}$ versus z for $\phi =$ 0.68 ({\color{black} $\boldsymbol \circ$}), $\phi =$ 0.71 ({\color{blue} $\boldsymbol \triangleleft$}), $\phi =$ 0.74 ({\color{brown} $\boldsymbol \triangledown$}), $\phi =$ 0.75 ({\color[rgb]{0.71,0.65,0.26} $\boldsymbol \diamond$}), $\phi =$ 0.76 ({\color{red} $\boldsymbol \blacksquare$}) and $\phi =$ 0.79 ({\color{green} $\boldsymbol \triangle$}). (B) $\text{log}(\tau_s(z) /  \tau_s^{bulk})$ as a function of z. The colors and symbols in (B) are identical to those in (A). In (A) and (B), the solid lines are exponential fits of the forms given in Eqns. \ref{XiPTS} and \ref{xidyn}, respectively. In (B) for $\phi = 0.76$ ({\color{red} $\boldsymbol \blacksquare$}), $\xi_{dyn}$ was extracted from the asymptotic slope. The dashed red line is a guide to the eye. (C) Point-to-set length scale, $\xi_{PTS}$, ({\color{red} $\boldsymbol \bullet$}) and dynamic length scale, $\xi_{dyn}$, ({\color{blue} $\boldsymbol \triangle$}). The error bars have been obtained from the exponential fits. The dotted black line indicates the mode coupling crossover $\phi_{MCT}$. (B-D) Representative 25-particle clusters of most mobile particles for $\phi =$ 0.74, $\phi =$ 0.76 and $\phi =$ 0.79 respectively. Core-like particles are shown in red and string-like particles are shown in light blue. Adapted from \cite{nagamanasa2015direct}.}
  \label{Figure29}
\end{figure}

\begin{figure}
\centering
  \includegraphics[width=\textwidth]{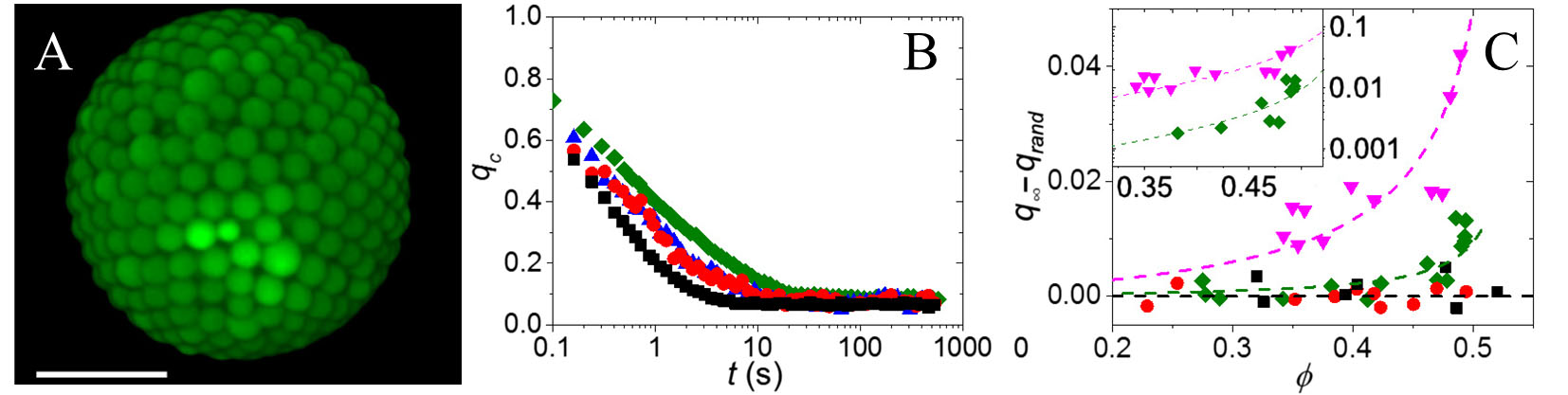}
  \caption{Experimental measurements of growing point-to-set correlations in three dimensions. (A) A 3D reconstruction of the binary colloidal suspension of fluorescent PMMA particles within an emulsion droplet of radius 5.4$\sigma_S$ from confocal microscopy. The volume fraction is $\phi = 0.565$. (B) Time evolution of the configurational overlap $q_c$ for $\phi = 0.41$ for various cavity radii R = 8.5$\sigma_S$ (olive), R = 16.5$\sigma_S$ (blue), R = 32.5$\sigma_S$ (red) and bulk (black). (C) Difference between the asymptotic value of the configurational overlap $q_{\infty}$ in the cavity geometry and its corresponding bulk value $q_{rand}$ for various cavity radii  R = 4.5$\sigma_S$ (magenta), R = 8.5$\sigma_S$ (olive), R = 32.5$\sigma_S$ (red) and bulk (black). The dashed curves are guides to the eye. The inset shows the same data on a semi-log plot. Adapted from \cite{zhang2016structures}.}
  \label{Figure30}
\end{figure}

From the preceding paragraphs, it is evident that the two most important predictions of RFOT that can be directly tested in experiments are the presence of a growing static point-to-set length and a crossover in the morphology of CRRs from string-like to compact form. The former has been verified several times in a variety of simulated glass-formers and for various pinning geometries. The first indirect numerical evidence for the latter in the form of non-monotonic temperature evolution of dynamic correlations was provided by Kob, Rold\'{a}n-Vargas and Berthier \cite{kob2012non}. Recently, both of these predictions have been directly verified in colloid experiments by Nagamanasa et al. \cite{nagamanasa2015direct}, using holographic optical tweezers (HOT) to realize the amorphous wall geometry. The experimental realization of any pinning geometry requires several colloidal particles to be held in place simultaneously, ideally over timescales much longer than $\tau_{\alpha}$. A facile way to meet this requirement is to use a HOT set-up, which facilitates the generation of multiple optical traps at desired locations \cite{dufresne2001computer,curtis2002dynamic,grier2003revolution,spalding2008holographic}. The key component of the HOT set-up is a spatial light modulator (SLM), which basically consists of an array of pixels. Each pixel in the array comprises of a liquid crystalline material whose polarization can be controlled by applying a local electric field. This field at each point in the array is determined by the input pattern that is fed to the SLM. The polarized liquid crystalline elements modulate the phase within parts of the incident laser beam in such a way that the interference pattern formed after reflection from the SLM corresponds to the spatial Fourier transform of the input pattern \cite{reicherter1999optical,liesener2000multi}. A suitable combination of lenses then allows the desired configuration of traps to be created in the focal plane of the microscope objective. In this manner, a variety of trap configurations can be created. In their experiments, Nagamanasa et al. generated the input pattern by identifying particle coordinates lying within a strip parallel to the X-axis (Fig. \ref{Figure28}). 
To verify the predictions of RFOT, the authors closely followed the procedure adopted in \cite{kob2012non}. To extract the static point-to-set correlation length $\xi_{PTS}$, they divided the field of view into a grid with mesh size 0.25$\sigma_S$, where $\sigma_S$ is the diameter of the smaller colloids in the binary mixture of polystyrene particles used the in the experiments \cite{nagamanasa2015direct}. Next, they defined the configurational overlap $q_c(t,z)$ at a distance $z$ from the wall as
\begin{equation}
q_c(t,z) = \frac{\sum_{i(z)}\langle n_i(t)n_i(0)\rangle}{\sum_{i(z)} \langle n_i(0)\rangle}
\end{equation} 
where $\langle  \rangle$ denotes time averaging, $n_{i}(t) = 1$ if cell $i$ in the grid is occupied by a particle at time $t$ and $n_{i}(t) = 0$ otherwise. As expected from simulations \cite{kob2012non}, at long times, $q_c(t,z)$ approaches its asymptotic value $q_{\infty}(z)$. Moreover, $q_{\infty}(z)$ decays exponentially with $z$ for all area fractions $\phi$ considered (Fig. \ref{Figure29}A), which allows the point-to-set length to be defined via the relation \cite{scheidler2002growing,berthier2012static,kob2012non}
\begin{equation}
q_{\infty}(z) - q_{rand} = B\:\text{exp}(-z/\xi_{PTS})
  \label{XiPTS}
\end{equation}
Here, $q_{rand}$ is the mean overlap between two uncorrelated configurations, and corresponds to the probability of occupation of a cell. $\xi_{PTS}$ was observed to grow with $\phi$ (Fig. \ref{Figure29}C), a result that constitutes the first experimental evidence for growing point-to-set correlations in glass-forming liquids. 
In a recent experimental study \cite{zhang2016structures}, Zhang and Cheng have provided experimental evidence of a growing point-to-set length scale in three dimensions using the spherical cavity pinning geometry. The cavity geometry is difficult to implement using optical tweezers, especially in 3D, which prompted the authors to adopt a novel approach to realize it in practice. They first emulsified the colloidal suspension in an aqueous gelatin solution at 70$^{\circ}$C. This resulted in a layer of particles getting trapped at the oil-water interface. On cooling, the aqueous phase solidified into a gel, which pinned the particles at the interface in an amorphous configuration, thus forming a spherical cavity whose radius is set by the size of the emulsion droplet. The colloidal suspension itself comprised of fluorescent PMMA particles of two different sizes suspended in a mixture of decalin and cyclohexyl bromide, which matches the density as well as refractive index of the particles. Fig. \ref{Figure30}A shows a 3D reconstruction of the suspension inside a cavity, obtained from a stack of confocal microscopy images. Next, Zhang and Cheng examined the time evolution of the configurational overlap in a small $3.7\sigma_S \times 3.7\sigma_S$ region at the centre of the cavity for various cavity radii, and observed that the decay of the overlap is slower for smaller cavities (Fig. \ref{Figure30}B). Moreover, the asymptotic value of the overlap for small droplets increases much more rapidly with the volume fraction $\phi$, compared to that for large droplets (Fig. \ref{Figure30}C). This observation points towards the existence of a growing static point-to-set correlation length in 3D as well. From the overlap profiles, the authors concluded that the length scale $\xi_{PTS} \geq 8.5\sigma_S$ for $\phi \approx$ 0.47. This value is larger than $\xi_{PTS} \approx 4\sigma_S$ observed in the 2D experiments in the amorphous wall geometry (Fig. \ref{Figure29}C). This difference could be a reflection of the 3D versus 2D nature of these experiments. Another plausible explanation is that the two pinning geometries probe different length scales in the system. Specifically, Cammarota and Biroli have argued \cite{biroli2014fluctuations} that the cavity geometry probes the mosaic length scale associated with domain size within RFOT, whereas the wall geometry probes domain surface fluctuations. The length scale extracted from the cavity geometry is expected to increase faster on approaching the glass transition compared to the one extracted from the wall geometry. The experimental results of Zhang and Cheng \cite{zhang2016structures} and Nagamanasa et al. \cite{nagamanasa2015direct} are consistent with this prediction, although it is worth noting that the prediction itself is expected to hold in a dynamical regime that is much more deeply supercooled than that investigated in these experiments. Finally, we note that the growth in $\xi_{PTS}$ observed in colloid experiments is much stronger than that observed in simulations over a comparable dynamical range \cite{berthier2012static,kob2012non}. While this may be an intrinsic feature of colloidal systems, it could also be due to the lack of averaging over multiple realizations of the quenched disorder in these experiments.  

In addition to static correlations, the wall geometry provides valuable information on the relaxation dynamics. To examine dynamic correlations, Nagamanasa et al. \cite{nagamanasa2015direct} computed the self-overlap $q_s(t,z)$, defined as
\begin{equation}
q_s(t,z) = \frac{\sum_{i(z)}\langle n_{i}^{s}(t)n_{i}^{s}(0)\rangle}{\sum_{i(z)} \langle n_{i}^{s}(0)\rangle}
\end{equation} 
This quantity is similar to the configurational overlap, with the important distinction being that the cell occupation number $n_{i}^{s}(t)$ retains the particle label. This means that the product $n_{i}^{s}(t)n_{i}^{s}(0)$ is 1 only if cell $i$ is occupied by the \textit{same} particle at times 0 and $t$. Due to this constraint, the self-overlap function is analogous to the self-intermediate scattering function $F_s(q,t)$, evaluated at a wave vector that corresponds to the cell size. Like $F_s(q,t)$, $q_s(t,z)$ also decays to zero at long times and one can therefore extract a timescale $\tau_s(z)$ from it at various distances from the wall. The variation of $\tau_s$ with $z$ (Fig. \ref{Figure29}B) allows one to define a dynamic length scale $\xi_{dyn}$ through the relation \cite{scheidler2002cooperative,kob2012non}
\begin{equation}
\text{log}(\tau_s(z)) = \text{log}(\tau_s^{bulk}) + B_s \text{exp}(-z/\xi_{dyn})
\label{xidyn}
\end{equation}
Here, $\tau_s^{bulk}$ is the relaxation time far away from the wall. Interestingly, $\xi_{dyn}$ was found to exhibit a striking non-monotonicity in the vicinity of the mode coupling crossover (Fig. \ref{Figure29}C), in concord with simulations \cite{kob2012non}. The authors of \cite{kob2012non} have argued that unlike other previously defined dynamic correlation lengths, $\xi_{dyn}$ is sensitive to the shapes of CRRs, since the amorphous wall breaks the translational symmetry of space. The non-monotonicity in $\xi_{dyn}$ is then attributed to a change in the shape of CRRs from string-like to compact form. To explore whether the non-monotonicity is indeed related to the shapes of CRRs, Nagamanasa et al. analysed the shapes of clusters of mobile particles \cite{nagamanasa2015direct} for corresponding data sets in the absence of the pinned wall. In particular, they examined the proportion of particles within CRRs that are organized in string-like or compact form. They found that the non-monotonicity in $\xi_{dyn}$ is indeed associated with the increased compaction of CRRs (Fig. \ref{Figure29}D-F), as predicted by the Fuzzy Sphere Model \cite{stevenson2006shapes}. Strictly speaking, within RFOT, activated hopping becomes relevant only at temperatures below the mode-coupling crossover and the transition from string-like to compact CRRs therefore occurs for $T < T_c$ or $\phi > \phi_c$. Nonetheless, the experimental results constitute the first direct evidence for the qualitative change in morphology of CRRs predicted within the framework of RFOT. 

\subsection{Geometric frustration-based approaches}
A different class of thermodynamic theories of the glass transition posit the onset of more physically transparent forms of geometric order compared to the rather abstract mosaic picture of RFOT. A hallmark feature of these approaches is the competition between the proliferation of local structural motifs and geometric frustration, which prevents long-ranged correlations between them. These theories are important from the viewpoint of colloid experiments because they emphasize real space aspects of the structure and dynamics of glass-forming liquids. Theoretical research motivated by the notion of geometric frustration can broadly be divided into two distinct perspectives. The first stems from the pioneering work of Frank and Kasper in the context of crystalline alloys \cite{frank1952supercooling,frank1958complex}. Frank showed that for a set of 13 particles of a monoatomic system interacting via the Lennard Jones potential, the ground state configuration is an icosahedron. This suggests that icosahedra should proliferate in such a system at low temperatures. However, icosahedra are incapable of tiling Euclidean space, owing to the presence of five-fold symmetry. This tradeoff between locally preferred structural order and global tiling of space was exploited by Tarjus and coworkers to develop a thermodynamic theory of glass-formation \cite{kivelson1995thermodynamic,tarjus2005frustration}. The authors postulated the existence of a critical point at a temperature $T^{*} \geq T_m$ which is avoided due to the presence of frustration $K$. This critical point is associated with the ordering of the liquid into a reference crystalline state composed of locally preferred structural units and frustration stems from the inability of these units to tile space. Below $T^{*}$, the system relieves the frustration-induced strain by breaking up into `frustration-limited' domains of average size $R_D$. In addition to $R_D$, the system has a second correlation length $\xi_0$, which diverges at $T^{*}$ in the absence of frustration as $\xi_0 \sim [(T^{*}-T)/T^{*}]^{-\nu}$. Structural relaxation corresponds to the reorganization of these domains. The authors have shown that the activation barrier for relaxation of these domains scales as $(R_D/\xi_0)^2$. Moreover, for the class of spin systems considered by the authors, $R_D \propto \xi_0^{-1}K^{-1/2}$. Thus, the dramatic increase in relaxation time upon cooling results from a combination of increasing domain size and decreasing correlation length. The foregoing arguments imply an inverse dependence of the activation barrier on frustration, which implies that frustration can tune the fragility of the glass-forming liquid. However, the above discussion does not provide insight into how the frustration $K$ itself might be tuned. Tarjus and coworkers have shown that this can in fact be done by curving space \cite{sausset2008tuning}. Icosahedra are incapable of tiling Euclidean space, but they can tile curved space. Curvature therefore relieves frustration and leads to a reduction in $K$. Testing the frustration-limited domain theory in experiments has been difficult, since there is no facile way to tune the degree of frustration in atomic and molecular glass-forming liquids. Colloidal systems on the other hand provide a promising alternative, since the structure and dynamics of 2D colloidal liquids can in principle be investigated on curved surfaces. Indeed, various defect structures in colloidal crystals on surfaces with positive as well as negative Gaussian curvature have already been observed \cite{bausch2003grain,irvine2010pleats,irvine2012fractionalization}.

\begin{figure}
\centering
  \includegraphics[width=0.6\textwidth]{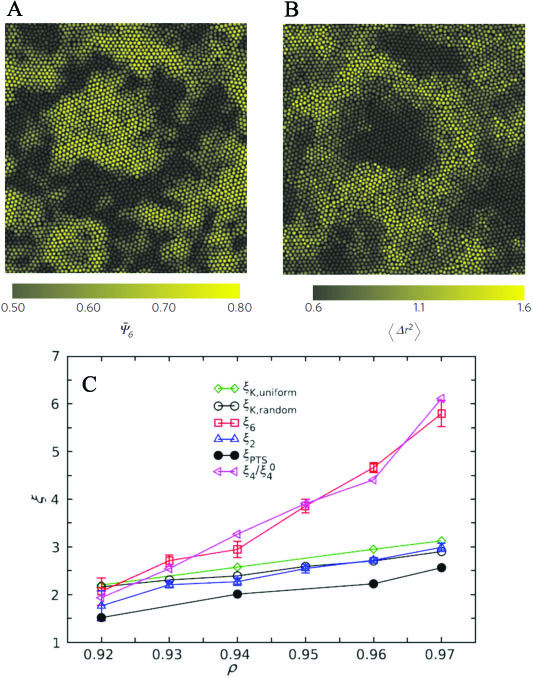}
  \caption{Correlation between Bond-orientational order and dynamical heterogeneity. (A) Spatial distribution of the local bond-orientational order parameter $\psi_6^{j}$ averaged over $\tau_{\alpha}$ for a simulated 2D hard sphere liquid with a polydispersity of 9\%. (B) Mean squared displacement $\langle \Delta r^2(t \rangle)$ for $t = 10\tau_{\alpha}$ for the same liquid. In (A) and (B), area fraction $\phi =$ 0.74. Adapted from \cite{tanaka2010critical}. (C) Comparison of various structural length scales with the dynamic four-point correlation length $\xi_4$ for a simulated 2D hard sphere liquid with polydispersity of 11\%. Adapted from \cite{russo2015assessing}.}
  \label{Figure31}
\end{figure}

Geometric frustration can play an important role in glassy dynamics even if the locally preferred structure is consistent with the symmetry of the crystalline space and can therefore tile space perfectly. This idea was put forward by Tanaka in a series of papers \cite{tanaka1999two,tanaka2005twoI,tanaka2005twoII,tanaka2005twoIII} and demonstrated in simulations \cite{tanaka2010critical} as well as experiments on granular \cite{watanabe2008direct,tanaka2010critical} and colloidal \cite{leocmach2012roles} glass-formers. As an illustrative example, consider the case of a 2D glass-former composed of polydisperse hard spheres. The reference crystalline state for this system for large area fractions is the triangular lattice, which has six-fold symmetry. Tanaka's two order parameter model postulates that such a system has two types of ordering on increasing the density. Crystallization is associated with the onset of translational as well as hexatic bond-orientational order. On the other hand, the presence of polydispersity frustrates long-ranged translational order but allows the growth of bond-orientational order. As a result, the system exhibits correlated regions of high local bond-orientational order that grow on approaching the glass transition. Such ordering is termed as medium ranged crystalline order (MRCO) \cite{tanaka2010critical}. Since the supercooled liquid is ergodic, these ordered regions eventually break and re-form elsewhere in the system. However, they are extremely long-lived and persist over timescales as long as 10$\tau_{\alpha}$. Local hexatic bond-orientational order is characterized by the order parameter $\psi_6^{j} = \sum_{k=1}^{n_j} \text{e}^{i6\theta_{jk}}$, where $j$ is the particle index, $k$ runs over $n_j$ nearest neighbors of particle $j$ and $\theta_{jk}$ is the angle between $(\mathbf{r_k} - \mathbf{r_j})$ with the X axis. Fig. \ref{Figure31}A shows the local order parameter field averaged over $\tau_{\alpha}$.
Correlated regions with high local order are clearly visible. These ordered regions have a profound impact on the dynamics, as can be seen from the corresponding spatial distribution of the mean squared displacement over a duration of 10$\tau_{\alpha}$ (Fig. \ref{Figure31}B). The mean squared displacement is perfectly anti-correlated with local order, demonstrating that ordered regions are associated with immobile particles. This suggests that the static correlation length associated with hexatic order should be comparable to the dynamic four-point correlation length $\xi_4$ which measures correlations between immobile particles. To measure the hexatic correlation length, Tanaka and coworkers computed the hexatic correlation function
\begin{equation}
g_6(r) = \frac{1}{2 \pi r\Delta r \rho(N-1)} \sum_{j \neq k} \delta(r - |\mathbf{r_k}-\mathbf{r_j}|)\psi_6^{j}{\psi_6^{k}}^{*}
\end{equation}
where $\rho$ is the density, $N$ is the total number of particles and $\delta(r-x) =$ 1 if $x \in [r,r+\Delta r]$ and 0 otherwise. The hexatic correlation length $\xi_6$ is then defined using the Ornstein-Zernike relation $g_6(r)/g(r) \propto r^{-1/4}\text{exp}(-r/\xi_6)$, where $g(r)$ is the radial pair correlation function \cite{tanaka2010critical}. Tanaka and coworkers indeed observed in experiments on granular media \cite{tanaka2010critical} as well as simulations \cite{russo2015assessing} that $\xi_6$ and $\xi_4$ are comparable and grow at the same rate on approaching the glass transition (Fig. \ref{Figure31}C). Moreover, $\xi_6$ grows much faster compared to other static lengths, such as the point-to-set length for various pinning geometries and the two-point structural correlation length associated with the exponential decay of the peaks of $g(r)$, making it the only length scale that scales with $\xi_4$. Based on this observation, Tanaka and coworkers have argued that $\xi_6$ is the only structural length that can explain the dramatic slowdown of dynamics, and is therefore the most relevant static length scale for the glass transition \cite{russo2015assessing}. The greatest drawback of the paradigm of Tanaka and coworkers is that it cannot be generalized easily to different types of glass-formers. Hexatic order is by no means universal and simulations and colloid experiments have indeed found that different glass-formers are best characterized by the proliferation of different structural motifs such as icosahedra \cite{coslovich2007understanding}, fcc crystallites \cite{leocmach2012roles} and 11 membered bicapped square antiprisms \cite{speck2012first,malins2013lifetimes} that may or may not tile Euclidean space. Moreover, Tanaka and coworkers have themselves shown that the correspondence between $\xi_6$ and $\xi_4$ breaks down even for relatively simple systems such as binary glass-forming liquids \cite{tanaka2010critical}.

Regardless of the lack of universality, the approach of Tanaka and coworkers is very appealing, since it establishes a direct connection between easily detectable real space geometric features and dynamics, unlike in RFOT, where structural correlations are subtle. Defining a universal length scale that is independent of the nature of the proliferating structural motif will therefore make a far more compelling case for the frustration-based approach. One potential candidate is the length scale $\xi_s$ associated with the two-body contribution to the liquid's structural entropy. This entropy is given by \cite{nettleton1958expression,mountain1971entropy}
\begin{equation}
s_2 = -\frac{\rho}{2} \int d\mathbf{r} [g(\mathbf{r})\text{ln}(g(\mathbf{r})) - (g(\mathbf{r})-1)]
\end{equation} 
Generalization of $s_2$ to multi-component liquids is straightforward \cite{tanaka2010critical}. To define $\xi_s$ Tanaka and coworkers defined the local version of $s_2$ and averaged it over 10$\tau_{\alpha}$, to obtain for each particle, the local average structural entropy $\bar{s_2}^{j}$. The spatial correlation of $\bar{s_2}^{j}$ is given by 
\begin{equation}
g_{s2}(r) = \frac{1}{2 \pi r\Delta r \rho(N-1)} \sum_{j \neq k} \delta(r - |\mathbf{r_k}-\mathbf{r_j}|)\bar{s_2}^{j}\bar{s_2}^{k}
\end{equation}
$\xi_s$ can then be extracted using the equation $g_{s2}(r)/g(r) \propto r^{-1/4}\text{exp}(-r/\xi_s)$. Since $\xi_s$ only requires the radial pair correlation function as the input, it can be computed without prior knowledge of the incipient local order. 

\begin{figure}
\centering
  \includegraphics[width=\textwidth]{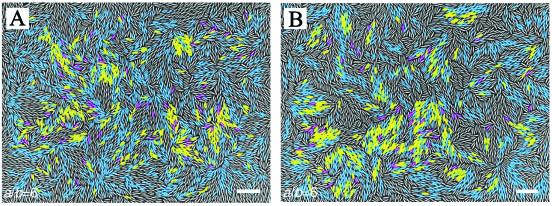}
  \caption{Spatial correlations between glassy particles and dynamically slow particles for translational (A) and rotational (B) degrees of freedom for ellipsoids of aspect ratio 6 at area fraction 0.74. In (A-B), glassy particles are shown in blue and dynamically slow particles are shown in magenta. Particles that are slow as well as glassy are shown in yellow. Adapted from \cite{zheng2014structural}.}
  \label{Figure32}
\end{figure}

\begin{figure}
\centering
  \includegraphics[width=0.7\textwidth]{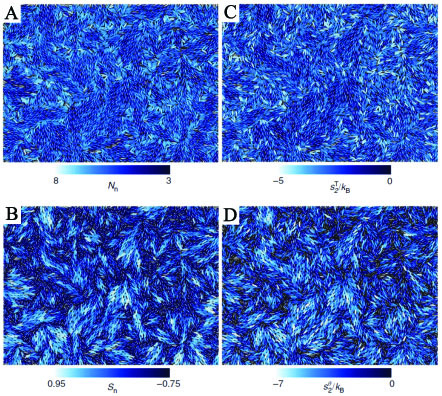}
  \caption{Spatial correlations between translationally and rotationally glassy particles and those possessing low structural entropy for colloidal ellipsoids of aspect ratio 6 at area fraction 0.74. The spatial distributions of (A) Number of nearest neighbors $N_n$, (B) local nematicity $S_n$ and the translational (C) and rotational (D) contributions to the local two-body structural entropy, $s_{2i}^{T}$ and $s_{2i}^{\theta}$, respectively. Adapted from \cite{zheng2014structural}.}
  \label{Figure33}
\end{figure}

A connection between structural entropy and dynamics was also observed in suspensions of colloidal ellipsoids by Han and coworkers \cite{zheng2014structural}. The authors first identified translationally and rotationally least mobile particles as those that traversed a distance of less than half the cage size over $\tau_{\alpha}$. To examine the spatial correlations between slow dynamics and structure, the authors also defined translationally and orientationally `glassy' particles, as follows. Particles with greater than or equal to 6 nearest neighbors, i.e. $N_n \geq$ 6, were defined as translationally glassy particles, since a large number of nearest neighbors indicates strong caging. To quantify orientational `glassiness', the authors invoked the local order parameter $S_n = \sum_{j=1}^{N_n} \text{cos}(2\Delta\theta_j)/N_n$, where $\Delta\theta_j$ is the difference in orientation between a given particle and its $j^{th}$ nearest neighbor. Accordingly, particles with $S_n \geq$ 0.8 were termed orientationally glassy. Fig. \ref{Figure32} shows the spatial correlation between glassy particles (blue) and dynamically slow particles (magenta). 
Particles that are both slow and glassy are shown in yellow. The large fraction of yellow particles indicates that translationally as well as orientationally glassy particles are strongly correlated in space. The authors then investigated whether glassy particles are also associated with low structural entropy. Towards this end, they computed the local two body contribution to the structural entropy in a manner analogous to that in \cite{tanaka2010critical}. In the case of ellipsoids, this entropy can be decomposed into translational and rotational components. For a given particle $i$, these components are given by
\begin{align}
s_{2i}^{T} = -\pi k_B\rho \int_{0}^{\infty} [g_i(r)\text{ln}(g_i(r)) - g_i(r)+1]rdr \\
s_{2i}^{\theta} = -\frac{1}{2}k_B\rho \int_{0}^{\infty} g_i(r)dr \int_{0}^{2\pi}g_i(\theta|r)\text{ln}(g_i(\theta|r))d\theta
\end{align}
Here, $g_i(r)$ is the local radial pair correlation function for the centres of mass of the ellipsoids and $g_i(\theta|r)$ is the distribution of angular differences $\theta$ between the long axes of ellipsoid $i$ and its neighbors, for a centre of mass separation of $r$. The authors showed that translationally and orientationally glassy particles are spatially correlated with particles with low $s_{2i}^{T}$ and $s_{2i}^{\theta}$, respectively, as shown in Fig. \ref{Figure33}. The authors further claim that the static length scales extracted from spatial correlations of glassiness and structural entropy are proportional to the dynamic four-point correlation length $\xi_4$, in concord with \cite{tanaka2010critical}. 

Despite these successes, the question of whether the length scale obtained from $s_2$ scales with $\xi_4$ or not, however, still remains debatable. In particular, Royall and coworkers have shown that the two length scales may get decoupled on approaching the glass transition \cite{dunleavy2012using}. Further, there are concerns that since $s_2$ is sensitive to the peakedness of $g(r)$, which may be influenced by the extent of cage rattling, $\xi_s$ can encode dynamical information and is therefore not a purely structural length scale \cite{dunleavy2012using}. Nonetheless, it is a worthwhile pursuit to investigate these questions in detail in order to find a universal static length scale that is compatible with the frustration scenario.

\section{Dynamical crossovers in glass-forming liquids}
From the preceding section, it is clear that several distinct theories of the glass transition have gathered support from experiments on colloids and granular media. However, a vast majority of these experiments focus on one theoretical framework or the other and do not compare predictions of different frameworks. As a result, there is no consensus on the correct theory of glass formation in spite of the growing body of accumulated experimental data. For that matter, it is not clear whether a `correct' theory of glass formation even exists. It is possible, and in the light of existing experimental data, quite probable, that multiple distinct relaxation processes are simultaneously at work during glass formation. The real goal then is to infer which of these processes are likely to dominate close to $T_g$ or $\phi_g$. At first sight, this appears to be a hopeless task, since both simulations and colloid experiments are generally limited to the dynamical range corresponding to $T \geq T_c$, and atomic experiments cannot give microscopic insights into the dynamics at $T_c \geq T \geq T_g$. Over the last few years, however, a number of dynamical crossovers have been observed in the vicinity of $T_c$ or $\phi_c$ and these offer a potential solution to the seemingly intractable problem of distinguishing between predictions of competing theories. The key idea is that dynamical crossovers are likely to be associated with changes in the dominant mechanism or structural relaxation. Since competing theories espouse distinct mechanisms, one can investigate whether the observed crossovers demarcate different dynamical regimes over which different theories are valid. Below, we shall describe some of the observed dynamical crossovers and discuss results from recent colloid experiments that employed these crossovers to critically assess the validity of RFOT and facilitation. 

\subsection{The failure of MCT}  
Perhaps the best known dynamical crossover associated with glass formation is the MCT transition, which is avoided by the presence of activated hopping events. Despite its considerable utility in explaining the shape of relaxation functions as well as novel qualitative predictions for complex glass-formers, it is now evident that MCT, at least in its idealized form, cannot be a complete theory of the glass transition. Although colloid experiments and simulations are difficult to perform in the vicinity of the mode coupling crossover, there is sufficient evidence in the literature to demonstrate that ergodicity persists below $T_c$ \cite{kob2012non} or above $\phi_c$ \cite{brambilla2009probing,nagamanasa2015direct}. This clearly shows that the divergence in relaxation time predicted by MCT does not in fact occur, and the putative transition is therefore rounded off into a crossover. In the context of colloids, the experiments of Brambilla et al. \cite{brambilla2009probing} were crucial since up to that point, structural relaxation in colloidal glass-forming liquids was well-described by MCT, and the need to look beyond it was not obvious. These findings have been questioned in a recent paper by Poon and coworkers \cite{zaccarelli2015polydispersity}, who suggest using simulations that the persistence of ergodicity can be attributed to polydispersity, rather than the predominance of activated hopping. Their argument relies on the difference in mobility among particles occupying the tail of the size distribution and those that constitute the peak. It is rather puzzling therefore, that the authors do not extend their arguments to experiments and simulations on binary glass-formers, for which the disparity in mobility is only amplified. There are also theoretical considerations that support the possibility that the experiments of Brambilla et al. indeed demonstrate the failure of MCT. As discussed before, according to RFOT, below the MCT transition, the liquid's energy landscape breaks into an exponentially large number of minima. In the mean field limit, the activation barriers between minima diverge and the system is frozen. In finite dimensional real world systems, however, the barriers are finite and ergodicity is restored below $T_c$ by activated hops between different minima. Recent experiments and simulations are consistent with this picture \cite{kob2012non,nagamanasa2015direct}. It is therefore likely that the experiments of Brambilla et al indeed correspond to the failure of MCT. 

\subsection{The breakdown of the Stokes-Einsten relation}  
In terms of the dominant mechanism of relaxation, the failure of MCT implies that the dominant mechanism of relaxation changes from the local stress-mediated flows envisioned by MCT to activated hopping events that involve an increasingly large number of particles with decreasing temperature. However, activated events are known to be relevant even above $T_c$, which suggests that multiple relaxation mechanisms must be present even in this mildly supercooled regime. A related crossover within the $T^{*} \geq T \geq T_c$ regime, where $T^{*}$ is the onset of glassy dynamics, is the decoupling of viscosity and diffusion, often referred to as the breakdown of the Stokes-Einstein relation (SER). The Stokes-Einstein relation $D\eta =$ constant, where $D$ is the self-diffusion coefficient and $\eta$ is the viscosity, is a manifestation of the fluctuation dissipation theorem and is obeyed in the liquid phase ($T>T_m$). In the supercooled regime, however, one typically finds instead the relation $D \sim \eta^{-1+\omega}$, with $\omega >$ 0. The breakdown of SER is an important phenomenological observation and nearly every major theory offers an explanation for its observation. Numerical studies have shed further light on the breakdown of SER by elucidating the fundamental role of spatial heterogeneity in particle mobility. Evidence from simulations suggests that mobile particles are associated with diffusion, whereas nonmobile ones determine viscosity. A separation in the characteristic timescales associated with the correlated motion of the mobile and immobile particles is therefore thought to result in the breakdown of SER. These observations seem to indicate that the SER breakdown is associated with the emergence of dynamical heterogeneity. 

\begin{figure}
\centering
  \includegraphics[width=0.7\textwidth]{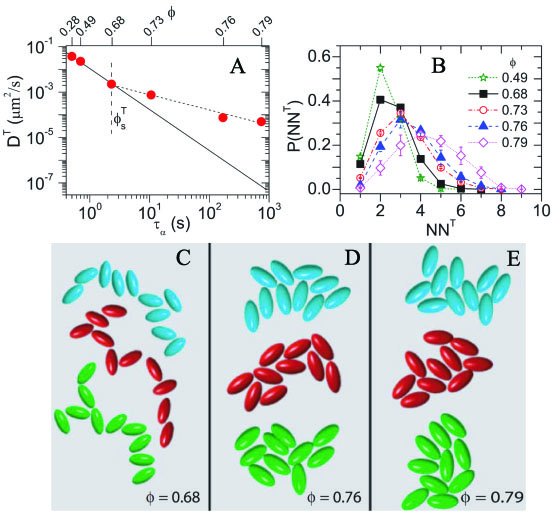}
  \caption{Correlation between the breakdown of SER and the shapes of CRR in suspensions of colloidal ellipsoids. (A) Translational self-diffusion coefficient $D^{T}$ vs relaxation time $\tau_{\alpha}$. The area fraction $\phi_s^{T}$ corresponding to the SER breakdown is denoted by a dashed vertical line. (B) The distribution of number of translationally mobile nearest neighbors of translationally mobile particles, $P(NN^{T})$ for various $\phi$. (C-E) Representative translational CRRs containing $N =$ 10 particles for $\phi =$ 0.68 (C), $\phi =$ 0.76 (D) and $\phi =$ 0.79 (E). Adapted from \cite{mishra2015shape}.}
  \label{Figure34}
\end{figure}

Numerical \cite{sengupta2013breakdown} and experimental \cite{mishra2015shape} studies have shown that in general, this is not the case. In an important numerical work, using five different glass-formers, Flenner, Staley and Szamel have shown that the SER breakdown is in fact associated with a characteristic temperature $T_s$ or volume fraction $\phi_s$, which lies between the onset of glassy dynamics and the mode coupling crossover \cite{flenner2014universal}. Thus, the SER breaks down after, rather than at the onset of heterogeneous dynamics. More importantly, the authors have demonstrated that for $T<T_s$, the four-point susceptibility $\chi_4$ and the four-point dynamic correlation length $\xi_4$ are related by $\chi_4 \propto (\xi_4)^3$. This implies that the breakdown in SER is accompanied by the emergence of compact clusters of immobile particles. A crossover in the shapes of cooperatively rearranging regions (CRRs) from string-like to compact form is expected within RFOT. However, RFOT predicts this change to occur beyond $T_c$, and not $T_s$. A strong correlation between the breakdown of SER and the change in morphology of CRRs was recently observed in colloid experiments by Mishra and Ganapathy \cite{mishra2015shape} who investigated a 2D glass-former composed of ellipsoids. Fig. \ref{Figure34} shows their results for translational relaxation in dense suspensions of polystyrene ellipsoids interacting via short-ranged repulsive interactions. 
The SER is clearly violated beyond $\phi_s \sim$ 0.68 (Fig. \ref{Figure34}A). To quantify the shapes of CRRs, the authors computed the distribution $P(NN)$ of the number of mobile nearest neighbors of mobile particles \cite{mishra2015shape}. Upto $\phi =$ 0.68, this distribution is peaked at $NN =$ 2, signifying that mobile particles typically have two mobile neighbors, a scenario expected for string-like CRRs. For $\phi >$ 0.68, the distribution evolves towards larger values of $NN$, suggesting that CRRs become increasingly compact (Fig. \ref{Figure34}B). Representative CRRs for various $\phi$ are shown in Fig. \ref{Figure34}C-E. While the experimental results appear to be consistent with the simulations, there is an important difference. The numerical work of Flenner et al. \cite{flenner2014universal} showed a change in morphology for clusters of immobile particles, whereas the experiments of Mishra and Ganapathy \cite{mishra2015shape} showed it for clusters of mobile particles. This distinction merits further research and analysis. 

In a recent work, Szamel and coworkers have examined the breakdown of the SER in strong rather than fragile glass-formers \cite{staley2015reduced}. In particular, they have shown that the scaling of the four-point susceptibility $\chi_4$ with the dynamic correlation length $\xi_4$ changes across the temperature $T_s$ corresponding to the SER breakdown. Once again, this presumably signals a change in the morphology of CRRs. A comparative analysis of fragile and strong glass-formers should be possible in colloidal systems. As mentioned before, Weitz and coworkers have already demonstrated that the fragility of colloidal glass-forming liquids can be changed by tuning the particle softness \cite{mattsson2009soft}. Deformation due to inter-particle contacts lends directionality to colloid interactions, which decreases the fragility. Despite these observations, strong glass-formers have not been explored in depth using colloids. Thermoresponsive PNIPAm particles are perhaps the most promising candidates for initiating studies that quantify dynamic correlations as a function of particle softness \cite{mattsson2009soft}. Nonetheless patchy colloids with non-spherical shapes or anisotropic interactions \cite{jiang2014orientationally} also offer interesting possibilities in elucidating non-trivial aspects of dynamics in strong glass-formers. 

\subsection{Anisotropic relaxation and the non-monotonic evolution of dynamic correlations} 
\begin{figure}
\centering
  \includegraphics[width=0.8\textwidth]{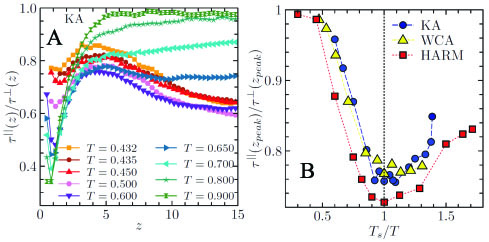}
  \caption{Anisotropic relaxation in the presence of an amorphous wall. (A) The relaxation anisotropy $\tau^{\parallel}(z)/\tau^{\perp}(z)$ vs $z$ for various temperatures for the Kob-Andersen binary mixture. (B) The peak value of relaxation anisotropy as a function of temperature for the Kob-Andersen (KA), Weeks-Chandler-Andersen (WCA) and harmonic sphere (HARM) systems. Adapted from \cite{hocky2014crossovers}.}
  \label{Figure35}
\end{figure}

It is worth comparing the results of Flenner et al. with the non-monotonic evolution of dynamic correlations observed in simulations \cite{kob2012non} as well as and colloid experiments \cite{nagamanasa2015direct}. A crucial observation is that the dynamic correlation length $\xi_{dyn}$ evaluated in the presence of a frozen wall exhibits a peak at a temperature (or area fraction) intermediate between the onset of glassy dynamics and the mode coupling crossover. It is therefore tempting to postulate that the non-monotonicity is associated with $T_s$ (or $\phi_s$). In the case of colloid experiments, due to uncertainties in the measurement of volume fractions, it is not possible to pinpoint whether the maximum in $\xi_{dyn}$ occurs at $\phi_s$ or $\phi_c$. Once again, the experiments have demonstrated a change in morphology of clusters of mobile particles. Further analysis of correlations between immobile particles is therefore in order. Further insight into the connection between the results of Flenner et al. and the non-monotonic evolution of $\xi_{dyn}$ was provided by Hocky et al. \cite{hocky2014crossovers} who analysed the anisotropy in structural relaxation imposed due to the presence of an amorphous wall. They found that the non-monotonicity in $\xi_{dyn}$ is not a universal feature. While it exhibits a maximum for some model glass formers such as harmonic spheres, it merely seems to saturate for others, such as the Kob-Andersen binary mixture. The authors then examined anisotropy in structural relaxation as a function of the distance $z$ from the amorphous wall. In agreement with previous simulations \cite{scheidler2004relaxation}, the authors found that relaxation parallel to the wall is slower than relaxation perpendicular to the wall, and hence the corresponding time $\tau^{\parallel}(z)$ is greater than $\tau^{\perp}(z)$. The variation of $\tau^{\parallel}(z)/\tau^{\perp}(z)$ with $z$ on the other hand has some intriguing features. Fig. \ref{Figure35}A shows this variation for the Kob-Andersen binary mixture. 
Interestingly, upon lowering the temperature, $\tau^{\parallel}(z)/\tau^{\perp}(z)$ appears to saturate to values less than 1 at low temperatures, implying that the influence of the wall is felt arbitrarily long distances from the wall. This is rather counter-intuitive, given that the relaxation time approaches its bulk value over a distance of $\sim \xi_{dyn}$. The second surprise is the development of a peak in $\tau^{\parallel}(z)/\tau^{\perp}(z)$ at a distance $z = z_{peak} \approx 4\sigma$. Most strikingly, the evolution of the peak value $\tau^{\parallel}(z_{peak})/\tau^{\perp}(z_{peak})$ with temperature exhibits at a temperature very close to $T_s$ for a variety of simulated glass-formers. These observations are testable in colloid experiments and further real space analysis in terms of particle rearrangements will play an instrumental role in obtaining a deeper understanding of the $\tau^{\parallel}(z)/\tau^{\perp}(z)$ profiles. 

\subsection{Using crossovers to probe changes in relaxation mechanisms}
From the point of view of solving the glass transition problem, it is of utmost importance to investigate whether the observed crossovers are associated with changes in the dominant mechanism of relaxation. Recent simulations by Royall and coworkers hint at such a possibility \cite{dunleavy2015mutual}. The authors identify two distinct populations of particles based on whether they are likely to be displaced from their initial positions at `early' or `late' times. These populations differ in their local density as well as local structural order. Interestingly, the authors compare the `early' movers in their simulations to mobile particles and the `late movers' to immobile particles. Since mobile and immobile particles are maximally correlated over timescales that get increasingly decoupled on approaching the glass transition, the identification of two populations of particles based on mobility \cite{dunleavy2015mutual} may be associated with the breakdown of SER \cite{starr2013relationship}. In addition to the connection to SER, Royall and coworkers show that the `late' moving particles are strongly correlated away from the glass transition whereas the `early' moving particles are strongly correlated close to the glass transition. This led the authors to conclude that there is a change in the relaxation mechanism on approaching the glass transition, even within the dynamical regime accessible to simulations and colloid experiments. However, the authors were not able to discern which relaxation mechanism dominates close to the glass transition.

Although different theories propose distinct relaxation mechanisms, each with their associated physical origins, from an experimental perspective, it is useful to divide these mechanisms into two categories: collective hopping and facilitation. While geometric frustration-based models and RFOT differ significantly in terms of their physical content, the primary relaxation mechanism for both involves the cooperative rearrangement of domains whose size grows on approaching the glass transition. Such processes cannot be distinguished based on dynamics alone. Facilitation on the other hand is a completely different process which can at least in principle be dynamically distinguishable from collective hopping. A rational strategy would therefore be to first compare the relative importance of facilitation and collective hopping on approaching the glass transition and then proceed towards capturing the finer distinctions between various cooperative processes. Below, we review recent experiments that have adopted this strategy and demonstrated that facilitation is dominated by collective hopping beyond the mode coupling crossover.

\subsubsection{The mobility transfer function}
\begin{figure}
\centering
  \includegraphics[width=0.5\textwidth]{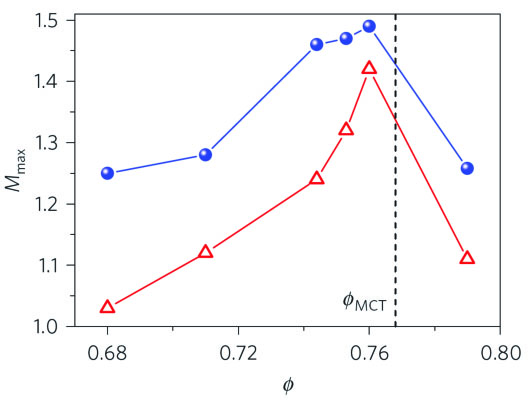}
  \caption{The maximum value of the mobility transfer function,$M_{max}$ as a function of area fraction $\phi$ for large (red triangles) and small (blue circles) particles in a binary colloidal glass-former composed of polystyrene particles. Adapted from \cite{nagamanasa2015direct}.}
  \label{Figure36}
\end{figure}

As mentioned earlier, the maximum value $M_{max}$ of the mobility transfer function quantifies the degree of facilitation in a glass-former. As long as facilitation is the dominant mechanism of relaxation, this value is expected to increase on approaching the glass transition. Elmatad and Keys have shown that this is indeed the case for the kinetically constrained East model, for which facilitation is by construction the dominant mechanism of relaxation \cite{elmatad2012manifestations}. They have further shown that if one introduces an activated hopping process that is capable of superseding facilitation, $M_{max}$ first increases with $1/T$ and then decreases close to the glass transition. Such a non-monotonicity in $M_{max}$ indicates a crossover from a high temperature regime dominated by facilitation to a low temperature regime associated with activated hopping. To determine the role of facilitation in their experiments, Nagamanasa et al. computed $M_{max}$ (Fig. \ref{Figure36}) for their colloidal system and compared its evolution with $\phi$ with that of $\xi_{dyn}$ (Fig. \ref{Figure29}C). 
Quite remarkably, $M_{max}$ exhibits a peak at the same $\phi$ as $\xi_{dyn}$. This suggests that a change in the shapes of CRRs is also accompanied by diminishing facilitation. It also implies that facilitation is unable to capture the compaction of CRRs on approaching the glass transition anticipated within RFOT. On the other hand, the diminishing importance of facilitation is entirely consistent with RFOT. According to RFOT, facilitation is a secondary relaxation process that emerges from the non-linear interaction between activated hopping and mode coupling \cite{bhattacharyya2008facilitation} and whose role in stuctural relaxation becomes increasingly peripheral below the mode coupling crossover \cite{stevenson2010universal}. 

\subsubsection{Localized excitations and the shapes of CRRs}
\begin{figure}
\centering
  \includegraphics[width=0.7\textwidth]{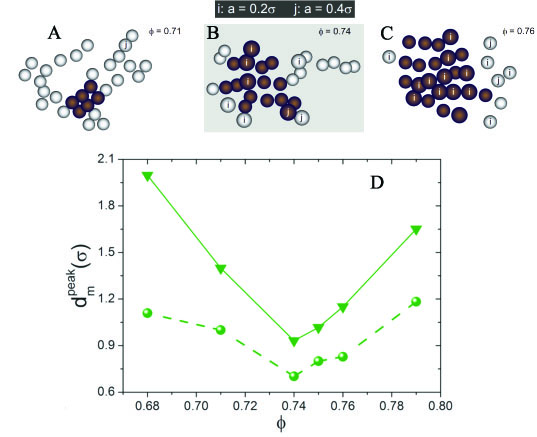}
  \caption{Expulsion of excitations from core-like regions of CRRs. (A-C) Illustrations of representative CRRs containing $N =$ 30 particles for $\phi =$ 0.71 (A), $\phi =$ 0.75 (B) and $\phi =$ 0.76 (C). In (A-C), particles belonging to the core are shown in brown and those belonging to the shell are shown in white. Excitations of size $a =$ 0.2$\sigma$ that overlap with the CRRs are denoted by ``i'' and those of size $a =$ 0.4$\sigma$ are denoted by ``j''. (D) $d_m^{\text{peak}}$ vs $\phi$ for $a =$ 0.4$\sigma$ for CRRs of size 25-35 particles (triangles) and 10-15 particles (spheres).}
  \label{Figure37}
\end{figure}

The foregoing discussion implies that RFOT-like activated hopping dominates facilitation close to the glass transition. However, one must bear in mind that the mobility transfer function is a rather indirect measure of facilitation since it considers mobility correlations over timescales that are typically much larger than those associated with excitation dynamics. To determine the relative importance of facilitation, therefore, one must examine the spatial organization of localized excitations within CRRs to determine whether they can or cannot generate the observed change in morphology from string-like to compact form. This strategy was recently employed to analyse the partitioning of excitations into string-like and compact regions of CRRs \cite{gokhale2016localized} using data from the colloid experiments of \cite{nagamanasa2015direct}. The starting point of the study is the fact that CRRs are in general composed of a string-like shell and a compact core \cite{stevenson2006shapes,nagamanasa2015direct}. Since excitations are carriers of mobility, the authors defined CRRs to be clusters of mobile particles. Within the DF theory, excitations are the building blocks of structural relaxation. As a result, if a particle is mobile over a time interval $\Delta t$, it is likely to be associated with an excitation. Further, if facilitation is the dominant process, excitations are equally likely to be associated with the string-like shell and compact core of CRRs. This can be quantified by computing the fraction of excitations associated with the stringy shell-like regions of CRRs (See Fig. \ref{Figure37}A-C for illustrations). 
The authors found that for $\phi <$ 0.76, this is indeed the case, and facilitation is therefore an important relaxation process. For $\phi \geq$ 0.76, however, excitations preferentially occur within the shell-like regions, suggesting that facilitation may be dominated by another relaxation process \cite{gokhale2016localized}. In order to understand the preferential occurrence of excitations in the shell, or equivalently, the depletion of excitations from the core, The authors examined the spatial organization of excitations within the core-like regions of CRRs. Towards this end they computed the average minimum distance of excitations from the centre-of-mass of the CRR core, $d_m^{\text{peak}}$, and observed its variation with $\phi$ (Fig. \ref{Figure37}D). $d_m^{\text{peak}}$ first decreases with $\phi$ and then increases significantly for CRRs containing 25-35 particles. While the initial decrease is consistent with facilitation, the subsequent increase is not. Further, the increase in $d_m^{\text{peak}}$ is attenuated  for CRRs containing 10-15 particles. Since smaller CRRs are predominantly string-like in character, this observation is consistent with the selective partitioning of excitations into string-like regions. The authors have argued that the results shown in Fig. \ref{Figure37}D can be satisfactorily explained by invoking the presence of a second relaxation process that corresponds to the spontaneous activated hopping of compact groups of particles. These relaxation events qualitatively resemble the primary relaxation events envisioned in RFOT \cite{kirkpatrick1989scaling}. The authors conclude that facilitation satisfactorily describes glassy dynamics for volume fractions lower than the mode coupling crossover. However, it is eventually dominated by collective hopping close to the glass transition.

\subsubsection{Connections between MCT, facilitation and RFOT}
RFOT is a complex overarching theoretical framework that has been significantly expanded to account for several phenomenological observations that it was not equipped to handle in its original formulation. As a result, it incorporates MCT as well as facilitation within its rich tapestry. Its treatment of MCT is quite convincing, especially considering the mathematical correspondence between schematic MCT equations and the dynamics of $p-$spin models \cite{berthier2011theoretical}. Its treatment of facilitation is less compelling, especially in the light of recent experimental evidence. As mentioned before, facilitation can be incorporated within RFOT as a secondary relaxation process \cite{bhattacharyya2008facilitation}. Specifically, activated hopping influences diffusion in the surrounding region in a manner that can be described as a facilitation effect. Within this description, facilitation is naturally associated with the ramified string-like shell of CRRs, which diminishes in size on approaching the glass transition \cite{stevenson2006shapes,stevenson2010universal}. Data from colloid experiments are not entirely consistent with this picture. First, if facilitation is always associated with the stringy shell of CRRs, one would expect the fraction of excitations associated with shell-like regions of CRRs $F_{S}^{a}$ to stay more or less constant at a large value for all $\phi$, which is not the case \cite{gokhale2016localized}. Secondly, for $\phi <$ 0.76, the $\phi$ as well as $a$ dependence of $F_{S}^{a}$ observed in \cite{gokhale2016localized} is completely consistent with the hierarchical nature of facilitated dynamics described in \cite{keys2011excitations}. Moreover, the maximum of the mobility transfer function $M_{max}$ increases with $\phi$ in this regime \cite{nagamanasa2015direct}, suggesting that facilitation is the dominant relaxation mechanism \cite{elmatad2012manifestations}. Lastly, the DF theory can predict the existence of re-entrant glass-transitions in colloidal ellipsoids from the $\phi$ dependence of the concentration of excitations over a dynamical range corresponding to $\phi < \phi_c$ \cite{mishra2014dynamical} (Fig. \ref{Figure27}). Collectively, these facts suggest that facilitation is the dominant relaxation mechanism from low to moderate $\phi$. As such, the experimental observations indicate a competition between two independent mechanisms of relaxation. Nonetheless further studies are necessary to determine whether the difference between the role of facilitation described in RFOT and that inferred from colloid experiments is semantic or conceptual.

The consensus from colloid experiments appears to be that MCT and the DF theory are valid over a nearly identical dynamical range from the onset of glassy dynamics to the mode coupling crossover. This makes one wonder whether MCT and facilitation essentially embody the same physics. On the face of it, the two theories could not be further apart. MCT is a first principles theory whereas the DF theory is phenomenological. Dynamical heterogeneities cannot be tackled within the original `local cage' paradigm of MCT, although extensions such as inhomogeneous MCT do predict the divergence of the four-point susceptibility. The DF theory on the other hand uses dynamical heterogeneity as its foundation. Nonetheless, correspondences between MCT and facilitation have been observed, particularly in spin models \cite{sellitto2005facilitated,sellitto2010dynamic,sellitto2012cooperative,sellitto2013disconnected}. One therefore wonders whether such correspondences exist even in particulate glass-formers.  

\subsubsection{The role of geometric frustration}
Thus far, the evidence from colloid experiments has shown that neither MCT nor DF can account for structural relaxation close to the glass transition. Moreover, recent experiments indicate that relaxation proceeds via collective hopping of compact clusters of particles. As mentioned at the beginning of the section, it is impossible to tell on the basis of dynamics alone whether such collective hopping supports the RFOT scenario or the frustration-based one. Moreover, this regime is exceedingly difficult to access experimentally. To demystify the role of geometric frustration, therefore, one can adopt two different approaches. The first approach is develop new analysis protocols within the $\phi < \phi_c$ regime in order to ascertain the relative importance of local structure in governing the dynamics. Specifically, both facilitation and MCT satisfactorily explain glassy dynamics within this regime. It is therefore interesting to see whether these processes are influenced by local structure, or simply coexist with the relaxation mechanism envisioned by frustration-based models. It would be particularly interesting to investigate specific glass-formers such as polydisperse hard spheres. For this glass-former, $\xi_6$ and $\xi_4$ exhibit identical scaling, suggesting that local structure plays an important role in dynamic arrest. Moreover, ordered regions are anti-correlated with particle displacements \cite{tanaka2010critical} (Fig. \ref{Figure31}), suggesting that local order influences the spatial occurrence of excitations. The second strategy would be to devise alternate protocols for approaching the glass transition, so that theoretical predictions can be tested in the context of unconventional control variables. Indeed, as we shall see in the forthcoming section, this approach is so appealing that new routes to glass formation have been developed from the perspective of geometric frustration, RFOT as well as facilitation and provide a promising way forward in solving the glass transition problem.

\section{Alternate routes to glass formation}
The primary reason why the glass transition problem remains unsolved is that supercooled liquids fall out of equilibrium long before the putative thermodynamic or dynamic transition underlying glass formation is reached. This has led to the development of novel theoretical approaches aimed at accessing the glass transition by varying control parameters other than temperature or density. One of these approaches, namely random pinning has been realized in colloid experiments \cite{gokhale2014growing}. Others, however, appear too abstract and obscure to be of interest to an experimentalist, at least at first glance. Nonetheless, we believe that these alternate routes to glass formation will occupy a central role in advancing our understanding of the glass transition by providing new grounds on which to examine the merits and demerits of various competing theories. Moreover, devising innovative experimental protocols to realize these theoretical constructs poses an exciting challenge in itself. The discussion in this section is more theoretical in nature than in previous sections, owing to the relative lack of experimental observations. However, the same dearth of observations presents new opportunities to colloid experimentalists to ply their art towards filling the existing gap between theory and experiment. We will therefore discuss in detail, these alternate routes to glass formation as well as the challenges and opportunities they represent for experimentalists In the forthcoming sections. 

\subsection{Random pinning}
\begin{figure}
\centering
  \includegraphics[width=\textwidth]{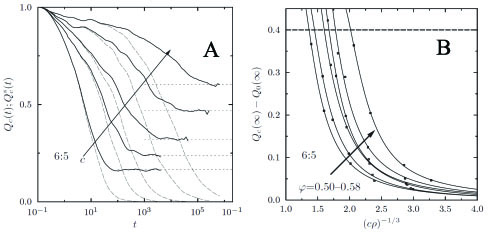}
  \caption{Effect of random pinning on overlap functions in a simulated binary hard sphere glass-former with particle size ratio 6:5. (A) Configurational overlap $Q_c(t)$ (solid curves) and self-overlap $Q_c^{s}(t)$ for various densities of pinned particles $c$ for volume fraction $\varphi =$ 0.55. The horizontal dashed lines represent the asymptotic value of the configurational overlap, $Q_c(\infty)$. (B) $Q_c(\infty) - Q_0(\infty)$ as a function of the mean separation between pinned particles for various $\varphi$. Adapted from \cite{charbonneau2013decorrelation}.}
  \label{Figure38}
\end{figure}

In our discussion on point-to-set correlations, we mentioned that $\xi_{PTS}$ can be extracted using various pinning geometries. One of the most important among these is the so called random pinning geometry. While extracting the point-to-set length using the random pinning geometry, one freezes a subset of particles whose positions are chosen at random from an equilibrium configuration of the glass-forming liquid and monitors the relaxation of the remaining free particles. As the density of pinned particles increases, the mean separation between them decreases. This imposes constraints on the configurations that the liquid can adopt, which leads to an increase in the asymptotic value of the configurational overlap $Q_c(\infty)$ relative to its value $Q_0(\infty)$ in the absence of pinning (Fig. \ref{Figure38}A). 
Further, much as in the case of the amorphous wall \cite{kob2012non,nagamanasa2015direct} the self-overlap decays to zero over increasingly longer times with increasing $c$ (Dashed curves in Fig. \ref{Figure38}A). With increasing volume fraction $\phi$, the length scale $\xi_{PTS}$, which demarcates regimes of high and low configurational overlap moves to larger values (Fig. \ref{Figure38}A), as expected \cite{berthier2012static}.  

\begin{figure}
\centering
  \includegraphics[width=\textwidth]{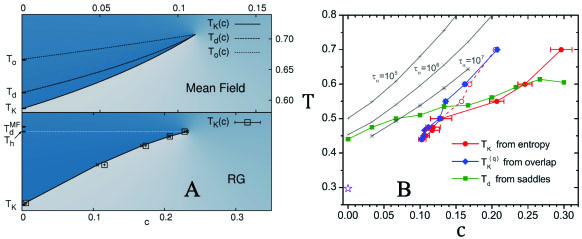}
  \caption{Random pinning glass transition. (A) Theoretical phase diagram of glass-forming liquids in the c-T plane using mean field as well as renormalization group techniques, within the framework of RFOT. The dark blue regions corresponds to the liquid phase whereas the white region corresponds to the glass phase. Adapted from \cite{cammarota2012ideal}. (B) Numerical phase diagram for the random pinning glass transition. Adapted from \cite{ozawa2015equilibrium}.}
  \label{Figure39}
\end{figure}

The idea that pinned particles constrain local configurations in a glass-forming liquid was exploited by Cammarota and Biroli \cite{cammarota2012ideal}. Working within the framework of RFOT, the authors predicted the existence of an ideal glass transition induced by randomly freezing a subset of particles in an equilibrium configuration of the glass-forming liquid. In particular, their theory predicts that the relaxation time increases with the density $c$ of pinned particles at temperature $T>T_K$ and diverges at a critical pinning density $c_K(T)$. Further, using mean field as well as renormalization group techniques, the authors mapped the phase diagram in the $c-T$ plane (Fig. \ref{Figure39}A). The phase diagram predicts that in the absence of pinning, the system exhibits a conventional RFOT type ideal glass transition at $T_K$. At higher temperatures, a finite fraction $c_K(T)$ is necessary to induce the glass transition. The $c>c_K(T)$ regime corresponds to an equilibrium glass phase. The physical basis for this so called `random pinning glass transition' (RPGT) is as follows. Freezing the position of a particle in the liquid restricts the number of metastable local configurations accessible to the liquid. This results in a configurational entropy loss $Y$ per pinned particle. Hence, in the presence of pinned particles, the configurational entropy of the liquid is lowered relative to its unconstrained value. For small pinning densities, this value is given by $s_c(T,c) \approx s_c(T) - cY$. In general, one expects $s_c(T,c)$ to decrease monotonically with $c$ and vanish at a temperature dependent critical value $c_K(T)$. Given the central importance of the vanishing of the configurational entropy within RFOT, it follows that the glass transition line in the $c-T$ plane is determined by the condition $s_c(T,c_K(T)) =$ 0, or equivalently $s_c(T_K(c),c) =$ 0 (Fig. \ref{Figure39}A). Initial numerical work yielded conflicting results, with some simulations supporting the existence of an RPGT \cite{kob2013probing} whereas others opposing it \cite{chakrabarty2015dynamics}. Recently, a numerically obtained equilibrium phase diagram that closely matches the theoretical predictions of Cammarota and Biroli has been reported \cite{ozawa2015equilibrium} (Fig. \ref{Figure39}B).

\begin{figure}
\centering
  \includegraphics[width=0.8\textwidth]{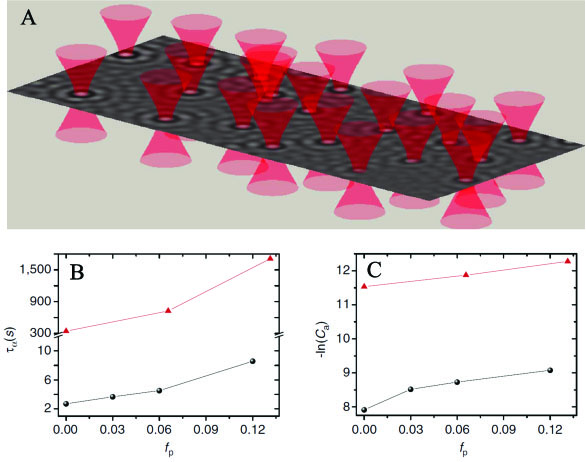}
  \caption{Experimental realization of the random pinning geometry. (A) Schematic of the trapping potentials (shown in red) created by the holographic optical tweezers. The underlying image represents a small portion of the field of view for $f_{p}$ = 0.06 and $\phi =$ 0.71. The image has been generated by averaging over $\sim 15\tau_{\alpha}$. Pinned particles appear bright due to their low mobility and high overlap with initial positions. (B)Structural relaxation time $\tau_\alpha$ for free particles as a function of $f_{p}$ for $\phi =$ 0.71 (filled black spheres) and $\phi =$ 0.74 (filled red triangles). Notice the break in the $\tau_{\alpha}$ axis.(C) Concentration of excitations $c_{a}$ as a function of $f_{p}$ for $\phi =$ 0.71 (filled black spheres) and $\phi =$ 0.74 (filled red triangles). Adapted from \cite{gokhale2014growing}.}
  \label{Figure40}
\end{figure}

The influence of random pinning on dynamics is also not well understood. In particular, whether or not the evolution of the four-point susceptibility $\chi_4(t)$ supports the existence of an RPGT is debated \cite{jack2013dynamical,kob2014nonlinear}. On the experimental front, Gokhale et al. have realized the random pinning geometry using holographic optical tweezers (See Fig. \ref{Figure40}A for a schematic) and provided the first direct evidence for a growth in relaxation time with increasing fraction of pinned particles $f_p$ \cite{gokhale2014growing} (Fig. \ref{Figure40}B). 
However, the authors found no discernible trend in the evolution of the four-point susceptibility $\chi_4(a,t)$ with $f_p$, which is inconsistent with the predictions of RFOT \cite{jack2013dynamical}. To investigate whether other relaxation processes govern the dynamics of pinned glass-formers, the authors tested predictions of the DF theory in the context of random pinning for the first time. They observed that consistent with the facilitation picture, the concentration of excitations decreased with increasing $f_p$ (Fig. \ref{Figure40}C) with a concomitant increase in facilitation volumes \cite{gokhale2014growing}. However, the observed variation of $\chi_4(a,t)$ with $f_p$ is inconsistent even with the DF approach \cite{chandler2006lengthscale, berthier2011theoretical}. These issues need to be addressed in detail in future experimental work. 

\subsection{Phase transitions in coupled replicas}
We owe the existence of yet another thermodynamic route to the glass transition to the ingenuity of Franz and Parisi \cite{franz1997phase}. The central idea was to study glass formation in systems comprising of two coupled replicas of the same glass-former. This approach has its roots in spin glass physics, where the concept of replicas plays a vital role. Spin glasses have been extensively studied theoretically as well as experimentally and are generally better understood compared to structural glasses \cite{karmakar2013random}. A detailed discussion of spin glasses is outside the scope of the present review and we direct the interested reader to the relevant literature in the field \cite{mezard1987spin,mydosh1993spin,castellani2005spin}. Nonetheless, to understand the approach of Franz and Parisi, it is necessary to dwell on the nature of the order parameter for the spin glass transition. Consider the model of Edwards and Anderson \cite{edwards1975theory}, whose Hamiltonian is given by $H = -\sum_{(i,j)}J_{ij}S_iS_j$. Here, $(i,j)$ are indices corresponding to nearest neighbor Ising spins $S_i$ and $S_j$, respectively, on a $d-$dimensional lattice and the quenched random couplings $J_{ij}$ are drawn from a Gaussian distribution $P(J_{ij})$ with mean $J_0$ and variance $J^2$. The model exhibits a spin glass phase for sufficiently low $T$ and $J_0$. 

\begin{figure}
\centering
  \includegraphics[width=0.85\textwidth]{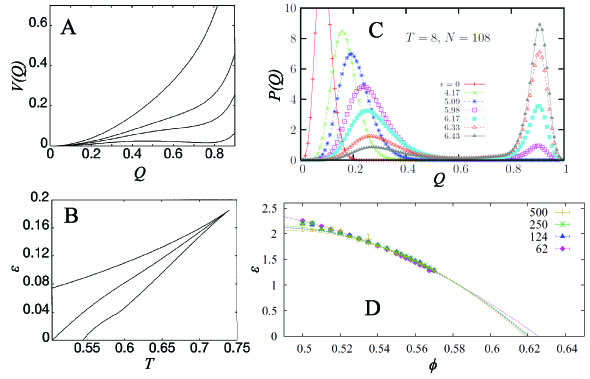}
  \caption{Phase transition in coupled replicas. (A) The Franz-Parisi potential $V(Q)$ for various temperatures and (B) Phase diagram in the $\epsilon-T$ plane for the $p-$spin model with $p =$ 4. Adapted from \cite{franz1997phase}. (C) Evolution of the overlap distribution $P(Q)$ for various $\epsilon$ for a simulated binary mixture of harmonic spheres. Adapted from \cite{berthier2013overlap}. (D) Phase diagram in the $\epsilon-\phi$ plane for a simulated binary mixture of hard spheres. Adapted from \cite{parisi2014liquid}.}
  \label{Figure41}
\end{figure}

For the case of $J_0 =$ 0, the net magnetization in the thermodynamic limit is zero, since the couplings $J_{ij}$ are evenly distributed about 0. As a consequence, magnetization cannot serve as the order parameter for the spin glass transition. However, since the spin glass phase is characterized by the freezing of spin fluctuations, spins remain correlated in time over arbitrarily long durations. Thus, in the spin glass phase, configurations at two widely separated time points maintain a high overlap with each other, unlike those in the paramagnetic phase in which spins decorrelate rapidly. The configurational overlap $Q$ therefore serves as the order parameter and is defined as
\begin{equation}
Q = \lim_{t \rightarrow \infty} \sum_{i=1}^{N} \langle S_i(t_0)S_i(t_0 + t) \rangle_{t_0}
\end{equation}
Unlike spin glasses, structural glasses do not possess quenched disorder \cite{karmakar2013random}. Nonetheless, the structural glass transition is also characterized by ergodicity breaking and the configurational overlap can therefore distinguish between the amorphous and supercooled liquid phases. Intuitively, a field that is conjugate to Q can promote a state with large Q and thereby induce a glass transition at higher temperatures. To demonstrate such an effect, Franz and Parisi considered a system comprising of two copies, or replicas $x$ and $y$ of the same glass-former. In the absence of coupling, the two copies are governed by their respective Hamiltonians $H(x)$ and $H(y)$. An asymmetric coupling is then introduced via the field $\epsilon$ in the following way. The Hamiltonian of $y$ is unaffected by $x$. On the other hand, the Hamiltonian of $x$ is perturbed to 
\begin{equation}
H_{\epsilon}(x|y_i) = H(x) - \epsilon Q_{xy_i}
\end{equation}
where $Q_{xy_i}$ is the overlap between the configuration in $x$ with a reference configuration $y_i$ drawn from the equilibrium ensemble of configurations of $y$ governed by $H(y)$. The thermodynamic properties of $x$ are then obtained by averaging over independent configurations $y_i$ \cite{franz1997phase}. Franz and Parisi first applied this procedure to the infinite ranged spherical $p-$spin model with $p =$ 4, which exhibits a random first-order transition and therefore shares many similarities with structural glass-formers. By introducing a Landau free energy, or `effective potential' $V(Q)$, they first analysed its behavior with decreasing temperature. At high temperatures, $V(Q)$ has a single minimum at $Q =$ 0 corresponding to the ergodic liquid phase (Fig. \ref{Figure41}A). 
Upon cooling, a secondary minimum at large $Q$ appears at a characteristic temperature $T_c$. This $T_c$ corresponds to the temperature at which the free energy landscape of the liquid breaks up into an exponentially large number of metastable minima. Thus, the emergence of a secondary minimun in $V(Q)$ is synonymous with the mode coupling crossover anticipated within RFOT. A second valuable insight is that for $T<T_c$, the difference between $V(Q)$ at the two minima yields the configurational entropy $s_c(T)$ \cite{franz1997phase}. Berthier and Coslovich have recently utilized this fact to develop a novel numerical protocol for quantifying $s_c(T)$ \cite{berthier2014novel}. Finally, below a second characteristic temperature $T_K$, the secondary minimum at high $Q$ becomes the global minimum, thereby signalling the transition to the non-ergodic ideal glass phase. $T_K$ therefore corresponds to the Kauzmann temperature. 

Applying a field $\epsilon$ biases the system towards large $Q$, which pushes both $T_c$ and $T_K$ to higher values. Franz and Parisi have shown that this results in a first order line on the phase diagram in the $\epsilon-T$ plane that terminates at a critical point (Fig. \ref{Figure41}B). These predictions have been tested in simulations of binary mixtures of soft \cite{berthier2013overlap} as well as hard spheres \cite{parisi2014liquid}. In atomistic simulations, Berthier has plotted the order parameter distribution at a fixed temperature for various $\epsilon$ and shown that as the first order line is traversed, the distribution evolves from an approximate Gaussian peaked at low $Q$ to a bimodal form near the coexistence point and then back to a nearly Gaussian form, but peaked at high $Q$ \cite{berthier2013overlap} (Fig. \ref{Figure41}C). These features are hallmarks of a first order phase transition. Parisi and Seoane have charted the phase diagram for the binary hard sphere system in the $\epsilon-\phi$ plane (Fig. \ref{Figure41}D) and determined the volume fraction corresponding to the mode coupling crossover ($\phi \approx$ 0.56) as well as the ideal glass transition ($\phi \approx$ 0.62) from extrapolated fits. This results is particularly important from the point of view of colloid experiments, for which $\phi$ and not $T$ is the control parameter. A plausible way to realize replica coupling in colloid experiments is to organize multiple harmonic traps using holographic optical tweezers into an equilibrium configuration of the liquid. This arrangement of traps is analogous to the quenched configuration with which the system is biased to have a large overlap with. The effect of the field $\epsilon$ can then be mimicked by the stiffness of the optical traps, which can be tuned readily by modulating the laser power. For large trap stiffness, particles are unable to escape from their traps, and the system resides in the reference configuration for long times. At the other extreme, if the laser is switched off, the system is completely unconstrained ($\epsilon =$ 0), and will eventually decorrelate from the reference configuration. This suggests that at least in principle, one can span a wide range of $\epsilon$ and even cross the expected phase transition in the $\epsilon - \phi$ plane. It would be fascinating to see whether similar experimental protocols can indeed investigate the ergodicity breaking transition predicted in \cite{franz1997phase} using colloids. 

\subsection{Dynamical transitions in trajectory space}
The random pinning glass transition as well as the glass transition in coupled replicas are thermodynamic phase transitions induced by imposing constraints on the number of configurations accessible to the liquid \cite{biroli2013perspective}. By contrast, the DF theory provides the basis for a purely dynamic phase transition to the amorphous state by biasing particle trajectories towards low mobility. According to the facilitation approach, glass-forming liquids are composed of mobile as well as immobile regions and structural relaxation is facilitated by the diffusion of these mobile regions throughout space. Increasing glassiness is a consequence of the decreasing concentration of mobile regions, or excitations. This can also be viewed as an increased proliferation of immobility. Motivated by these ideas, Chandler, Garrahan and coworkers argued \cite{merolle2005space,jack2006space,garrahan2007dynamical} that the glass transition can be viewed as a dynamic order-disorder transition from an `active' (mobile) state to an`inactive' (immobile) phase. It is easy to see that such a transition is not thermodynamic in nature, because the relevant microstates are not configurations, but trajectories. Moreover, in an ergodic liquid, mobile and immobile regions frequently inter-convert, which leads to a uniform mobility at long times. A true phase transition to the amorphous state can be said to have occurred only if some particle trajectories remain immobile over infinitely long times. This is a non-equilibrium extension of the thermodynamic limit for equilibrium phase transitions. Just as equilibrium phase transitions cannot occur in system of finite size, phase transitions in trajectory space cannot occur for finite observation times. Further insight into the nature of the proposed dynamic transition can be gained from the observed phenomenology of glass formation. For instance, knowing that the very existence of dynamical heterogeneity implies a coexistence between mobile and immobile phases, one can expect the putative dynamic transition to be of first order. 

While the initial theoretical work of Chandler, Garrahan and coworkers was confined to kinetically constrained spin models (KCMs), they successfully extended these ideas to the realm of atomistic glass-formers \cite{hedges2009dynamic} using computer simulations of the Kob-Anderson binary mixture of particles interating via the Lennard-Jones potential \cite{kob1994scaling}. Since the two phases separated by this transition differ in mobility, a natural choice of the order parameter is the dynamical activity  
\begin{equation}
K[X(t)] = \Delta t \sum_{t=0}^{t_{obs}} \sum_{j=1}^{N} |\mathbf{r_j}(t+\Delta t)-\mathbf{r_j}(t)|^2
\end{equation}  
Here, $\mathbf{r_j}(t)$ denotes the position of particle $j$ at time $t$, $t_{obs}$ is the observation time and $N$ is the number of particles in the system. The dynamical activity is a functional of $X(t)$, where $X(t)$ represents the time evolution of a point in configuration space. The thermodynamic limit in space-time corresponds to $N \rightarrow \infty$ and $t_{obs} \rightarrow \infty$. The time interval $\Delta t$ is chosen to be the average time taken by a particle to move by one diameter in the liquid phase \cite{hedges2009dynamic} and the sum over $t$ is performed in discrete steps of width $\Delta t$. It is obvious that K[X(t)] decreases with temperature, which is the usual control variable for the glass transition. The insight from Chandler and coworkers was that even at a temperature corresponding to the ergodic supercooled liquid phase, a glass transition can be induced by introducing a field $s$ conjugate to $K[X(t)]$. Essentially, $s$ alters the Botlzmann weight associated with activity $K[X(t)]$ through an exponential term of the form $\text{exp}(-sK[X(t)])$. Clearly, positive $s$ favors small values of $K[X(t)]$, i.e. it biases the system towards immobility. 

\begin{figure}
\centering
  \includegraphics[width=\textwidth]{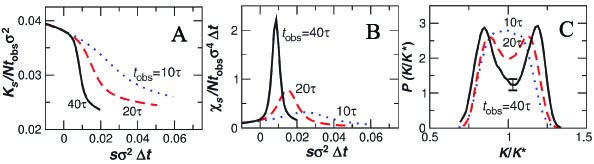}
  \caption{Dynamic order-disorder transition in trajectory space in a simulated Kob-Andersen binary mixture. (A-B) Ensemble averaged order parameter $K_s$ (A) and susceptibility $\chi_s$ (B) normalized by $Nt_{obs}$, as a function of the fictitious field $s$ for various observation times. (C) Order parameter distribution $P(K/K^{*})$ at the coexistence point $s = s^{*}$ for various observation times. In (A-C), $k_BT/\epsilon =$ 0.6, where $k_B$ is the Boltzmann constant and $\epsilon$ is the energy parameter of the Lennard-Jones potential used in the simulations. Adapted from \cite{hedges2009dynamic}.}
  \label{Figure42}
\end{figure}

Within this so called `s-ensemble', the authors showed that glass formation exhibits several hallmark features of a first order transition, albeit in trajectory space. In particular, the ensemble averaged order parameter $K_s = \langle K[X(t)]\rangle_s$ exhibits a crossover from high to low mobility with increasing $s$ that becomes increasingly sharper for longer observation times (Fig. \ref{Figure42}A). This sharpening is analogous to the evolution of finite size effects in equilibrium phase transitions. Near the coexistence regime one should expect large fluctuations in the order parameter, which are captured by the susceptibility
\begin{equation}
\chi_s = -\frac{\partial K_s}{\partial s} = \langle (K[X(t)] - K_s)^2 \rangle_s
\end{equation}
$\chi_s$ indeed exhibits a peak that becomes sharper with $t_{obs}$ and can therefore be used to locate the transition point $s = s^{*}$ (Fig. \ref{Figure42}B). The authors note that unlike in KCMs, where $s^{*} =$ 0, it may have a finite value for atomistic glass-formers \cite{hedges2009dynamic}. In fact, for KCMs with softened kinetic constraints, which are expected to resemble realistic glass-formers, it has been shown that $s^{*}$ is finite for finite temperatures and moreover, the first order line in the $T-s$ plane ends in a critical point \cite{elmatad2010finite}. The authors further demonstrated that at the coexistence point $s^{*}$, the order parameter distribution is bimodal, as one would expect in a conventional first order transition (Fig. \ref{Figure42}C). 

\begin{figure}
\centering
  \includegraphics[width=0.7\textwidth]{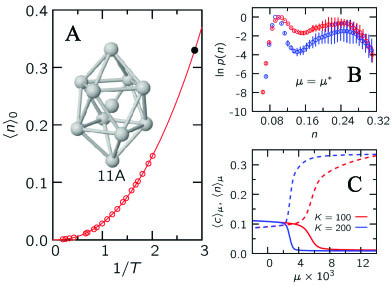}
  \caption{Space-time phase transition driven by tuning local order. (A) Average number of motifs of type 11A, $\langle n \rangle_0$ vs $1/T$. The geometry of 11A clusters is also shown. (B) The distribution of number of 11A clusters $p(n)$ at the coexistence chemical potential $\mu^{*}$ for observation times K = 100 units (red) and 200 units (blue). (C) The average number of 11A clusters (dashed curves) and the concentration of excitations (solid curves) for observation times K = 100 units (red) and 200 units (blue) in the $\mu-$ensemble. Adapted from \cite{speck2012first}.}
  \label{Figure43}
\end{figure}

The foregoing discussion focuses solely on dynamical aspects of glass formation. However, as we discussed in the section on geometric frustration, local structural order can be strongly anti-correlated with mobility. This suggests the possibility that the dynamical transition in trajectory space may in fact be driven by a field that couples to some form of local structural order, rather than the dynamical activity. This intuitive idea was cleverly exploited by Speck, Malins and Royall to demonstrate numerically that dynamic space-time phase transitions can also be driven by tuning structural order \cite{speck2012first}. First, the authors identified locally ordered 11-membered bicapped square anti-prism clusters, termed 11A, using the topological cluster classification scheme \cite{royall2008direct} and showed that in concord with \cite{coslovich2007understanding}, they increase in number on approaching the glass transition (Fig. \ref{Figure43}A).  
Since these clusters are anti-correlated with particle mobility, increasing the number of 11A clusters should have the same role as decreasing the activity $K[X(t)]$. Equivalently, instead of applying $s$, one can increase the chemical potential $\mu$, which alters the Boltzmann weight associated with the number of 11A clusters through the form $\text{exp}(\mu n)$. The authors then examined the behavior of the structural order parameter $\langle n \rangle_{\mu}$ as well as the dynamic order parameter $\langle c \rangle_{\mu}$ in the $\mu-$ensemble. Here, $\langle c \rangle_{\mu}$ is the concentration of excitations, which has been shown to behave like the order parameter for the dynamic order-disorder transition driven by the field $s$ \cite{speck2012constrained}. As seen from Fig. \ref{Figure43}B-C the results for probability distribution of the order parameter as well as the evolution of the order parameter with $\mu$ are qualitatively similar to the results of Chandler and coworkers for the $s-$ensemble (Fig. \ref{Figure42}A \& C). These findings show that space-time phase transitions can be induced by fields that couple to purely structural order parameters, thereby suggesting that local order can lead to the dramatic slowdown of dynamics.

Realization of the field $s$ in experiments is a challenge that is both fascinating and formidable in equal measure. The reason $s$ is much harder to generate experimentally, compared to the static fields employed in random pinning and replica coupling, is that it couples to trajectories rather than configurations. In particular, a positive $s$ biases trajectories towards low mobility. To generate such a field, it is necessary to identify relaxation events and suppress them \textit{before} they take place. It is not immediately clear how this can be achieved in practice. However, we speculate that if precursors to relaxation events can be reliably identified, it might be possible to realize the $s-$ensemble in experiments. For instance, regions of high local Debye-Waller factors, measured over $\tau_{\beta}$ are likely to undergo cage rearrangements over $t^{*}$ \cite{widmer2008irreversible,candelier2010spatiotemporal} and can therefore be used as precursors of relaxation events. In practice, therefore, it should be possible to identify particles with large Debye-Waller factors and selectively pin them using optical tweezers for a short duration. The magnitude of $s$ can be controlled by varying the frequency with which particles are pinned. This procedure requires real time evaluation of local Debye-Waller factors, from which the desired trap configuration is selected. Importantly, the trap configurations must be created before the particles escape from their cages, and hence, this method is likely to be successful only at large volume fractions, for which $\tau_{\beta} << t^{*}$. The processing time can be reduced significantly by applying the field to a subset of the particles in the field of view. This is tantamount to applying a spatially inhomogeneous version of $s$. We also note that Flindt and Garrahan have shown for the East model that information about space-time singularities for finite $s$ in the limit $t_{obs} \rightarrow \infty$ can be obtained by evaluating higher order cumulants of $K[X(t)]$ over short times \cite{flindt2013trajectory}. Thus, all aspects of trajectory space transitions may not even require $s-$ensemble to be realized in experiments. We hope that the ideas discussed above will pave the way for experimental exploration of trajectory space phase transitions. 

\section{Future directions}
The evolution of experiments on the colloidal glass transition can be neatly divided into three distinct periods. The 1990s were dominated by dynamic light scattering experiments aimed at testing predictions of MCT. The 2000s were defined by the increasing dominance of confocal microscopy and the literature in this period is dominated by investigations of dynamical heterogeneity, without particular emphasis on testing predictions of specific theories. On the other hand, experiments in the last few years have predominantly focussed on structural as well as dynamic features in the light of various approaches such as RFOT, facilitation and geometric frustration-based models. As the field stands today, it is well-established that MCT, despite its many successes cannot be the correct theory of the glass transition. Further, every viable theory of glass formation has managed to incorporate broad features of dynamical heterogeneity into its framework, which makes it difficult to distinguish between competing approaches. Further, as we have discussed at some length, nearly every major theory of glass formation has garnered at least some support from experiments on colloids or granular media. Over the last two decades, two factors have in our opinion contributed significantly to a change in the role of colloid experiments in understanding glass formation. First, numerous experiments aimed at addressing a wide range of condensed matter physics problems including glass formation, nucleation, plasticity, epitaxy and friction have repeatedly demonstrated that dense colloidal suspensions are highly useful models to gain intuition about the microscopic underpinnings of a variety of physical processes. As a result, it is no longer surprising if experiments on colloids, particularly those interacting via hard-sphere or screened Coulomb interactions, reproduce the results of numerical simulations. Such studies are no doubt invaluable, since they provide direct experimental evidence for new phenomena, but they do not add much to our physical understanding of glass formation. The second factor is the recent explosion in the development of new theoretical and numerical constructs designed to decode glass formation. Many of these are simply impossible to realize in conventional colloid experiments, although they are clearly vital from the perspective of glass physics. These factors demonstrate that future colloid experiments need to be increasingly innovative and critical in nature. We believe that in the years to come, this approach will chiefly follow two related paths. The first involves the development of new data analysis protocols that can make use of experimentally accessible dynamical crossovers at or below the putative MCT transition volume fraction to compare and contrast the predictions of different theories. The second involves the development of novel techniques that can allow abstract concepts such as replica coupling and the $s-$ensemble to be realized in experiments. With these promising future directions in mind, we list some key open questions that need to be resolved in the near future in order to ascertain the validity of various theoretical formulations. 

\section{Configurational entropy and local order}
\begin{figure}
\centering
  \includegraphics[width=0.5\textwidth]{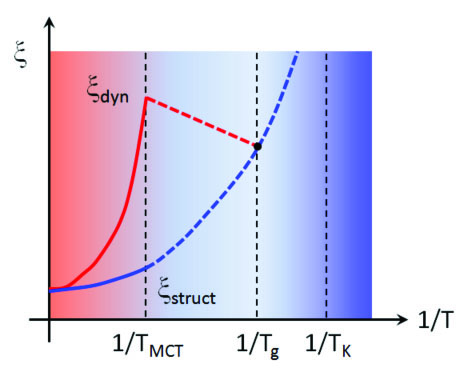}
  \caption{Schematic showing the typical profiles of growing dynamic and structural length scales. Adapted from \cite{royall2015role}.}
  \label{Figure44}
\end{figure}

Within RFOT, the configurational entropy $s_c$ is perhaps the most important quantity associated with glass formation. As mentioned before, the ideal glass transition at $T_K$ is signalled by a vanishing of $s_c$. Indeed, the correspondence between the VFT temperature $T_0$ and $T_K$ extracted from extrapolation of $s_c$ still remains one of the most striking experimental observations in favor of RFOT. RFOT further predicts that the vanishing of $s_c$ should be accompanied by the divergence of a static `mosaic' length scale that signal the onset of long-ranged amorphous order. However, this amorphous order cannot be characterized in terms of a local order parameter, which obfuscates its interpretation in terms of real space structure. In this respect, the geometric approach of Tanaka and coworkers is far more transparent, and provides a direct visual link between the spatial fluctuations of structural order and dynamical heterogeneity. For polydisperse hard spheres, this is reflected in the proportionate growth of $\xi_6$ and $\xi_4$. On the other hand, $\xi_{PTS}$, a measure of the `mosaic' lengthscale, typically shows a much slower growth. Tanaka and coworkers have used this fact to argue that $\xi_6$ and not $\xi_{PTS}$ is the relevant static length scale associated with glass formation, a claim that has also been endorsed by Langer \cite{langer2014theories}. The most obvious objection to this claim is that the correspondence between $\xi_6$ and $\xi_4$ holds only for the special case of polydisperse hard spheres and cannot be generalized. Locally preferred structures must be identified anew for every single glass-former. Even when the appropriate locally preferred structure is identified, the associated static correlation length does not always grow as rapidly as $\xi_4$ \cite{malins2013identification}. Moreover, in the dynamical regime accessible to simulations and colloid experiments, the mosaic picture is obscured by secondary relaxation processes and $\xi_{PTS}$ is therefore not expected to grow at the same rate as $\xi_4$ even within RFOT. The current status of the relationship between structural and dynamic length scales can be best represented by a schematic (Fig. \ref{Figure44}) \cite{royall2015role}. 
From a cumulative analysis of growing length scales described in the literature, the general consensus is that dynamic length scales grow faster than static or structural ones. Moreover, the value of the dynamic length scales at $T_c$ is similar to that at $T_g$. This suggests the possibility that dynamic length scales may exhibit a flattening or even a decrease between $T_c$ and $T_g$. Indeed, the simulations of Kob et al. \cite{kob2012non} and the experiments of Nagamanasa et al. \cite{nagamanasa2015direct} provide strong support to this possibility. On the other hand, theories based on geometric frustration do not present any arguments to explain the observed behavior of dynamic length scales.   

The disparity between $\xi_6$ and $\xi_{PTS}$ in the case of polydisperse hard spheres merits greater scrutiny. Tanaka et al. have reasoned that the proliferation of hexatic bond-orientational order is a reflection of the fact that particles sacrifice configurational entropy and gain vibrational entropy with increasing density \cite{tanaka2010critical}. This is intuitively plausible, and even consistent with the notion of decreasing $s_c$ on approaching the glass transition. However, RFOT suggests that this reduction in $s_c$ should be adequately captured by the growth of $\xi_{PTS}$, which is clearly not the case. We suggest two possible scenarios to reconcile these findings. The first possibility is that $\xi_{PTS}$ simply does not capture the reduction in configurational entropy over a regime in which the mosaic state is not well-defined and hence, $\xi_{PTS}$ and $\xi_6$ should be comparable only in the regime well beyond the mode coupling crossover. The other possibility is that crystallization interferes with glass formation well before the ideal glass transition is reached. This would imply that the Kauzmann entropy crisis is averted by the presence of a lower bound on the degree of supercooling, rather than the presence of a thermodynamic phase transition. Tanaka et al. have stressed quite correctly that increasing $\xi_6$ does not by itself signal crystallization, since polydispersity prevents translational ordering. Nonetheless, the degree of polydispersity (9\%-11\%) used provides rather weak frustration to crystallization and the putative glass phase is necessarily characterized by long-ranged hexatic order. One cannot rule out that such a phase might be possible to generate by perturbing the crystalline phase, in a manner such that the strain produced by polydispersity is relieved by the formation of dislocations. This would then imply that the system samples a part of phase space associated with the crystal, in contradiction with the usual assumption that crystallization is bypassed. In this situation, the rapid growth of $\xi_6$ compared to $\xi_{PTS}$ may be indicative of the fact that crystallization outcompetes glass formation. To test this scenario, it is necessary to perform systematic experiments that study the evolution of $\xi_6$ and $\xi_{PTS}$ with $\phi$ for samples with increasing polydispersity. Both lengthscales have been measured in experiments \cite{tanaka2010critical,nagamanasa2015direct} and the proposed experiments are therefore quite feasible. With increasing polydispersity, one expects frustration to crystallization to increase, which should push the lower metastable limit closer to the ideal glass transition. This would in turn be manifested as a diminishing discrepancy between $\xi_6$ and $\xi_{PTS}$. 

For glass-formers in which the preferred local structure is the icosahedron or the bicapped square antiprism, the proliferating order itself competes with crystallization, which should lead to a scenario different from that for polydisperse hard spheres. For such systems, which include binary mixtures of particles, it would be interesting to tune the frustration by curving space. For 2D systems, this can in principle be achieved in experiments by confining the particles to an air-liquid \cite{zahn1999two} or liquid-liquid \cite{loudet2005capillary} interface. Such experiments could yield valuable results concerning the validity of the theory proposed by Tarjus and coworkers. Another important aspect is the evolution of dynamical heterogeneities on approaching the glass transition. RFOT predicts that cooperative rearrangements adopt a compact form close to the glass transition and are string-like in the mildly supercooled regime. Geometric frustration-based approaches do not anticipate such a crossover. It would therefore be interesting to examine whether this change in morphology occurs, particularly in the low frustration limit for which $\xi_6$ and $\xi_{PTS}$ are decoupled. 

Finally, we note that while different thermodynamic approaches such as random pinning \cite{cammarota2012ideal}, replica coupling \cite{franz1997phase} and the $\mu -$ensemble \cite{speck2012first} all induce a glass transition from an ergodic liquid to a glass, the nature of the glass formed by these approaches is rather different. In random pinning, each pinned site restricts the possible local configurations that neighboring particles can adopt, which leads to a decrease in $s_c$. Replica coupling similarly constrains the system to explore the part of the energy landscapes that is close to a specified configuration. Effectively, this is similar to reducing the configurational entropy. Importantly, neither approach favors the development of specific forms of structural order but amorphous order of the form encoded by $\xi_{PTS}$ is expected to grow, owing to the constraints imposed on the set of allowed configurations. By contrast in the $\mu -$ensemble, local order is deliberately promoted, but no connection to the lowering of $s_c$ is explicitly made. As a result, it would be fascinating to measure $\xi_{PTS}$ and $s_c$ in the $\mu -$ensemble in simulations and compare their behavior with that for random pinning and replica coupling. Further, in the context of random pinning and replica coupling, an important prediction is the existence of a critical point in the low $\phi$ regime at sufficiently high density of pinned particles $c$ or field $\epsilon$, respectively. This critical point corresponds to the point of intersection of curves corresponding to the mode coupling crossover and the ideal glass transition in the $T-c$ or $T-\epsilon$ plane. Experimental verification of the existence of such a critical point would definitely lend more credibility to the underlying theories. It would also be interesting to examine, at least numerically, whether such a critical point exists in the $\mu -$ensemble. On the experimental front, from previous discussions it is clear that of the three procedures, random pinning is the simplest to realize in experiments \cite{gokhale2014growing}. We have argued in the preceding section that replica coupling may also be realizable using holographic optical tweezers. In principle, optical manipulation of colloids may help in realizing an experimental analogue of the $\mu -$ensemble. For instance, it should be possible to organize particles into clusters of a specified geometry. If these clusters are thermodynamically favored, they should persist for long times and influence relaxation. It might therefore be possible to induce a glass transition by tuning the density of such clusters. We hope that this intuition will be tested in future colloid experiments.

\subsection{Does facilitation have a thermodynamic origin?}
As mentioned in a previous section, the theory of dynamical facilitation makes a very strong assumption that the entire phenomenology of glass formation can be attributed to dynamical effects. Moreover, through the sophisticated construct of the $s-$ensemble, it claims that if glass formation is at all associated with a phase transition, it is one that occurs in trajectory space, in contrast to thermodynamic ones that occur in configuration space. Since the theory is purely dynamic in its formulation, it makes no predictions regarding structural evolution. Indeed, one of its foundational principles is that structural evolution is not a necessary condition for glassy dynamics. Regardless of the assumptions of the DF theory, one cannot escape the fact that structural changes do occur in real world glass-formers. These changes are often minuscule, subtle, and difficult to detect, but they are undeniably present. It is therefore worthwhile to ask the question whether facilitation itself is influenced by these structural changes. In fact, in terms of answering whether the glass transition is purely dynamic or thermodynamic in origin, this is probably the most important question, since at present, the DF theory is perhaps the only viable dynamic theory of glass formation.

In conventional phase transitions, dynamics is almost always a slave to structure. Even in the context of glass formation, this argument has been offered to explain the so-called trajectory space phase transition observed by Chandler and coworkers \cite{hedges2009dynamic}. For instance, Berthier has shown in the context of replica coupling that the glass transition driven by the field $\epsilon$ is accompanied by a sudden drop in diffusivity \cite{berthier2013overlap}, suggesting that the thermodynamic phase transition drives the dynamic one. Further, using the $\mu -$ensemble, Speck et al. have shown that the dynamic transition can also be driven by a field that couples to local structural order \cite{speck2012first}. However, there are worthy counter-arguments to these assertions. Garrahan has shown that for the East model, for which facilitation is the de facto relaxation mechanism, the phase diagram in the $T-\epsilon$ plane indeed has a first order line ending in a critical point, but its behavior in the $\epsilon \rightarrow$ 0 limit is very different from that for $p-$ spin models \cite{franz1997phase}, which form the basis for RFOT \cite{kirkpatrick1989scaling}. In particular, the glass transition in the East model at $\epsilon =$ 0 only occurs at $T =$ 0, whereas for $p-$spin models it occurs for $T = T_K$. The validity of these arguments depends on whether real world glass-formers are better approximated by East-like models or $p-$ spin models. For the $\mu -$ensemble, the authors of \cite{speck2012first} have already noted that while the influence of $\mu$ on the dynamic order parameter is similar to that of $s$, the effect on structure is significantly different. In particular, in the $\mu-$ensemble, the inactive glassy phase has a much greater number of 11A clusters. Clearly, the two ensembles sample vastly different configurations and hence, it is not clear whether the glass formed in the $s \rightarrow$ 0 limit is identical to that formed in the $\mu \rightarrow$ 0 limit. Hence, one cannot discern whether facilitation is indeed a slave to structure.

A completely different approach was adopted by Candelier et al. \cite{candelier2010spatiotemporal} to examine whether the spatial occurrence of excitations, or `cage jumps' as the authors called them, is correlated with local structure. Towards this end, the authors employed a numerical construct known as the isoconfigurational ensemble \cite{widmer2004reproducible,widmer2006predicting}. In the isoconfigurational ensemble, the initial configuration of particles is kept fixed across all copies of the system. However, the velocities are selected randomly from the Maxwell-Boltzmann distribution. To investigate whether the chosen configuration influences subsequent dynamics the authors computed local Debye-Waller factor averaged over the isoconfigurational ensemble. As expected from previous studies \cite{widmer2006free,widmer2008irreversible}, the Debye-Waller map is heterogeneous, with some regions having significantly higher Debye-Waller factors than others. The authors found that the spatial organization of cage-jumps is strongly correlated with regions of high local Debye-Waller factor. It has therefore been suggested that analysis of local Debye-Waller factors may provide a purely structural method for identifying excitations \cite{candelier2010spatiotemporal}.  

Some intuition can be gained from experimental studies of facilitation in randomly pinned colloidal glass-formers \cite{gokhale2014growing}. These studies showed that facilitation plays an important role in structural relaxation in randomly pinned glass formers for low $\phi$. However, as discussed before, the peak of the mobility transfer function, a reliable quantifier of the degree of facilitation \cite{elmatad2012manifestations}, decreases at large $\phi$, suggesting that facilitation is superseded by other relaxation processes close to the glass transition. This claim is further supported by data on the partitioning of excitations between core-like and shell-like regions of CRRs. Since the high $\phi$ regime has not been explored in randomly pinned colloidal glass-formers, it is not yet known how important facilitation is in that regime. However, these studies collectively demonstrate that different relaxation mechanisms might be important in different dynamical regimes. Thus, facilitation might dominate dynamics in the random pinning glass transition in the low $\phi$ or high $T$ regime, whereas it might fail in the low $T$ or high $\phi$ regime. Similar studies in the context of replica coupling and the $\mu -$ensemble will shed further light in the relevance of facilitation for structural relaxation on approaching the glass transition. An interesting possibility is that even in a regime where facilitation is the dominant mechanism of structural relaxation, one cannot rule out the possibility that it may be influenced by thermodynamic changes. In the context of random pinning for instance, each pinned particle contributes towards a lowering of the configurational entropy. In doing so, it creates a low mobility zone around itself, in which excitations are extremely unlikely to occur. Thus, the spatial occurrence of excitations depends on the nature of the quenched disorder. This may lead to a fascinating scenario in which facilitation is the dominant mechanism of relaxation, although glass formation is driven by an underlying thermodynamic phase transition. 

\subsection{The growth of static length scales}
An important challenge faced by thermodynamic theories of the glass transition is to provide a link between the evolution of structure and the slowdown of dynamics. This usually amounts to identifying a static length scale that grows on approaching the glass transition. In this context, an important theoretical result by Montanari and Semerjian is that the finite temperature divergence of the relaxation time must be accompanied by a diverging static length scale \cite{montanari2006rigorous}. They have also proven that this static length must grow at least as slowly as $[\text{ln}(\tau_{\alpha})]^{1/d}$, where $d$ is the spatial dimension. This divergence is so slow, that over the typical window of 5-6 decades of relaxation time accessible to simulations and colloid experiments, and for three dimensional systems, the relevant static length scale need not grow by more than a factor of $\sim$ 1.8. We have already discussed two candidates for the putative static length scale governing the glass transition in great detail: $\xi_{PTS}$, whose growth is predicted by RFOT \cite{biroli2008thermodynamic} and $\xi_6$, which is an outcome of growing medium ranged crystalline order \cite{tanaka2010critical}. While $\xi_{PTS}$ grows much slower than $\xi_6$, it comfortably satisfies the Montanari-Semerjian inequality \cite{charbonneau2013decorrelation}. Other static correlation lengths have been proposed and evaluated in simulations. While these lengths are not motivated by particular theoretical frameworks, it would be quite interesting to compute them in experimental systems and compare their growth on approaching the glass transition with $\xi_{PTS}$ and $\xi_6$. We briefly discuss some of these length scales below. 

\subsubsection{Patch correlation length} 
The basic concepts underlying the patch correlation length are best understood in the context of a polycrystalline material. Consider a polycrystal composed of domains of size $R_d$ on average. The orientation of crystallographic axes is uniform within a domain and varies discontinuously across adjacent domains. Consider a patch of size $R$ centred on some arbitrary point within the crystal. The idea is to enumerate the distinct types of patches of size $R$ present in the polycrystal and their associated probabilities of occurrence. It is easy to see that for $R<R_d$, the most numerous patches are those located within the bulk of the crystalline domains. These can easily be transformed into each other through appropriate rotations. Hence, for $R<R_d$, the number of distinct patches, quantified by the patch entropy $S(R)$ does not scale with the patch volume $V(R)$. For $R>R_d$ the contribution to $S(R)$ from interfaces between patches become important and hence, $S(R) \propto V(R)$. At the average domain size $R_d$, there is a crossover from sub-extensive to extensive dependence of $S(R)$ on $V(R)$. $R_d$ therefore captures the extent of spatial correlation within the system. Kurchan and Levine argued \cite{kurchan2011order} that the same intuition can be applied to glass-forming liquids. They established a procedure to identify the patch entropy and quantified the patch correlation length for simulated glass-formers. Further, they compared this length scale with $\xi_{PTS}$ and $\xi_6$. They found that the patch correlation length scales, $\xi_6$ as well as $\xi_{PTS}$ exhibit nearly identical scaling for their model system. The advantage of the patch correlation length is that unlike $\xi_6$, it is `order-agnostic', meaning that it does not require one to define a local order parameter. Indeed, as Sausset and Levine have shown, the presence of hexatic order can be inferred from multiplicity of patches of a given type \cite{sausset2011characterizing}. The method is simple enough to apply to experimental systems, particularly since it only requires time-averaged snapshots of particle configurations, which can easily be obtained from video microscopy experiments. It would be most interesting to compute this length for a slightly polydisperse hard sphere system, for which Tanaka and coworkers have shown that $\xi_6$ grows much faster than $\xi_{PTS}$ \cite{russo2015assessing}. 

\subsubsection{Crossover lengthscale from local plasticity to asymptotic elasticity}
A different proposal for growing static correlations was offered by Karmakar, Lerner and Procaccia \cite{karmakar2012direct}. It is known from experiments on atomic \cite{buchenau1984neutron} and colloidal \cite{chen2010low,ghosh2010density} systems that the density of states of glasses exhibits excess low frequency modes over and above the Debye contribution. These modes are associated with localized plastic rearrangements. Building on prior work \cite{dauchot2011athermal,hentschel2011athermal}, the authors of \cite{karmakar2012direct} demonstrated that the low frequency tail of the density of states can be expressed in terms of eigenvalues of the Hessian matrix, in the following form
\begin{equation}
P\Bigg( \frac{\lambda}{\lambda_D} \Bigg) \approx Nd \Bigg[ \hat{A}\Bigg( \frac{\lambda}{\lambda_D} \Bigg)^{\frac{d-2}{2}} + B(T)f_{pl}\Bigg( \frac{\lambda}{\lambda_D} \Bigg) \Bigg]
\end{equation}
Here, $\lambda_D$ is the Debye cut-off. The first term on RHS represents the Debye contribution whereas the second term corresponds to plastic modes. The central idea is that for small system sizes $N$, the minimum eigenvalue of the Hessian will be determined by the plastic modes whereas for large system sizes, it will be determined by elastic modes. The crossover between these two regimes corresponds to a structural length scale. Since the excess low frequency plastic modes emerge from disorder, the crossover length quantifies the extent of spatial correlations of disorder and is expected to grow on approaching the glass transition. The authors showed that this is indeed the case \cite{karmakar2012direct}. Moreover, the authors demonstrated that this length scale is identical to the one obtained from finite size scaling of configurational entropy \cite{karmakar2009growing}. Since methods for extracting the density of states of colloidal glass-forming from particle displacements have already been developed \cite{chen2010low,ghosh2010density}, estimating this length scale in a colloidal system should be straightforward. In an important computational study, Karmakar, Biroli and Procaccia have shown that the length scale obtained from the minimum eigenvalue of the Hessian scales identically to $\xi_{PTS}$ \cite{biroli2013comparison}, a result that begs experimental validation.

\subsubsection{Mutual information length}
Another static length of interest to experimentalists was defined by Royall and coworkers \cite{dunleavy2012using} using concepts from information theory \cite{cover2012elements}. The basic idea is that if structural correlations in the system extend over a lengthscale $\xi$, then knowing the structure within a small region of the system, one can, at least to some extent predict the structure of another region located within a distance $\xi$. In other words, prior knowledge of local structure in one region reduces the uncertainty in determining the local structure in a neighboring region. In information theoretic terms, the mututal information of two regions is large if they lie within a distance $\xi$ of each other. Formally, if X and Y are two regions in the glass-former, the mutual information between them is given by 
\begin{equation}
I(X,Y) = H(X) - H(X|Y) = H(Y) - H(Y|X)
\end{equation}  
Here, $H(Z) = -\sum_{z\in Z} p(z)\text{log}_2(z)$ is the Shannon entropy \cite{shannon1948note} of random variable $z$ defined over domain $Z$ and having a probability distribution $p(z)$. Similarly, $H(Z|W)$ is the Shannon entropy of the conditional probability distribution of the variable $z$ given $w \in W$. The Shannon entropy measures the uncertainty in drawing a particular value from a distribution. Broad distributions have a larger Shannon entropy compared to strongly peaked distributions. The above equation therefore measures the lowering in uncertainty, or equivalently gain in information about the configuration in patch X given the configuration of patch Y or vice versa. The authors computed $I(X,Y_d)$ for square patches X and Y separated by distance $d$ and defined the mutual information length $\xi_{mi}$ as
\begin{equation}
\xi_{mi} = \frac{\sum_d dI(X,Y_d)}{\sum_d I(X,Y_d)}
\end{equation} 
The authors offer qualitative arguments suggesting that $\xi_{mi}$ should scale as $\xi_{PTS}$ and the patch correlation length. Once again, it would be fascinating to test this conjecture for the polydisperse hard sphere system. In view of numerical data published so far, we surmise that for this system, $\xi_6$ and the patch correlation length will exhibit identical scaling, whereas $\xi_{mi}$ and $\xi_{PTS}$ will grow much slower. Experiments that compare these length scales will play a crucial role in determining whether different lengthscales embody fundamentally different physics or whether they are simply distinct probes of the same phenomenon.  

\subsection{Ultrastable glasses}
We end this section by a discussion on a burgeoning research avenue of that has attracted a lot of interest in recent years. This topic deals with the preparation and characterization of ultrastable glasses, i.e. glasses that have exceptionally high kinetic and thermodynamic stability. This section is different from the rest of the review in that it focuses on the non-ergodic amorphous phase rather than the supercooled liquid. From our discussion of alternate routes to glass formation, it is evident that different approaches can lead to the formation of qualitatively distinct glassy states. One hopes that by analysing the stability of these states, one might be able to gain insights into the underlying physics of glass formation. With this aim in mind, we provide a brief overview of experimental and numerical research on ultrastable glasses and suggest new colloid experiments that might further our understanding of this field. 

\begin{figure}
\centering
  \includegraphics[width=\textwidth]{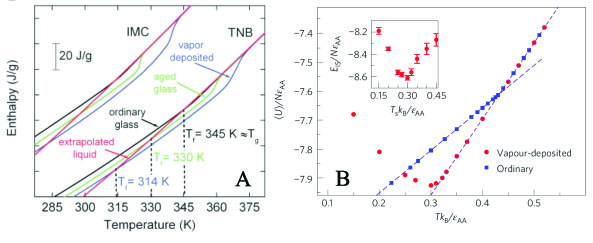}
  \caption{Ultrastable glasses. (A) Enthalpy vs temperature for glasses composed of 1,3-bis-(1-naphthyl)-2-(5-naphthyl)benzene (TNB) and indomethacin (IMC). Different curves correspond to vapor-deposited, aged and ordinary glasses as indicated. The extrapolated curve for the ergodic liquid is also shown. Adapted from \cite{swallen2007organic}. (B) Potential energy of ordinary glasses obtained by cooling at a rate of 3.33 $\times$ 10$^{-7}$ in Lennard-Jones units (blue) and glasses prepared by vapor-deposition at substrate temperatures corresponding to points on the X axis (red). Inset shows the inherent structure energy of vapor-deposited glasses as a function of substrate temperature. Adapted from \cite{singh2013ultrastable}.}
  \label{Figure45}
\end{figure}

Conventionally atomic and molecular glasses are prepared by the liquid phase sufficiently rapidly to bypass crystallization. However, glasses formed by this route are not very stable, meaning that they are kinetically trapped in high energy minima of the potential energy landscape. From a theoretical perspective on the other hand, it is highly desirable to prepare glassy states with low energy, since these states are likelier to shed light on long standing issues such as the Kauzmann entropy crisis. The problem is that particles in glasses prepared by cooling simply do not have enough energy to escape from one energy minimum and explore deeper ones. Transitions to minima with lower energy certainly occur during ageing, but this process is exceedingly slow and therefore of little practical relevance. Accessing low lying states therefore remained a considerable challenge. In a vital experimental breakthrough, Ediger and coworkers circumvented this hurdle by forming a molecular glass through vapor deposition rather than cooling \cite{swallen2007organic}. In particular, the authors vapor-deposited molecules on a substrate held at a fixed temperature $T$ below the conventional laboratory glass transition $T_g$, such that $T/T_g \approx 0.85$ and quantified the kinetic and thermodynamic stability of the resulting form using differential scanning calorimetry. From the temperature dependence of the heat capacity, the authors observed that the onset temperature was much higher for vapor-deposited glasses compared to ordinary ones. This implies that vapor-deposited glasses need to be heated to a higher temperature in order to escape from their vitreous state and are therefore more kinetically stable. To quantify thermodynamic stability, the authors defined a fictive temperature $T_f$ defined as the point of intersection between the extrapolated enthalpy of the supercooled liquid and the enthalpy of the glass. $T_f$ is therefore indicative of how low the glass is on its energy landscape. Based on $T_f$, the authors defined a figure of merit
\begin{equation}
\theta_{K} = \frac{T_g - T_f}{T_g - T_K}
\end{equation}
where $T_K$ is the Kauzmann temperature. $\theta_{K}=$ 0 corresponds to an ordinary glass whereas $\theta_{K}=$ 1 corresponds to the lowest position on the energy landscape. As shown in Fig. \ref{Figure45}A, even aged glasses possess a substantially higher $T_f$ and therefore lower $\theta_{K}$ compared to vapor-deposited glasses. 

Ediger and coworkers attributed the high kinetic and thermodynamic stability of vapor-deposited glasses to enhanced mobility near the surface of the growing glass film, which allows molecules to explore configuration space and thereby reach low energy states. Further, for high deposition rates, surface molecules do not get enough time to reorganize before getting embedded in the bulk of the glass by the incoming molecules. Also, at very low substrate temperatures molecular diffusivity is low and molecular motion is too slow to efficiently access stable states. This suggests a non-monotonicity in the dependence of thermodynamic stability of these so-called `ultrastable' glasses on the substrate temperature. Such non-monotonicity was indeed observed in computer simulations of vapor-deposited glasses \cite{singh2013ultrastable}. Interestingly, as shown in Fig. \ref{Figure45}B, the potential energy as well as the inherent structure energy (inset) of the simulated glass exhibit a minimum when the substrate is kept at the Kauzmann temperature. 

Another important finding is that ultrastable glasses deposited at substrate temperatures close to $T_K$ exhibit negligible hexatic bond-orientational order correlations. By contrast, ordinary glasses exhibits clusters of particles with high local bond-orientational order \cite{singh2013ultrastable}. If the maximum in stability at $T_K$ is related to the structure of the glass formed, one expects an RFOT-like scenario at deep supercooling. Further, the lack of bond-orientational order in ultrastable glasses raises concerns regarding whether $\xi_6$ captures the true amorphous correlations in supercooled liquids or is in fact influenced by crystal-like minima in the energy landscape. Interestingly, Jack et al. numerically generated extremely stable glassy states using the $s-$ensemble \cite{jack2011preparation}. As noted earlier in our comparison the $s$ and $\mu-$ensembles, locally preferred structures occur with much lower frequency in the $s-$ensemble \cite{speck2012first}. Combining these two observations, it appears as if vapor deposition, which allows structural reorganization, and the $s-$ensemble, which biases particle trajectories, are both efficient techniques to access deeper regions of the energy landscape. Further critical comparisons between glasses formed by these procedures are likely to reveal valuable clues about the correct theory of glass formation. In this context, it is worth discussing yet another approach for preparing ultrastable glasses, namely random pinning \cite{hocky2014equilibrium}. In this route, glasses are prepared by equilibrating a supercooled liquid at low temperatures and then crossing the liquid-glass boundary by pinning a fraction of particles within the equilibrium configuration of the liquid. The authors showed that on heating, these amorphous states melted at higher temperatures, confirming their kinetic stability. Moreover, the glass obtained on re-cooling had a higher energy than the one prepared by pinning particles. This shows that random pinning also provides a viable route to preparing thermodynamically stable glassy states. The preparation and characterization of ultrastable glasses is a promising avenue for future colloid experiments. In the context of vapor deposition, the major difference between colloidal and molecular systems is that the control variable is volume fraction, rather than temperature. However, the experiments of Ediger and coworkers \cite{swallen2007organic} have shown that the key factor is the balance between the deposition rate and surface mobility. The deposition process itself amounts to sedimentation of colloids and its rate can be controlled easily by increasing the concentration of colloidal particles or their density difference with the fluid in which they are suspended. Preliminary experiments should therefore first test the stability of glasses formed at various deposition rates by examining local structural and dynamic features. Finding the colloid analogue of the substrate temperature and thereby tuning surface mobility presents a more fascinating challenge that we hope will be tackled in the coming years. In the context of random pinning, preparation of glassy states is certainly possible, but studying the melting kinetics of such states would require innovative protocols. Using temperature-sensitive size-tunable poly N-isopropylacrylamide (PNIPAM) particles to mimic temperature quenches is a time-tested strategy routinely employed by colloid experimentalists over the years. However, these particles are extremely difficult to trap using optical tweezers, since their refractive index nearly matches that of water, the solvent in which they are typically suspended. A promising way of overcoming this difficulty is to use composite particles with a polystyrene core and a PNIPAm shell \cite{dingenouts1998observation,lu2006preparation,crassous2006thermosensitive}. Sustained efforts along these lines will certainly pave the way for novel and illuminating colloid experiments that provide definitive answers to the glass puzzle. 

We conclude our discussion on ultrastable glasses by underscoring their relevance to the connection between the jamming \cite{liu1998nonlinear,o2002random,van2009jamming} and glass transitions. Within an RFOT-like scenario, the energy landscape of glasses consists of smooth meta-basins separated by large energy barriers. Such glasses when compressed sufficiently undergo a jamming transition at which the system attains mechanical equilibrium. This jamming transition is characterized by critical scaling laws \cite{van2009jamming} and the emergence of soft vibrational modes \cite{o2002random,brito2009geometric}. These features suggest the existence of a multitude of small barriers in the energy landscape, a picture inconsistent with the smooth landscape that emerges within RFOT \cite{charbonneau2014fractal}. The resolution lies in the fact that deep within the stable glass phase, metabasins in RFOT-like systems undergo a roughening transition and develop a fractal hierarchy of basins within basins, analogous to the spin glass phase transition in the Sherrington-Kirkpatrick model \cite{sherrington1975solvable,parisi1980sequence}. This roughening transition was first discovered in spin models by Gardner \cite{gardner1985spin}. Charbonneau et al. have shown that in the limit of infinite dimensions, hard sphere glasses also undergo a Gardner transition that separates the RFOT-type equilibrium glass phase from the marginally stable jammed phase \cite{charbonneau2014fractal}. Detecting the Gardner transition in experiments is challenging because it is unclear whether it even exists in finite dimensions \cite{urbani2015gardner}. However, a recent study has detected signatures of the Gardner transition in numerical simulations \cite{charbonneau2015numerical}, albeit for the somewhat unrealistic mean field Mari-Kurchan hard sphere system \cite{mari2011dynamical}. The study suggests that perhaps the most straightforward way of detecting the Gardner transition in experiments is to observe the divergence of the beta relaxation time $\tau_{\beta}$, although this divergence may be rounded off in finite dimensions by activated hopping processes. Yet another numerical work suggests that fluctuations in elastic moduli diverge at the Gardner transition, which opens the pathway for rheological determination of the Gardner point \cite{biroli2016breakdown}. The feasibility of these suggestions is unclear, since all of these signatures are expected to be evident only for systems located deep in the stable glass phase, which is prohibitively difficult to prepare by conventional means. The optimistic view, however, is that the preparation of ultrastable glasses may help surmount this hurdle and facilitate the investigation of the Gardner transition in experiments. In an exciting recent development, the first experimental signatures of the Gardner phase have been observed, albeit in a non-equilibrium granular glass \cite{seguin2016experimental}. We expect these results to catalyse extensive efforts in detecting and characterizing the Gardner transition in colloid experiments.  

\section{Conclusions}
In summary, we have reviewed experimental research on the colloidal glass transition with a special emphasis on colloids as test beds to gauge the validity of competing theories of glass formation. For a number of years the foremost goal of colloid experiments had been to provide experimental evidence for various microscopic predictions of theoretical and numerical studies. As a result, these experiments have not only elucidated the nature of spatially heterogeneous dynamics but also provided support for various theoretical formulations such as the mode coupling theory, the random first-order transition theory, dynamical facilitation and geometric frustration-based models. In recent times, however, it has become increasingly evident that in order to further our understanding of glass formation, we need to go beyond testing individual theories in isolation and must adopt a critical comparative approach. The limited dynamical range available to colloid experiments can potentially limit the feasibility of this approach. However, the discovery of new dynamic crossovers in the vicinity of $\phi_c$ has allowed colloid experiments to circumvent this difficulty. A recent set of experiments have exploited the crossover in the morphology of CRRs to ascertain the relative importance of facilitation and collective hopping on approaching the glass transition. These results tentatively point towards a thermodynamic origin of the glass transition. The need for critical tests of various theoretical formulations have also fuelled the search for alternate routes to glass formation such as random pinning, the $s-$ensemble and replica coupling. We have reviewed certain landmark studies that have investigated these approaches \textit{in silico}. Experimental realization of these approaches and their subsequent utilization for critical assessment of competing frameworks provides some of the most daunting and exciting challenges for future colloid experiments. Although research on colloidal glasses has a long history, the field has witnessed a transition from an exploratory to a critical outlook only recently. Keeping in mind this burgeoning nature of critical experiments on the colloidal glass transition, we have concluded our article with a comprehensive overview of some of the most cutting edge open questions and research problems in the field, including the growth of static length scales and the preparation of ultrastable glasses. Through this section, we have aimed to highlight the fact that the canonical glass transition problem provides ample scope for colloid experimentalists to exercise their creative skills in instrumentation as well as data analysis. Our focus on theoretically oriented subject matter rather than experimental techniques is deliberate. Our objective behind this choice is twofold. On one hand, we have attempted to provide experimentalists with a summary of the conceptual problems that are most relevant to solving the glass transition problem. On the other, our exposition on experiments using holographic optical tweezers is aimed at acquainting theoretical and computational scientists with a detailed understanding of the potential of colloid experiments. Ultimately, we hope that our review will encourage greater interactions between theoreticians and experimentalists and will thereby culminate in a cohesive and holistic research effort aimed at obtaining a final resolution to the glass transition problem, if one indeed exists.    

\section*{Acknowledgements}
We thank K. Hima Nagamanasa for her invaluable contributions to our collaborative effort in this field. We also thank Chandan Mishra for fruitful collaborations. S.G. thanks the Council for Scientific and Industrial Research (CSIR), India,for a Shyama Prasad Mukherjee Fellowship and the Department of Science and Technology (DST), India for financial support. R.G. thanks the International Centre for Materials Science (ICMS) and the Sheikh Saqr Laboratory (SSL), Jawaharlal Nehru Centre for Advanced Scientific Research (JNCASR) for financial support and A.K.S. thanks DST, India, for support under the J.C. Bose Fellowship.

\bibliography{references_AdP}
\bibliographystyle{tADP}
\end{document}